\documentclass[twocolumn,showpacs,preprintnumbers,amsmath,amssymb]{revtex4}

\usepackage{graphicx}
\usepackage{dcolumn}
\usepackage{bm}

\newtheorem{property}{Property}

\begin{document}

\preprint{}

\title{Statistical thermodynamics for choice models on graphs}

\author{Arkadiusz Majka}
 \email{majka@icm.edu.pl}
\affiliation{Interdisciplinary Centre for Mathematical and Computational Modelling, University of Warsaw, Pawi\'nskiego 5a, PL-02-106 Warszawa}
\author{and\\ \vspace{3mm} Wojciech Wi\'slicki}
 \email{wislicki@fuw.edu.pl}
\affiliation{Interdisciplinary Centre for Mathematical and Computational Modelling,\\ University of Warsaw,~Pawi\'nskiego 5a, PL-02-106 Warszawa\\ and \\
A. So\l tan Institute for Nuclear Studies, Laboratory for High Energy Physics, Ho\.za 69, PL-00-681 Warszawa}

\date{\today}

\begin{abstract}
Formalism based on equilibrium statistical thermodynamics is applied to communication networks of decision making individuals.
It is shown that in statistical ensembles for choice models, properly defined disutility can play the same role as energy in statistical mechanics.
We demonstrate additivity and extensivity of disutility and build three types of equilibrium statistical ensembles: the canonical, the grand canonical and the super-canonical.
Using Boltzmann-like probability measure one reproduce the logit choice model.
We also propose using $q$-distributions for temperature evolution of moments of stochastic variables.
The formalism is applied to three network topologies of different degrees of symmetry, for which in many cases analytic results are obtained and numerical simulations are performed for all of them.
Possible applications of the model to airline networks and its usefulness for practical support of economic decisions is pointed out.
\end{abstract}

\maketitle

\section{\label{sec:level1}Introduction}

Models used in demand analysis of transportation and communication problems usually assume that demand represents the result of decisions of each individual in the population.
These decisions consist of choices made among finite sets of possibilities.
As an example of a sequence of choices, in the context of airline transportation demand, consider decision to be taken by potential air passenger.
If the passenger wills to fly from the origin port $O$ to destination port $D$, he has to choose among offers of carriers, accounting for many factors relevant for decision, as e.g. scheduled departure and arrival times and how they relate to his needs, flight durations, flight fares, numbers of stops and changes on the route $(OD)$, availability of seats in first class, probability of delays, declared and expected quality of service and many other aspects, too numerous to itemize.
In order to quantify such process of decision making, one incorporates discrete choice models in hope of better understanding and predicting behaviour of such complex system as transportation network and thus obtaining hints for marketing and revenue management decisions.

One of the key concepts used in decision theory, dating back to the XIX-th century economy by Jeremy Bentham and other representatives of so called utilitarians economy, is the concept of {\it utility}, which had been further evolving since that time and found its more rigorous utterance in the classical work of von Neumann and Morgenstern \cite{neumann1} and in neoclassical economic theory (cf. e.g. refs \cite{hennings1}, \cite{bierlaire1}). 
Utility function ascribes real numbers to alternatives and so defines the preference order on the choice set and puts the choice making process on more rational footing. 

An approach to communication based on discrete choice models has long history and abundant literature (cf. e.g. refs \cite{benakiva1}, \cite{louviere1}).
Utility function in neoclassical economic theory is defined axiomatically.
In order to use it in specific applications of choice models, one has to postulate a lot from outside of the theory and to find an effective way to either derive it from any fundamental or general theory or to parametrize it and estimate from data.
Derivation of the utility function from first principles represents major theoretical problem because of lack of fundamental theory of human behaviour.
Normally, one has to rely on partially justified models using power series approximations for utility functions, often restricted to linear or quadratic terms (cf. ref. \cite{dobson1}), and finding empirically relevant variables and estimators for parameters, if any data are available at all.
Apart from those problems, numerous conceptual difficulties arise when applying utility functions to quantify the behaviour of the choice makers and to predict them.
In particular, classical Bentham's approach of maximizing the overall utility, integrated over individuals, when taking market decisions, is often a subject of serious objections \cite{straffin1}.

Being aware of a long-lasting debate around the meaning and the role of utility functions in economy, social sciences and, more generally, in game theory, in this paper we try to incorporate it into economic games on somewhat different way.
First, extending from the concept of utility, we find an analogy between the utility function, as it is used in choice models, and the energy function for physical systems.
After specifying the system, consisting of an airline network and a set of passengers, and its state space, we discuss assumptions normally taken in theories of systems in thermal equilibrium and build quasistatic probability measures on the state space for equilibrium ensembles.
This allows us to derive complete classical thermodynamics for the communication network.
We consider the canonical and the grand canonical statistical ensembles for systems with fixed and variable numbers of clients.

As a possible extension of the model, we propose to consider also the network topology and the number of network vertices as a multidimensional random variable and formulate generalized canonical ensemble where the existence of any connection between sites may fluctuate and is represented by a binomial random variable.
In the frameworks of the canonical and the grand canonical formalisms, we find some analytical results for thermodynamics of particular network topologies and perform numerical simulations.
We study thermodynamical potentials, response functions and correlations between extensive variables in the system. 
The case of random topology is discussed in general terms only.
It needs more study of a couple of specific and self-contained problems and this, altogether with numerical results, is relegated to future work.
 
We also develop another approach, based on so called $q$-distributions or {\it escort} distributions, which allows us to monitor fluctuations and correlations of interesting characteristics of the network, their temperature evolution and phase relations in the network.

Building our formalism we keep in mind its specific application to networks of airline traffic and often use its terminology, although we believe that this framework and most of its features are general enough to be used for other network-related problems, as various kinds of telecommunication, communications in social groups, traffic routing, energy networks, etc.
On the other hand, the model we build does not account for many features of real airline markets, either because of lack of knowledge about utility functions or simplifications of the model at this stage.

\section{\label{sec:level2}Assumptions and fundamentals of the model}

\subsection{\label{ssec:level1}Axiomatics of utility}

In choice models considered here we assume that decision makers are individuals and that these individuals are independent of each other.
Such models are usually called {\it disaggregate}.
Assumption on independence of decisions between individuals is rather poorly justified and often just not true in reality.
But for many practical applications this assumption is not restrictive and depends on further details of the model, in particular on the utility function.
In many cases, also in this paper, one may redefine the notion of a decision maker and consider a group of individuals as the decision maker, without loss of generality, and only mutual independence of such defined decision makers is relevant.
This assumption can be further criticized because the ways decisions are taken by groups, even for very restricted class of consumer decisions considered here, certainly depend on the size and internal communication structure of groups and are qualitatively different than in case of individuals. 
Thus the utility function for aggregated decision makers may differ from that of individuals.
In order to give full account of these diffences in general case, which we do not pretend to do in this paper, one has to consider many additional effects, as internal negotiations inside groups, their dynamics, possible splits, compactness of groups etc.
However, at this stage of model development and for the case of limited number of variables relevant for utility, and for reasonably homogeneous groups, we feel it is fair to assume that utilities for all decision makers are independent and identically distributed stochastic variables.
In other words, we assume high degree of {\it decision coherence} of groups, in the sense that decisions of aggregated individuals do not differ significantly from that of real individuals.

Analysis of the choice requires knowledge of the complete set of options disponsible for the choice maker.
The set of options, or alternatives, $a,b,c,\ldots$ is called {\it the choice set} ${\cal C}=\{a,b,c,\ldots\}$.
The complete set of all possible options available to any individual is called {\it the universal choice set} and a subset of the universal choice set considered by a particular individual is called {\it the reduced choice set}. 
If ${\cal C}$ is the universal choice set then all possible reduced choice sets correspond to events and form what is called {\it $\sigma$-algebra over ${\cal C}$} in the probability calculus.
Hereon we assume that results of a choice performed by any decision maker are unique for given specification of the choice set.
This means that results of a choice are those elements of $\sigma$-algebra which correspond to elementary events of the probability theory.
Finally, we consider only {\it discrete choice models}, i.e. those for which their choice sets are finite and all options can be explicitly enumerated.

Each alternative of the choice set can be characterized by a set of {\it attributes}.
Generally, attributes are random variables in the sense used in probability theory, i.e. they are funcions defined on reduced choice sets with values in other sets.
We consider both directly observable attributes and functions of attributes.

In order to construct decision rules, the decision maker has to be able to compare alternatives.
In neocalassical economy \cite{bierlaire1} two alternatives $a$ and $b$ are comparable using {\it the preference-indifference} operator $\succeq$ which orders ${\cal C}$ linearly, i.e.
\begin{enumerate}
\item The $\succeq$ is reflexive:
\begin{eqnarray}\forall_{a\in{\cal C}} \quad a\succeq a, \label{eq1} \end{eqnarray}
\item The $\succeq$ is transitive:
\begin{eqnarray}\forall_{a,b,c\in{\cal C}} \quad (a\succeq b \; \land \; b\succeq c) \Rightarrow (a\succeq c), \\ \nonumber \label{eq2} \end{eqnarray}
\item Any two alternatives are comparable:
\begin{eqnarray}\forall_{a,b\in{\cal C}} \quad a\succeq b \; \lor \; b\succeq a. \label{eq3} \end{eqnarray}
\end{enumerate}
Since the choice set ${\cal C}$ is finite, the existence of the most preferred alternative $a^{\ast}$ is guaranteed
\begin{eqnarray}
\exists_{a^{\ast}}\; \forall_{a\in{\cal C}} \quad a^{\ast}\succeq a. 
\label{eq4}
\end{eqnarray}
Because of the properties (\ref{eq1}-\ref{eq3}) there exists finite random variable $U:{\cal C}\longrightarrow \mathbb{R}^1$, refered to as {\it the utility function}, such that
\begin{eqnarray}
\forall_{a,b\in {\cal C}} \quad a\succeq b \; & \Leftrightarrow & \; U(a)\geq U(b).
\label{eq5}
\end{eqnarray}
From eq. (\ref{eq4}) it follows that the most prefered alternative $a^{\ast}$ has the largest utility $U$
\begin{eqnarray}
\forall_{a\in{\cal C}} \quad U(a^{\ast})\ge U(a)
\label{eq6}
\end{eqnarray}
or the smallest {\it disutility} $\bar{U}=-U$.
Hence, important property of $\bar{U}$
\begin{property}
Disutility function is bounded from bottom.
\end{property}

\subsection{\label{ssec:level1}Specification of the system}

Our system ${\cal S}$ consists of a complex communication network ${\cal G}$, represented by a set of directed graphs, and set of decision makers.
Nodes of graphs in ${\cal G}$ represent airports and its directed edges correspond to air connections between them.
Each ordered pair $(OD)$ of the origin $O$ and destination $D$ ports and set of all directed paths, called {\it routes}, connecting $O$ and $D$, excluding loops, constitute the $k$-th {\it market} and is represented by the graph ${\cal G}_k$ ($k=1,\ldots,M$).
The whole network ${\cal G}$ is the sum of market graphs 
\begin{eqnarray}
{\cal G}={\cal G}_1\cup\ldots\cup{\cal G}_M.
\label{eq6_1}
\end{eqnarray}
Using graph-theoretic terminology, markets are represented by sets of Hamilton paths, i.e. no one site on the route is visited more than once. 
So defined markets can overlap.
The system is further specified by defining all features relevant for its state as sets of potential passengers (decision makers) for each market, resources of aircrafts, flight schedules, flight fares, etc.
{\it The state} of the system is defined using results of choices performed by all decision makers, which means that the state-space consists of all possible distributions of all decision makers for all markets of the network.

We assume quasistatic approximation, meaning that any time variation of the system as a whole is slow compared to the time of individual decision.
Moreover, specifying the state of the system, we accumulate decisions taken by individuals in time interval at least an order of magnitude shorter than the system time scale, but not necessarily short compared to individual decision time.
In airline practice, one day is normally a good time bin for accumulation of decisions.

In present model we do not embed the network in any metric space, such that only the topology, or connectivity scheme, is relevant and not the geometry.
Accounting for geometry may appear necessary along with sofistication of our model for utility.

For each market $k$ we define the total {\it market disutility} as a sum of disutilities of all passengers from this market $\bar{U}_{i_k}$ where $i_k$ stands for index of the $i$-th passenger in the $k$-th market $(i_k=1,\ldots,I_k)$
\begin{eqnarray}
\bar{U}_k=\sum_{i_k=1}^{I_k}\bar{U}_{i_k}.
\label{eq7}
\end{eqnarray}
Overall {\it network disutility} $\bar U$ is a sum of market disutilities over all $M$ markets
\begin{eqnarray}
\bar U=\sum_{k=1}^M \bar U_k.
\label{eq8}
\end{eqnarray}

States of the system are defined in discretized time, which means that for definitions of functions of states we use all individual decisions taken in given time interval, e.g. one day.

\subsection{\label{ssec:level1}Additivity and extensivity of the utility}

\begin{figure}[h]
\begin{center}
\includegraphics[width=80mm,height=55mm]{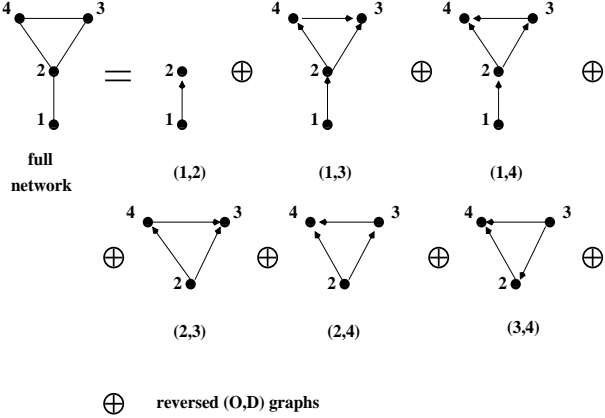}
\caption{\label{fig:fg0} Decomposition of the network into markets}
\end{center}
\end{figure}

In order to find the probability of the state of the network using utility function and incorporating classical Boltzmann-Gibbs arguments, the utility has to be {\it an additive stochastic variable}. 
Disutility $\bar{U}$ of the system is called additive with respect to any two subsystems $A$ and $B$ if their disutilities add up to $\bar{U}$, i.e. $\bar{U}=\bar{U}_{A}+\bar{U}_{B}$ (cf. e.g. ref. \cite{touchette1} for more detailed discussion).
Therefore, additivity requires the notion of the subsystem to be clarified first.
In our case, the subsystems cannot be defined by dividing the connectivity net only, because the smallest meaningful entities in the theory are the markets, represented by graphs ${\cal G}_k$ $(k=1,\ldots,M)$.
We decompose the system ${\cal S}$ into the sum of markets
\begin{eqnarray}
{\cal S}={\cal S}_1\oplus\ldots\oplus{\cal S}_M,
\label{eq8_1}
\end{eqnarray} 
where $\oplus$ means the set sum of corresponding graphs {\it and} individuals being ascribed to corresponding markets.
Fig.~\ref{fig:fg0} illustrates decomposition of an example system into markets.

Each market $(O,D)$ can be further split into routes $[O,C_1,\ldots,C_L,D]$ in the same way. 
For our example from Fig.~\ref{fig:fg0} decomposition of markets into routes is the following:
\begin{eqnarray}
(1,2) & = & [1,2] \nonumber \\
(1,3) & = & [1,2,3]\oplus [1,2,4,3] \nonumber \\
(1,4) & = & [1,2,4]\oplus [1,2,3,4] \nonumber \\
(2,3) & = & [2,3] \oplus [2,4,3] \nonumber \\
(2,4) & = & [2,4] \oplus [2,3.4] \nonumber \\
(3,4) & = & [3,4] \oplus [3,2,4] 
\label{eq8_2}
\end{eqnarray}
and reversed $(O,D)\rightarrow(D,O)$ pairs.

For subsystems defined in this way, from the definition of network disutility (\ref{eq7},\ref{eq8}), it follows by construction
\begin{property}
Network disutility is an additive random function of state.
\end{property}

In addition, also {\it the extensivity} of disutility, in the sense used in ref. \cite{touchette1}, can be easily justified.
If $n$ is the number of decision makers in the system, then thermodynamic limit of disutility exists
\begin{eqnarray}
\lim_{n\rightarrow\infty}\frac{\bar{U}}{n}<\infty
\label{eq8_3}
\end{eqnarray}
for $\bar{U}$ proportional to $n^{\alpha}$ ($\alpha\le 1$).
By definition, this is our case with $\alpha=1$, hence 
\begin{property}
Network disutility is an extensive random function of state.
\end{property}

From properties 1, 2 and 3 it follows, that network disutility function has all the same properties relevant for the construction of probability measure as the total energy function of finite physical system with no long-range interactions.
Therefore it can play the same role in construction of statistical ensemble of systems as hamiltonians do, viz. it can be used for definition of the exponential probability measure on the state-space.
For doing this we note that because of assumption of disaggregate choice model and additivity of disutilities, for any two subgroups of passengers, either belonging to different markets or from the same market, disutilities of subgroups are independent of each other.
Hence, we find 
\begin{property}
For any independent subgroups of decision makers, disutilities of subgroups are mutually independent random variables.
\end{property}
Validity of Boltzmann and Gibbs arguments, as applied to $\bar{U}$, follows from properties 1-4.
Hence, the probability of the state of the network for given value $u$ of the network disutility $\bar U$ is equal to
\begin{eqnarray}
{\cal P}(\bar U=u)\sim \exp(-\beta u),\;\;\;\;\;\;\;\;\;\;\beta>0.
\label{eq9}
\end{eqnarray}
An important consequence of this is that one can consider statistical ensemble of networks and statistical thermodynamics for them with the probability density (\ref{eq9}), and there is no need for non-extensive statistics with power-law stochastic measure.
At first sight, this seems counter intuitive because of apparent long-range correlations in the communication network which usually entails usage of non-extensive statistics (cf. e.g. ref. \cite{tsallis1}).
This is understandable when realizing that only sites, and not markets, are really correlated in this model.
Markets, representing elementary objects in the system, are uncorrelated since decision makers decide on routes only, as mentioned above. 
The choice of the market is not a subject of the game considered here.
Correlations between markets appear when additional constraints are taken into account, as e.g. finite transmittance of nodes.

In the following we restrict our model to equilibrium thermodynamics, or require existence of stationary states, which on turn requires assumption on existence of the first moment of disutility function
\begin{eqnarray}
\langle\bar U\rangle<\infty.
\label{eq10}
\end{eqnarray}
Otherwise, the equilibrium temperature $T_{eq}=1/\beta_{eq}$ would not exist.
We are not particularly concerned here neither in studying processes leading to equilibrium nor in all sufficient conditions for existence of equilibria in general.
Therefore we do not start our discussion of ensembles from defining transition probabilities and deriving probability densities from detailed balance. 

In our approach we do not incorporate random intensive variables. 
It has been recently shown \cite{beck1} that in the most general case of all variables of the system being stochastic, including intensive ones, one arrives to the class of ensembles with even more general proability densities than Levy and Tsallis power-law functions.
Our assumption purposefully limits generality of statistical ensembles considered here, since our original intent was to consider intensive variables of the system as simple steering parameters.

\section{\label{sec:level3}Equilibrium statistical ensembles}
 
Assuming existence of stationary regime for time evolution of complex network and using utility functions discussed above, it is straightforward to find probability measure on the state-space of the network.
As known from statistical thermodynamics, the specific form of the probability density is determined by the nature of constraints on the system, or the set of variables of merit which are allowed to be stochastic. 

\subsection{\label{ssec:level1}The microcanonical ensemble}

In the simplest case of disutilities being the same for all states of the system or, for non-discrete case, localized in very narrow interval $[\bar{U_0},\bar{U_0}+\delta\bar{U}]$, $(\delta\bar{U}/\bar{U_0}\ll 1)$, one normally assumes the probabilities for all states being the same and calls it {\it the principle of a priori equal probabilities}.
Ensemble of systems so defined is called {\it the microcanonical ensemble} and corresponds to the extremly tight, and rather unrealistic in our case, constraints imposed on the system, and on the set of decision makers in particular.
This means that the overall disutility is insensitive to the distribution of decision makers which could be interpreted as either lack of decision makers' sensitivity to the conditions of the network or as an extremly poor offer of the carrier, giving no choice to its customers.
This is rather trivial case and we do not discuss it furtheron.

\subsection{\label{ssec:level2}The canonical ensemble}

As the first non-trivial case consider {\it the canonical ensemble} of systems where the number $N$ of decision makers is fixed and the structure of connectivity, or the topology of the connections network, does not vary in time.
One defines {\it the partition function} or {\it the statistical sum} of the system
\begin{eqnarray}
Z(\beta) & = & \prod_{k=1}^M Z_k(\beta) \nonumber \\
         & = & \prod_{k=1}^M \sum_{j_k=1}^{J_k} e^{-\beta\bar{U}_{j_k}}.
\label{eq11}
\end{eqnarray}
The $Z_k(\beta)$ stands for the partition function of the $k$-th market and index $j_k$ runs over all $J_k$ distributions of passengers over routes belonging to the $k$-th market.
Note that each market itself is treated here as a canonical ensemble, which means that also numbers of clients $N_k$ ($k=1,\ldots,M$) belonging to each market is fixed.
This means that potential passengers from given origin node are at least decided as for their destination node, although they may not {\it a priori} know which route to fly there.

Due to additivity of disutility for each market (\ref{eq7}) the $Z_k$ can be rewritten in terms of explicit sums over $R_k$ routes for the $k$-th market
\begin{eqnarray}
Z_k(\beta) & = & Z(\beta,N_k) \nonumber \\
           & = & \sum_{j_1,\ldots,j_{N_k}=1}^{R_k} e^{-\beta\sum_{l=1}^{N_k}\bar{U}_{l,j_l}} \nonumber \\
           & = & \prod_{l=1}^{N_k}\sum_{j_l=1}^{R_k} e^{-\beta\bar{U}_{l,j_l}},
\label{eq12}
\end{eqnarray}
where $j_1,\ldots,j_{N_k}$ are passenger's indices running over $R_k$ routes and $l$ runs over passengers.
Assuming further the {\it non-subjectivity} of the utility, i.e. that utility for given route does not depend on the decision maker, $\bar{U}_{l,j_l}=\bar{U}_j$ ($l=1,\ldots,N_k$), the partition function for the $k$-th market can be written as
\begin{eqnarray}
Z_k(\beta) & = & \Big(\sum_{j=1}^{R_k}e^{-\beta\bar{U}_j}\Big)^{N_k} \nonumber \\
           & = & Z_k^1(\beta)^{N_k},
\label{eq13}
\end{eqnarray}
where $Z_k^1(\beta)$ stands for the partition function for the $k$-th market with one passenger on it.

The state of the system in the canonical ensemble is defined by specifying disutility for the complete set of distributions of passengers over all markets, i.e. $\Big\{\big\{\bar{U}_{j_k}\big\}_{j_k=1}^{J_k}\Big\}_{k=1}^M$.
Probabilities of states are 
\begin{eqnarray}
{\cal P}_{j_k}(\beta)=\frac{e^{-\beta\bar{U}_{j_k}}}{Z(\beta)}.
\label{eq14}
\end{eqnarray}
Noteworthly, the choice probability given by formula (\ref{eq14}) is the same as obtained in the framework of {\it the multinomial logit choice models}, usually using strong assumptions on the type of distributions, extreme value or Gumbel, for the stochastic components of the utility (cf. e.g. ref. \cite{train1}).

Mean disutility can be calculated using (\ref{eq14}), or by diffentiation over the Lagrange multiplier $\beta$:
\begin{eqnarray}
\langle\bar{U}\rangle & = & \sum_{k=1}^M \sum_{j_k=1}^{J_k} {\cal P}_{j_k}(\beta)\bar{U}_{j_k} \nonumber \\
                      & = & -\frac{\partial}{\partial\beta}\ln Z(\beta)
\label{eq15}
\end{eqnarray}
and if its value $\langle\bar{U}\rangle_{meas}$ is also known from measurements, their equality determines the equilibrium temperature $\beta_{eq}=1/T_{eq}$.
The entropy and disutility fluctuations are given by formulae
\begin{eqnarray}
S(\beta) & = & -\sum_{k=1}^M \sum_{j_k=1}^{J_k} {\cal P}_{j_k}(\beta) \ln {\cal P}_{j_k}(\beta) \nonumber \\
         & = & -\beta^2 \frac{\partial}{\partial\beta}\frac{1}{\beta}\ln Z(\beta) \nonumber \\
         & = & \beta\langle\bar{U}\rangle+\ln Z(\beta) 
\label{eq16}
\end{eqnarray}
and
\begin{eqnarray}
\mbox{{\it Var}}\,(\bar{U}) & = & \frac{\partial^2}{\partial\beta^2}\ln Z(\beta) \nonumber \\
                     & = & -\frac{\partial}{\partial \beta}\langle\bar{U}\rangle.
\label{eq17}
\end{eqnarray}

\subsection{\label{ssec:level3}The grand canonical ensemble}

In {\it the grand canonical} case, both disutility $\bar{U}$ and the number of decision makers $N$ represent random variables, but the connectivity of the network remains fixed.
The grand partition function or grand canonical sum is given by
\begin{eqnarray}
\Xi(\beta,\vec{\mu})=\sum_{N_1,\ldots,N_M}e^{\beta\sum_{k=1}^M\mu_k N_k}Z_{N_k}(\beta),
\label{eq18}
\end{eqnarray}
where $Z_{N_k}(\beta)=\prod_{k=1}^M Z_k(\beta)$ and $\vec{\mu}=(\mu_1,\ldots,\mu_M)$ is the vector of chemical potentials for markets, corresponding to the vector of passengers numbers $\vec{N}=(N_1,\ldots,N_M)$.
Here $Z_k$ is the canonical partition function for the $k$-th market (\ref{eq13}) but with $N_k$ being stochastic variable.
Following intuitive meaning of the chemical potential for physico-chemical systems, {\it a priori} one could think about different chemical potentials for different markets, as individual passengers from different markets may contribute unequally to the overall disutility of the system. 
Here we assume, however, {\it the hypothesis of chemical uniformity} of the system which states that chemical potentials of all markets are equal: $\mu:=\mu_1=\ldots =\mu_M$.
This hypothesis does not exclude multiphase systems from our considerations, so that our communication network corresponds to the case of a one-component, though not necessarily one-phase, chemical system.
In further sofistication of this formalism one should categorize passengers regarding professions, wealth, aims of travels etc, and consider different chemical potentials for different categories.
Those characteristics are in many cases correlated with markets but, obviously, the scopes of those features are not in one-to-one relations to the markets.
Formally, $\mu$ is the Lagrange multiplier corresponding to the constraint
\begin{eqnarray}
\langle N\rangle =\sum_{k=1}^M\langle N_k\rangle .
\label{eq19}
\end{eqnarray}

Following the same reasoning as in the canonical case, the grand partition function can be rewritten
\begin{eqnarray}
\Xi(\beta,\mu) & = & \sum_{N_1,\ldots,N_M}\prod_{k=1}^M e^{\beta\mu N_k}Z(\beta,N_k) \nonumber \\
               & = & \prod_{k=1}^M \frac{Z(\beta,N_k+1)e^{N_k}-e^{-\beta\mu}}{Z_k(\beta,1)-e^{-\beta\mu}} \nonumber \\
               & = & \prod_{k=1}^M \frac{[e^{\beta\mu}Z_k^1(\beta)]^{N_k+1}-1}{e^{\beta\mu}Z_k^1(\beta)-1}.
\label{eq20}
\end{eqnarray}

The state of the system in the grand canonical ensemble is defined by specifying disutility for the complete set of distributions of passengers over all markets and for all numbers of passengers, i.e. $\Big\{\big\{\{\bar{U}_{j_k}(N_k)\}_{j_k(N_k)=1}^{J_k(N_k)}\big\}_{N_k}\Big\}_{k=1}^M$.
Probabilities of states are
\begin{eqnarray}
{\cal P}_{j_k(N_k)}(\beta,\mu)=\frac{e^{\beta(\mu N_k-\bar{U}_{j_k(N_k)})}}{\Xi(\beta,\mu)}.
\label{eq21}
\end{eqnarray}
This reminds again the choice probability from the multinomial logit choice model, where $\beta\mu N_k$ plays the role of {\it the alternative specific constant} (cf. e.g. refs \cite{louviere1,train1}). 
Here, however, this constant is specific for the whole class of routes for the $k$-th market with $N_k$ individuals on it and for the particular state $j_k(N_k)$.

First moments of $\bar{U}$ and $N$ and their correlations can be found by differentiation of $\Xi$ over Lagrange multipliers $\beta$ and $\mu$ as follows:
\begin{eqnarray}
\langle\bar{U}\rangle & = & \sum_{k=1}^M\sum_{N_k}\sum_{j_k(N_k)=1}^{J_k(N_k)} {\cal P}_{j_k(N_k)}(\beta,\mu) \bar{U}_{j_k(N_k)} \nonumber \\
                      & = & -\Big(\frac{\partial}{\partial\beta}\ln \Xi(\beta,\mu)\Big)_{\mu\beta} \nonumber \\
\langle N\rangle      & = & \sum_{k=1}^M\sum_{N_k}\sum_{j_k(N_k)=1}^{J_k(N_k)} {\cal P}_{j_k(N_k)}(\beta,\mu) N_k \nonumber \\
                      & = & \frac{1}{\beta}\Big(\frac{\partial}{\partial\mu}\ln \Xi(\beta,\mu)\Big)_{\beta} \nonumber \\
\mbox{{\it Var}}\,(\bar{U}) & = & \Big(\frac{\partial^2}{\partial\beta^2}\ln \Xi(\beta,\mu)\Big)_{\mu\beta} \nonumber \\
                  & = & -\Big(\frac{\partial}{\partial\beta}\langle\bar{U}\rangle\Big)_{\mu\beta} \nonumber \\
\mbox{{\it Var}}\,(N) & = & \frac{1}{\beta^2}\Big(\frac{\partial^2}{\partial\mu^2}\ln \Xi(\beta,\mu)\Big)_{\beta} \nonumber \\
            & = & \frac{1}{\beta}\Big(\frac{\partial}{\partial\mu}\langle N\rangle\Big)_{\beta} \nonumber \\
\mbox{{\it cov}}(\bar{U},N) & = & -\frac{1}{\beta}\Big(\frac{\partial}{\partial\mu}\big(\frac{\partial}{\partial\beta}\ln\Xi(\beta,\mu)\big)_{\mu\beta}\Big)_{\beta} \nonumber \\
                           & = & \frac{1}{\beta}\Big(\frac{\partial}{\partial\mu}\langle\bar{U}\rangle\Big)_{\beta}
\label{eq22}
\end{eqnarray}
and the entropy is given by
\begin{eqnarray}
S(\beta,\mu) & = & -\mu\Big(\frac{\partial}{\partial\mu}\ln\Xi(\beta,\mu)\Big)_{\beta}-\beta^2\Big(\frac{\partial}{\partial\beta}\frac{1}{\beta}\ln\Xi(\beta,\mu)\Big)_{\beta\mu} \nonumber \\
             & = & -\beta\mu\langle N\rangle + \beta\langle\bar{U}\rangle +\ln\Xi(\beta,\mu).
\label{eq23}
\end{eqnarray}
Similarily to the canonical statistics, provided the first moments $\langle\bar{U}\rangle_{meas}$ and $\langle N\rangle_{meas}$ are known from measurements, the temperature $1/\beta$ and the chemical potential $\mu$ can be determined from eqns (\ref{eq22}).

\subsection{\label{ssec:level4}An outreach: possible extensions towards stochastic network topology -- the super-canonical ensemble}

For the network ensembles considered so far the network topology was fixed.
In real communication networks it is not so and the connection graph has to be considered as a stochastic object.
This can be realized by representing each pair of sites as a binary random variable and making the number of vertices random, and defining {\it the super-canonical ensemble} with new, multidimensional extensive random variable $\vec{L}=(L_1,\ldots,L_M)$, where $L_k$ represents the total path length for the $k$-th market $(OD)$
\begin{eqnarray}
L(OD)=\sum_{C_{\sigma(1)},\ldots,C_{\sigma(S)}}L[O,C_{\sigma_1},\ldots,C_{\sigma_S},D],
\label{eq23_1}
\end{eqnarray}
where $\sigma(1),\ldots,\sigma(S)$ are sequencies of permutations of sites corresponding to existing routes.
The $L_k$ is an analogy of the volume and we call it {\it the volume of the $k$-th market}, represented by the graph ${\cal G}_k$.
For example, for the network in Fig.~\ref{fig:fg0}, the volumes of markets are
\begin{eqnarray}
L(1,2) & = & L[1,2]=1 \nonumber \\
L(1,3) & = & L[1,2,3]+L[1,2,4,3]=2+3=5 \nonumber \\
L(1,4) & = & L[1,2,4]+L[1,2,3,4]=2+3=5 \nonumber \\
L(2,3) & = & L[2,3]+L[2,4,3]=1+2=3 \nonumber \\
L(2,4) & = & L[2,4]+L[2,3,4]=1+2=3 \nonumber \\
L(3,4) & = & L[3,4]+L[3,2,4]=1+2=3 
\label{eq23_2}
\end{eqnarray}
such that the volume of the whole market, including reversed $(O,D)\rightarrow(D,O)$ pairs, amounts to 40.

Then we define the super partition function
\begin{eqnarray}
 Y(\beta,\mu,p) & \nonumber \\
= \sum_{L_1,\ldots,L_M}\sum_{N_1,\ldots,N_M}\prod_{k=1}^M e^{-\beta(pL_k-\mu N_k)}Z_k(\beta) &
\label{eq23a}
\end{eqnarray}
where $p$ stands for the intensive variable, coupled to extensive variables $L_k$ and being direct analog to the pressure, and $Z_k(\beta)$ is the partition function given by eq. (\ref{eq13}).
Formally, $p$ is the Lagrange multiplier corresponding to the constraint
\begin{eqnarray}
\langle L\rangle =\sum_{k=1}^M \langle L_k\rangle
\label{eq23b}
\end{eqnarray}
where $L=L_1+\ldots +L_M$ will be called {\it the total volume of the network}.

The state of the system in the super-canonical ensemble is defined by specifying disutility for complete set of distributions of passengers over all markets and for all possible numbers of passengers, and for all possible random routes, i.e. $\Big\{\big\{\{\bar{U}_{j_k}(N_k,L_k)\}_{j_k(N_k,L_k)=1}^{J_k(N_k,L_k)}\big\}_{N_k,L_k}\Big\}_{k=1}^M$.
Probabilities of states are
\begin{eqnarray}
{\cal P}_{j_k(N_k,L_k)}(\beta,\mu,p)=\frac{e^{\beta(\mu N_k-pL_k-\bar{U}_{j_k(N_k,L_k)})}}{Y(\beta,\mu,p)}.
\label{eq23_3}
\end{eqnarray}
The analogy with multinomial logit choice model is again close, with the logit alternative specific constant being equal to $\beta(\mu N_k-pL_k)$.

For completness, we list the formulae for first moments and correlations of extensive variables, including the cummulant $\langle\langle \bar{U}NL\rangle\rangle$ of $\bar{U}$, $N$ and $L$ accounting for true triple correlations between them, irreducible to pairwise corrlations:
\begin{eqnarray}
\langle \bar{U}\rangle & = & -\Big(\frac{\partial}{\partial\beta}\ln Y(\beta,\mu,p)\Big)_{\beta\mu,\beta p} \nonumber \\ 
\langle N\rangle & = & \frac{1}{\beta}\Big(\frac{\partial}{\partial\mu}\ln Y(\beta,\mu,p)\Big)_{\beta,p} \nonumber \\
\langle L\rangle & = & -\frac{1}{\beta}\Big(\frac{\partial}{\partial p}\ln Y(\beta,\mu,p)\Big)_{\beta,\mu} \nonumber \\
\mbox{{\it Var}}\,(\bar{U}) & = & \Big(\frac{\partial^2}{\partial\beta^2}\ln Y(\beta,\mu,p)\Big)_{\beta\mu,\beta p} \nonumber \\
\mbox{{\it Var}}\,(N) & = & \frac{1}{\beta^2}\Big(\frac{\partial^2}{\partial\mu^2}\ln Y(\beta,\mu,p)\Big)_{\beta,p} \nonumber \\
\mbox{{\it Var}}\,(N) & = & \frac{1}{\beta^2}\Big(\frac{\partial^2}{\partial p^2}\ln Y(\beta,\mu,p)\Big)_{\beta,\mu} \nonumber \\
\mbox{{\it cov}}\,(\bar{U},N) & = & -\frac{1}{\beta}\Big(\frac{\partial}{\partial\mu}\big(\frac{\partial}{\partial\beta}\ln Y(\beta,\mu,p)\big)_{\beta\mu,\beta p}\Big)_{\beta,p} \nonumber \\
\mbox{{\it cov}}\,(\bar{U},L) & = & \frac{1}{\beta}\Big(\frac{\partial}{\partial p}\big(\frac{\partial}{\partial\beta}\ln Y(\beta,\mu,p)\big)_{\beta\mu,\beta p}\Big)_{\beta,\mu} \nonumber \\
\mbox{{\it cov}}\,(N,L) & = & -\frac{1}{\beta^2}\Big(\frac{\partial}{\partial p}\big(\frac{\partial}{\partial\mu}\ln Y(\beta,\mu,p)\big)_{\beta,p}\Big)_{\beta,\mu} \nonumber 
\end{eqnarray}
and
\begin{eqnarray}
\langle\langle\bar{U}NL\rangle\rangle & \nonumber \\
=\langle\bar{U}NL\rangle-\langle\bar{U}N\rangle\langle L\rangle-\langle\bar{U}L\rangle\langle N\rangle & \nonumber \\
-\langle NL\rangle\langle\bar{U}\rangle+\langle\bar{U}\rangle\langle N\rangle\langle L\rangle & \nonumber \\
=\frac{1}{\beta^3}\bigg(\frac{\partial}{\partial p}\Big(\frac{\partial}{\partial\mu}\big(\frac{\partial}{\partial\beta}\ln Y(\beta,\mu,p)\big)_{\beta\mu,\beta p}\Big)_{\beta,p}\bigg)_{\beta,\mu} &
\label{eq23_4p}
\end{eqnarray}
and for the entropy
\begin{eqnarray}
S(\beta,\mu,p) & \nonumber \\
 =\beta\langle\bar{U}\rangle-\beta\mu\langle N\rangle+\beta p\langle L\rangle + \ln Y(\beta,\mu,p). &
\label{eq23_4}
\end{eqnarray}

We note that randomization of the network topology includes a couple of specific issues and its complete treatment needs further clarification.
The set of extensive variables $\vec{L}$ is related to the network itself and not so much to the choice model, as it does not {\it a priori} reflect any preferences of decision makers.
Even statistical ensembles of bare graphs, being less complex objects with no individuals making their choice among graph's properties, are already complicated enough to be described with a few extensive variables and distinguishing interesting cases of the microcanonical, the canonical and the grand canonical, depending on the contruction procedures.
The concept of bare random networks, unrelated to decision theory in the sense used here, is a subject of interest since pioneering works of Erd\"os and R\'enyi \cite{erdos1}.
Various procedures of randomization of graphs with certain constraints on their characteristics, defining Hamilton paths and building probability measures for their ensembles are proposed e.g. in refs \cite{dorogovtsev1} and \cite{burda1} and summarized in refs \cite{dorogovtsev2} and \cite{barabasi1}.

\subsection{\label{ssec:level4}{Escort moments}}

There is a useful approach to study of temperature dependence of thermodynamic functions, incorporating  escort distribution \cite{beck2}, which for the probability measure $\mu$ is defined as 
\begin{eqnarray}
P_{q_i}=\frac{p_i^q}{\sum_{i=1}^K p_i^q}\;\;\;\;\;\;\;\;\;\;\;\;\;\;\; (i=1,\ldots,K)
\label{eq23aa}
\end{eqnarray}
where $p_i=\int_{\Delta_i}\,d\mu$, $\Delta_i$ is the $i$-th phase space cell from the set of non-overlaping cells $\{\Delta_i\}_{i=1}^K$ covering completly the whole phase space, and $q\in\mathbb{R}^1$.
From sum over cells one excludes those where $p_i=0$.

Using $q$-parametrized escort distributions, instead of ordinary probability density, one can efficiently probe those regions of the phase space where the probability measure is most concentrated (large positive $q$) or most rarified (large negative $q$).
It was shown \cite{wislicki1,majka1} that using R\'enyi entropies \cite{renyi2}
\begin{eqnarray}
I_q=\frac{1}{1-q}\ln\sum_{i=1}^K p_i^q
\label{eq23ad}
\end{eqnarray}
instead of ordinary Kolmogorov-Shannon entropies, and $q$-derivatives \cite{jackson}
\begin{eqnarray}
\frac{d_q F(\beta)}{d_q \beta}=\frac{F(q\beta)-F(\beta)}{\beta(q-1)}
\label{eq23ae}
\end{eqnarray}
instead of ordinary derivatives over inverse temperature, one effectively evolves the partition functions and all thermodynamic functions.
For example, for the partition functions $Z$ and $\Xi$ one gets formulae
\begin{eqnarray}
I_q=\frac{1}{1-q}[\ln Z(q\beta)-q\ln Z(\beta)] \nonumber \\
I_q=\frac{1}{1-q}[\ln \Xi(q\beta,q\mu)-q\ln \Xi(\beta,\mu)]
\label{eq23af}
\end{eqnarray}
which control evolution of thermodynamic potentials and response functions with temperature and chemical potential.

Generally, for any stochastic function of temperature $A(\beta)$, the temperature scaling is equivalent to modification of the averaging prescription
\begin{eqnarray}
\langle A(q\beta)\rangle=\langle A(\beta)\rangle_q
\label{eq23ab}
\end{eqnarray}
and by iterating this procedure
\begin{eqnarray}
\langle A(q_1q_2\ldots q_n\beta)\rangle=\langle\langle\ldots \langle A(\beta)\rangle_{q_n}\ldots\rangle_{q_2}\rangle_{q_1}
\label{eq23ac}
\end{eqnarray}
where $\langle ..\rangle_q$ stands for average over the escort distribution (\ref{eq23aa}).
This is understandable because any temperature dependence of macroscopic observables, being themselves mean values or functions of them, stays in the probability measure.
Modification of the measure $p_i^q$ in case of Boltzmann-like $p_i=P(\bar{U}=\bar{u}_i)\sim \exp(-\beta\bar{u}_i)$ is just temperature rescaling.
For $q\rightarrow 1$ one recovers classical thermodynamics.
In order to follow the temperature evolution of ordinary mean value or variance of $A$ without measuring it, one has to calculate corresponding $q$-moment for given temperature and vary $q$.
Obviously, the same applies to any moment of the random variable $A(\beta)$, if it exists.

The concept of escort moments is also useful for monitoring of desired regions of the state space, e.g. those where experimental information is known most precisely or where the data exhibit the most advantageous signal-to-noise ratio and systematic bias is small (cf. ref. \cite{majka1} for detailed discussion).

\section{\label{sec:level4}Applications}

\subsection{\label{ssec:level4}{The model: analytical results and numerical simulations}}

The formalism can be directly applied for study of relations between first moments of disutilities and market occupancies and their correlations and their dependencies on intensive variables.
For any communication network, direct numerical simulations and evaluation of partition functions and probabilities on the state space can be performed.
In many cases, however, this may be very demanding computationally because of factorial dependence of numbers of configurations on the graph size.
For particular network topologies analytical shortcuts can be found and also some qualitative features of the networks can be inferred by pure reasoning.

\begin{figure}[h]
\begin{center}
\includegraphics[width=80mm,height=60mm]{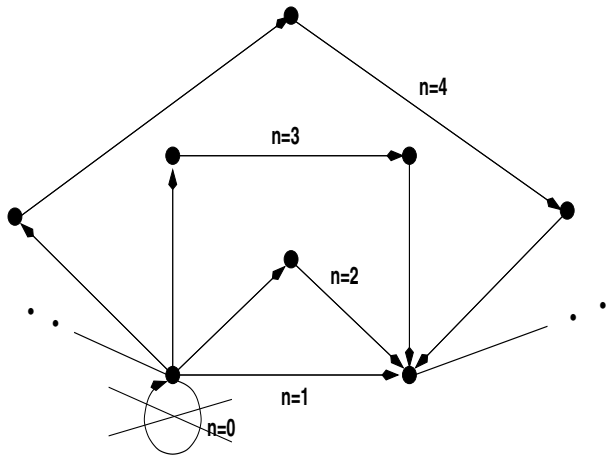} 
\caption{\label{fig:fg1} The maximum connectivity graph (a), the hub-and-spoke (b) and the spider-web (c).}
\end{center}
\end{figure}

We consider three network topologies (cf. Fig. \ref{fig:fg1}):
\begin{enumerate}
\item {\it The maximum connectivity} network, defined by a complete graph, where any two sites are directly connected by one-segment path (Fig. \ref{fig:fg1}a),
\item {\it The hub-and-spoke} network, where every site is only connected to the central hub (Fig. \ref{fig:fg1}b),
\item {\it The spider-web}, representing an intermediate case between the previous ones, where the hub-and-spoke structure is complemented by connections between spokes forming the ring-like structure (Fig. \ref{fig:fg1}c).
\end{enumerate} 

\begin{figure}[h]
\begin{center}
\includegraphics[width=60mm,height=40mm]{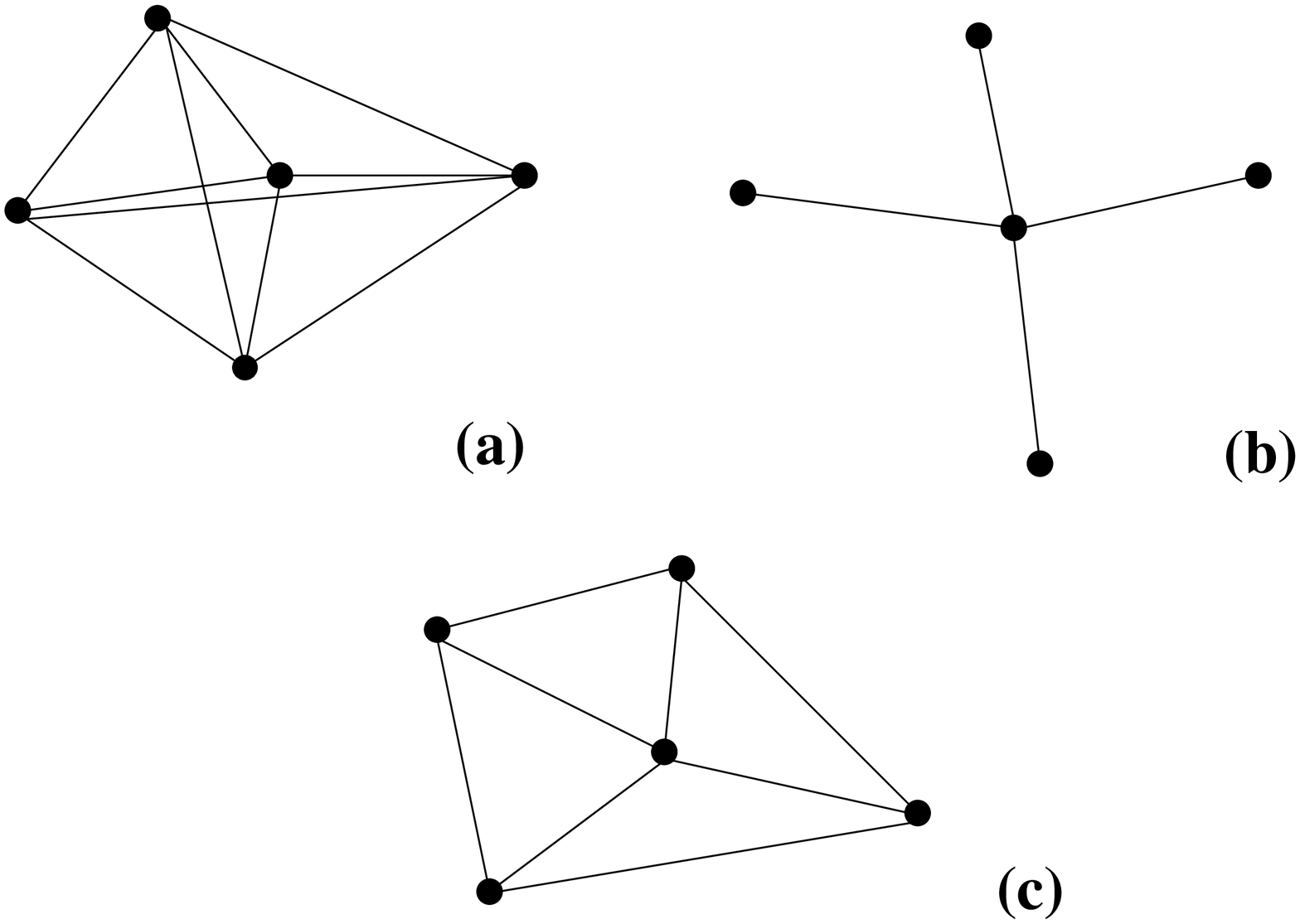}
\caption{\label{fig:fg2} The hierarchy of routes of lengths $n=0,1,\ldots$, corresponding to the energy spectrum of one-dimensional quantum harmonic osillator. Crossed $n=0$ path (tadpole) corresponds to exclusion of the no-trip alternative from our model.}
\end{center}
\end{figure}

For exploratory study of basic properties of the model we assume the passengers disutility being proportional to the number of segments of the $j$-th route, $n_j$ $(n_j=0,1,\ldots)$, disregarding any additional dependencies, i.e. $\bar{U}=v_0+n_j v$, where $v$ is a positive constant and $v_0$ corresponds to disutility of making no trip and we assume $v_0=0$ without loss of genarality.
This form corresponds to one of the basic features of the consumers disutility, which is his large reluctance to choose a route with large number of stops and changes.
Such utility function allows us to find some analytical results and, in addition, it exhibits clear physical analogy to the one-dimensional quantum harmonic oscillator, where the energy of its $n$-th state depends linearly on $n$, $E_n=\hbar\omega(n+\frac{1}{2})$ $(n=0,1,\ldots)$.
The no-trip disutility $v_0$ corresponds to vacuum energy $\hbar\omega/2$.
The hierarchy of states of the harmonic oscillator corresponds to the market consisting of non-forking routes of increasing lengths, as illustrated in Fig. \ref{fig:fg2}.

\begin{figure}[h]
\begin{center}
\begin{tabular}{cc}
\includegraphics[width=40mm,height=40mm]{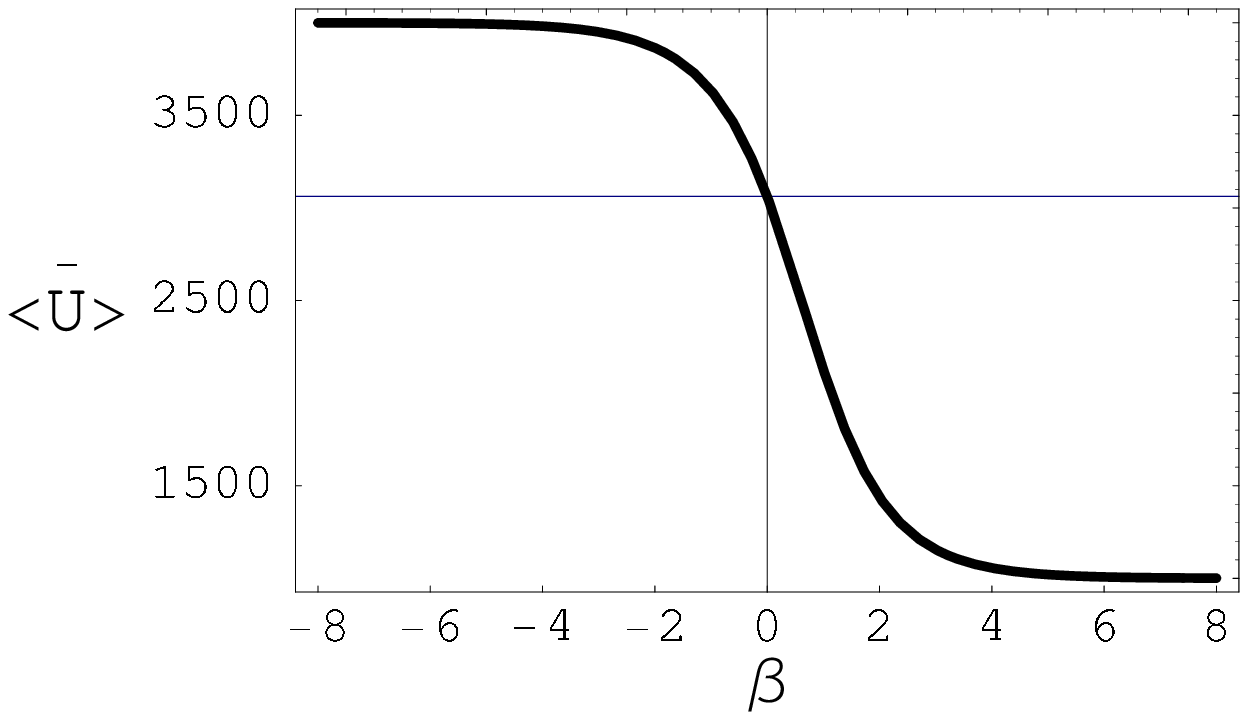} & \includegraphics[width=40mm,height=40mm]{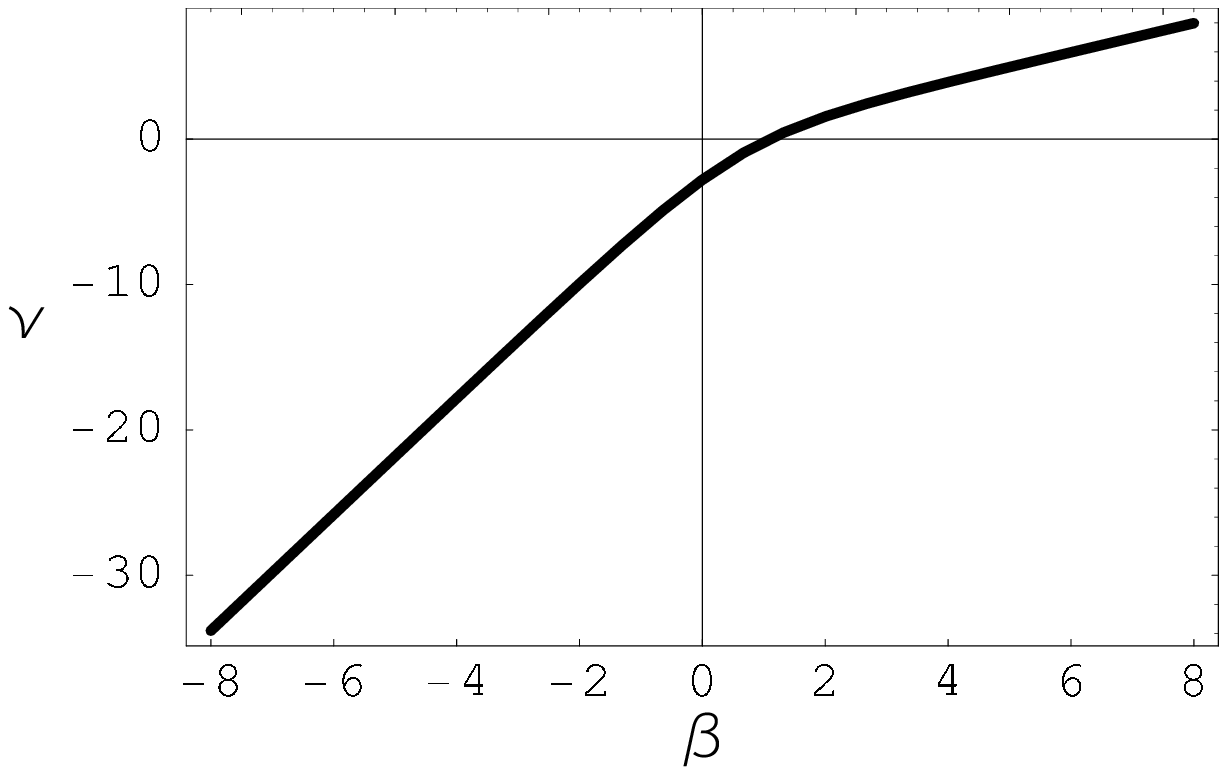} \\
\includegraphics[width=40mm,height=40mm]{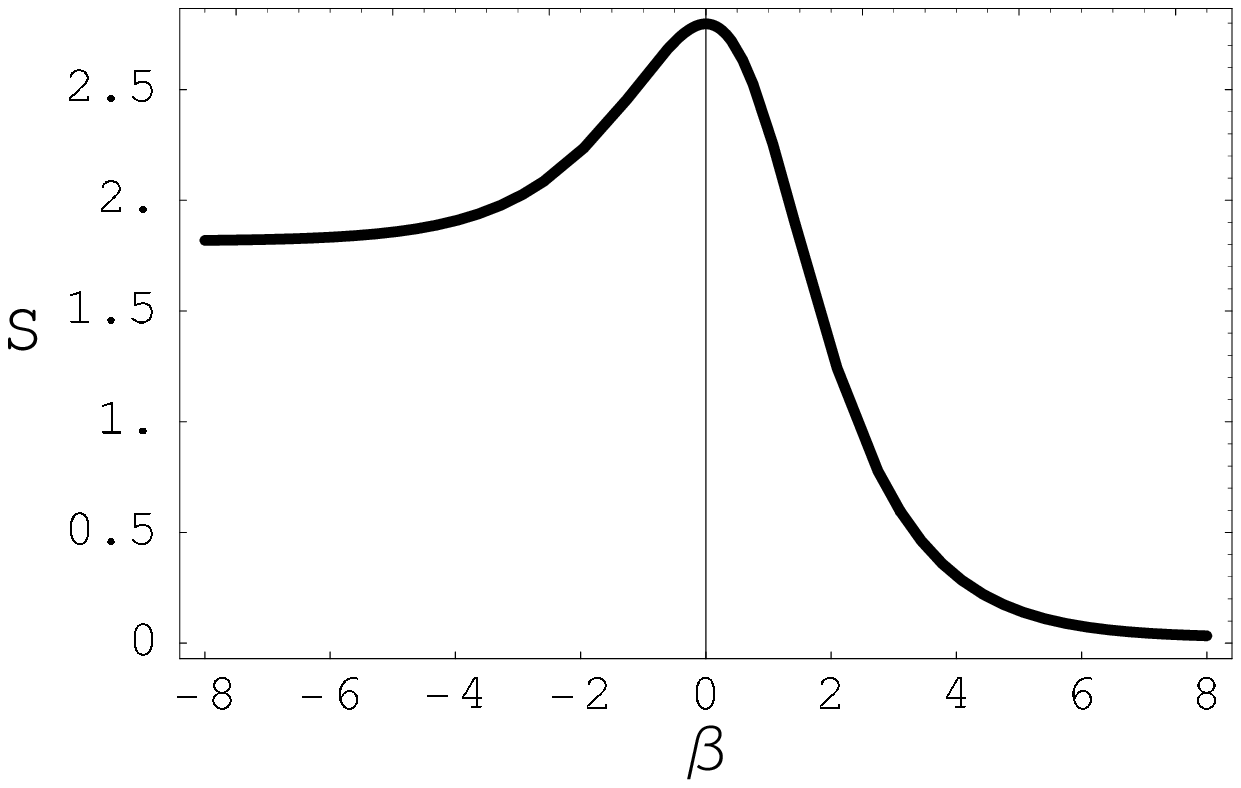} & \includegraphics[width=40mm,height=40mm]{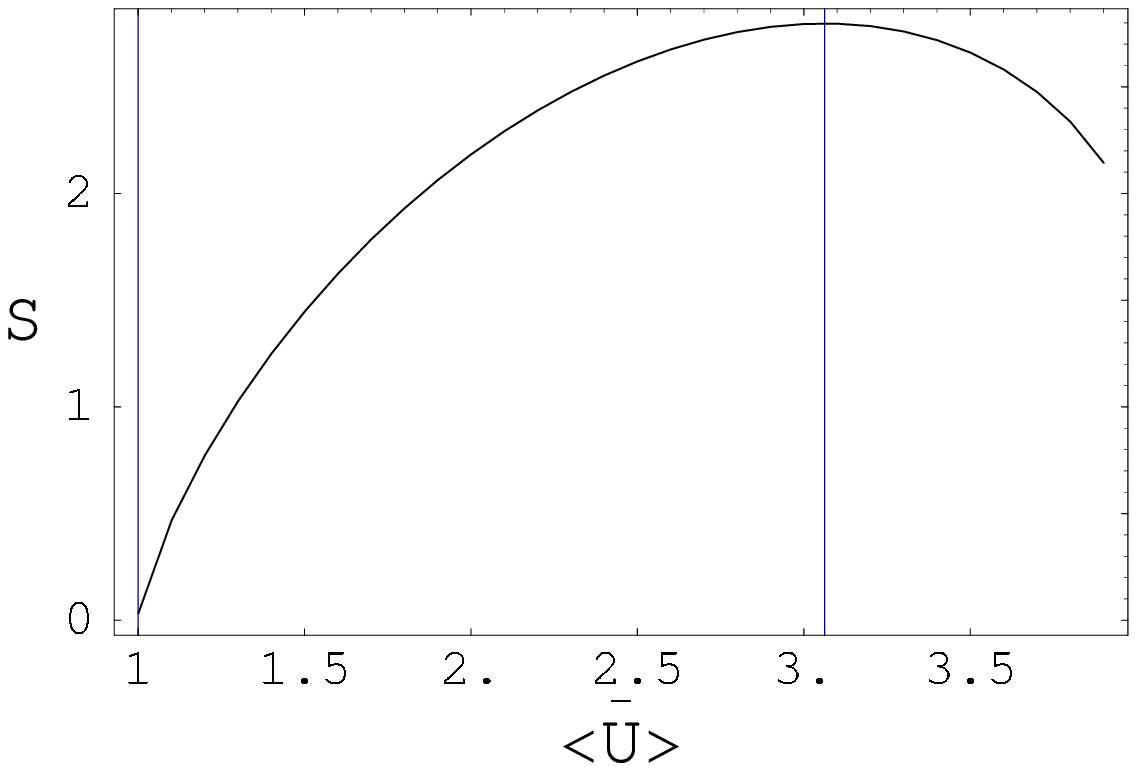}
\end{tabular}
\caption{\label{fig:fg3a} Mean disutility $\langle\bar{U}\rangle$ {\it vs.} inverse temperature $\beta$ (upper left), the $\nu=\beta\mu$ {\it vs.} $\beta$ (upper right), the entropy $S$ {\it vs.} $\beta$ (lower left) and $S$ {\it vs.} $\langle\bar{U}\rangle$ (lower right) for the maximum connectivity graph with 5 vertices.
In the lower right panel, the values of $\bar{U}=3.062$ corresponding to $\beta>0$ extend to the left and those for $\beta<0$ - to the right from the $\beta=0$ point of $\langle\bar{U}\rangle=3.062$, indicated as vertical line. Values of entropies are divided by 1000. Curves correspond to analytical results and points are from numerical simulation.}
\end{center}
\end{figure}

\begin{figure}[h]
\begin{center}
\begin{tabular}{cc}
\includegraphics[width=40mm,height=40mm]{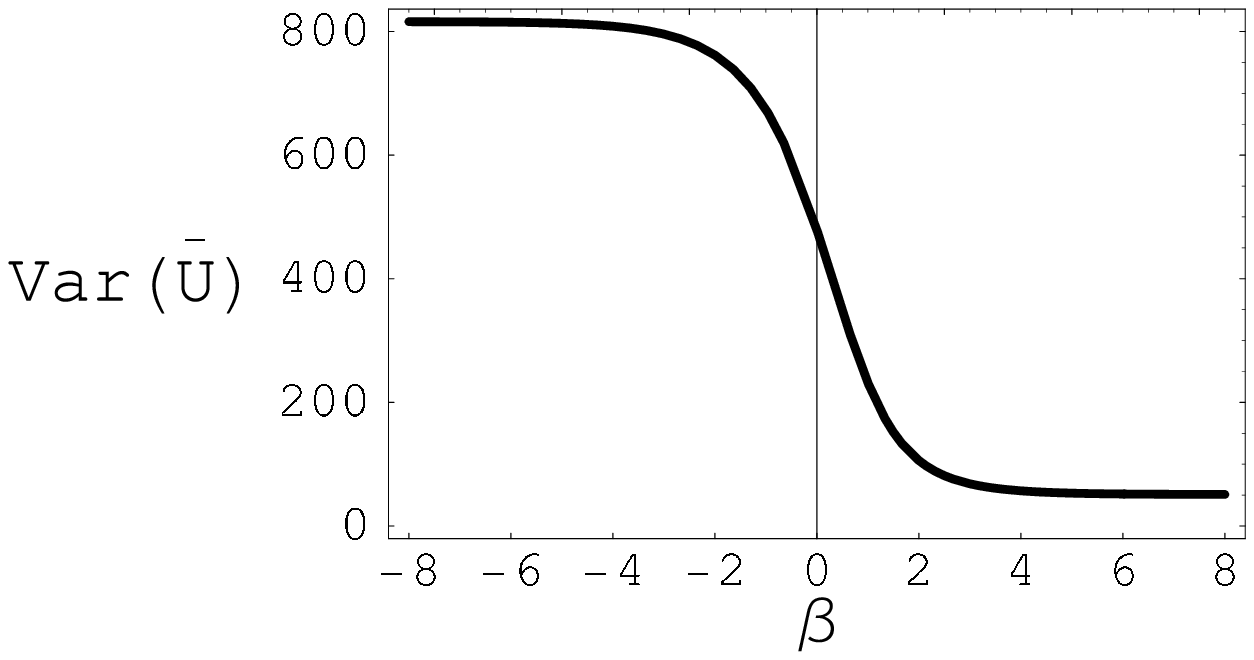} & \includegraphics[width=40mm,height=40mm]{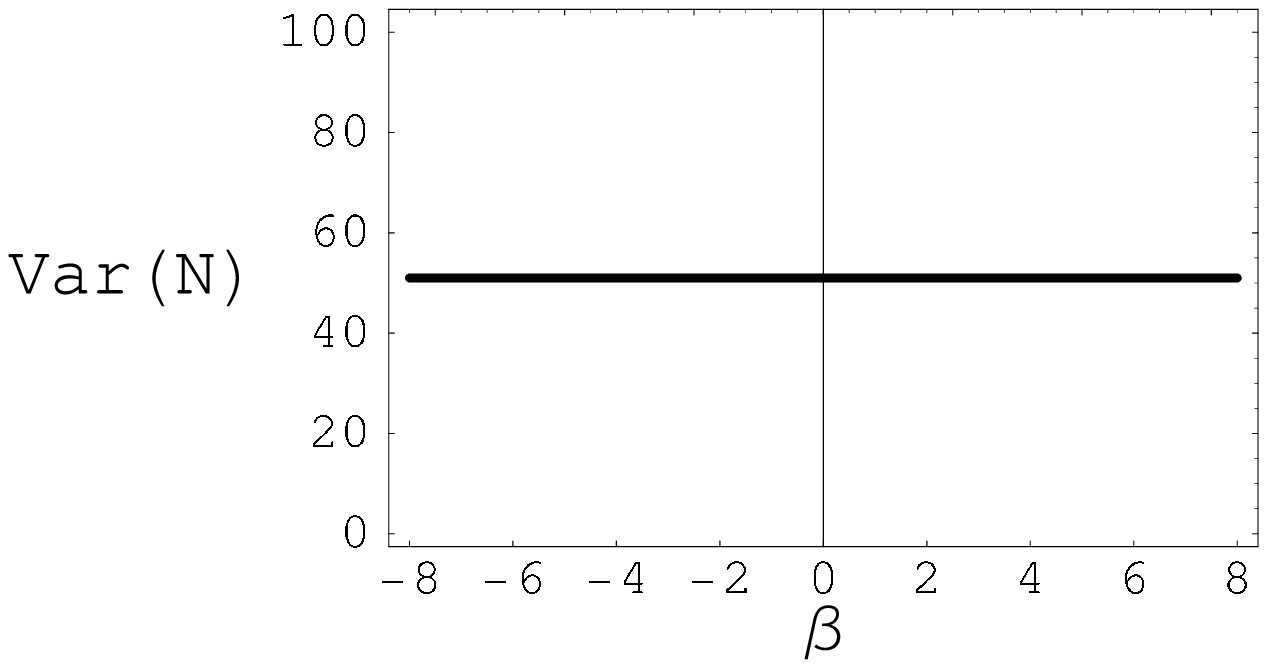} \\
\includegraphics[width=40mm,height=40mm]{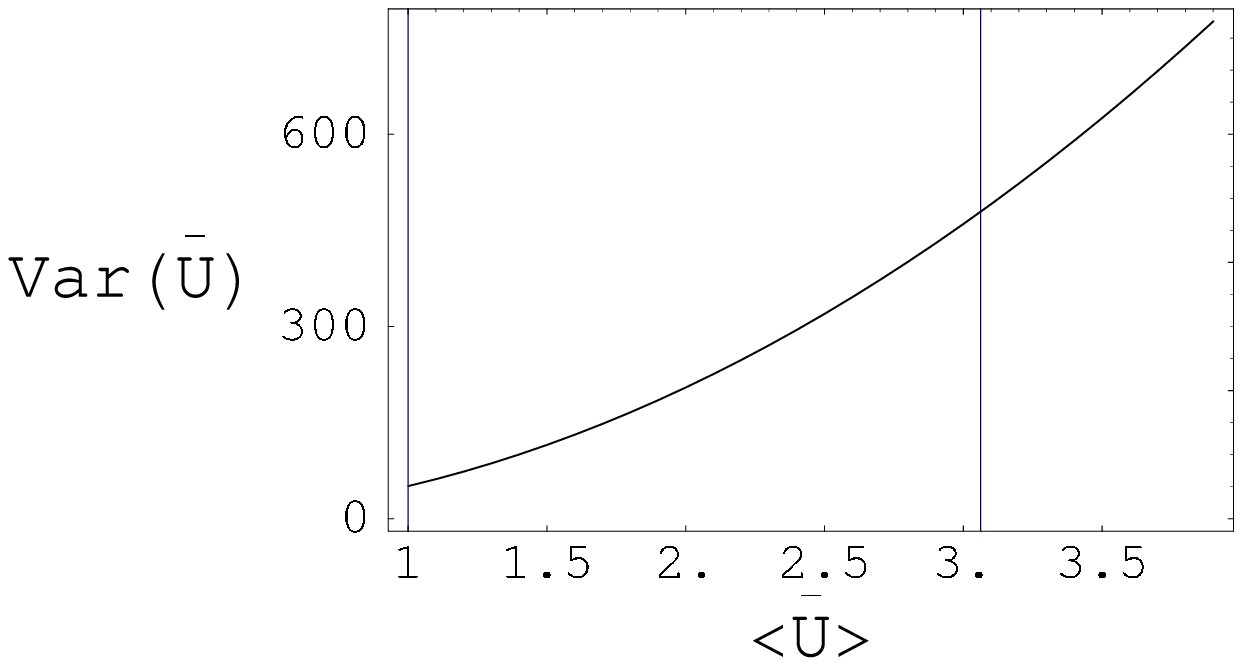} & \includegraphics[width=40mm,height=40mm]{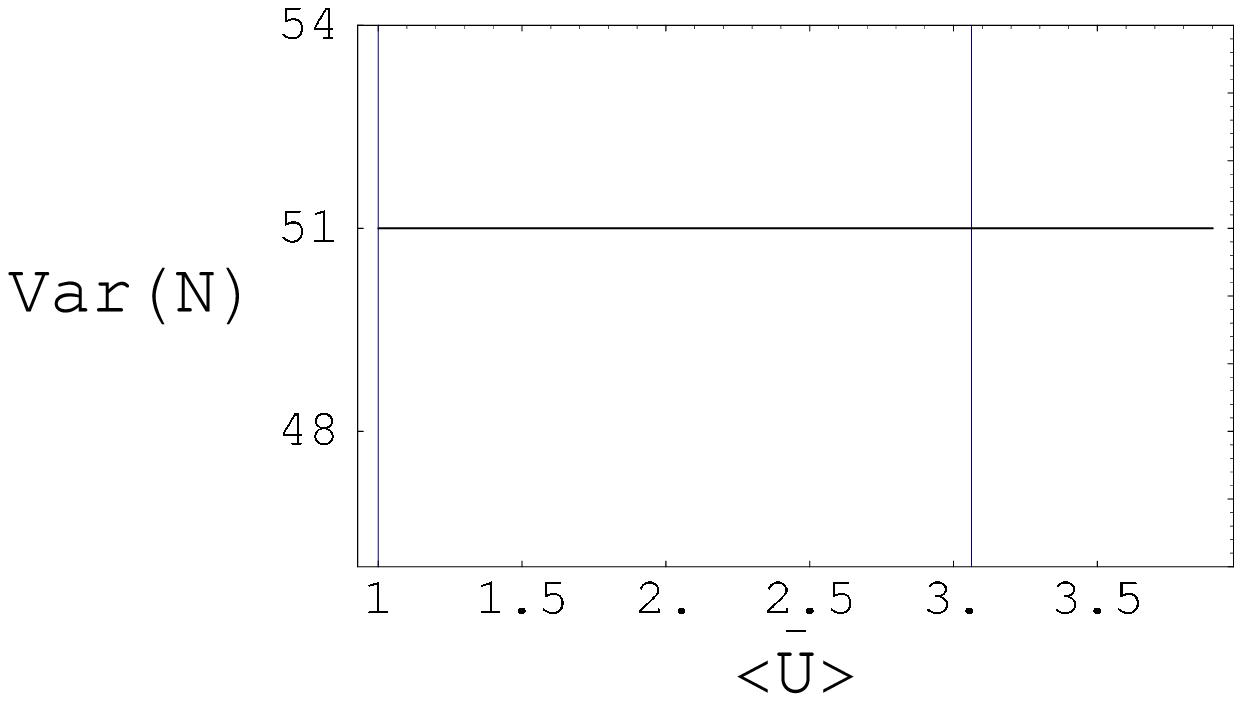} \\
\includegraphics[width=40mm,height=40mm]{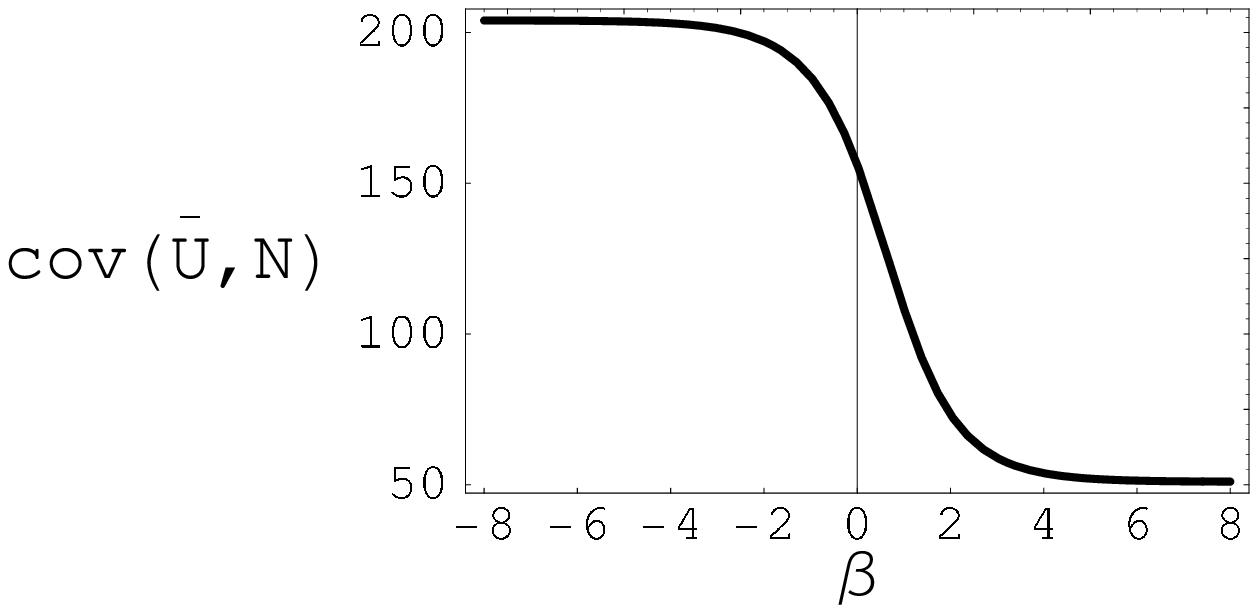} & \includegraphics[width=40mm,height=40mm]{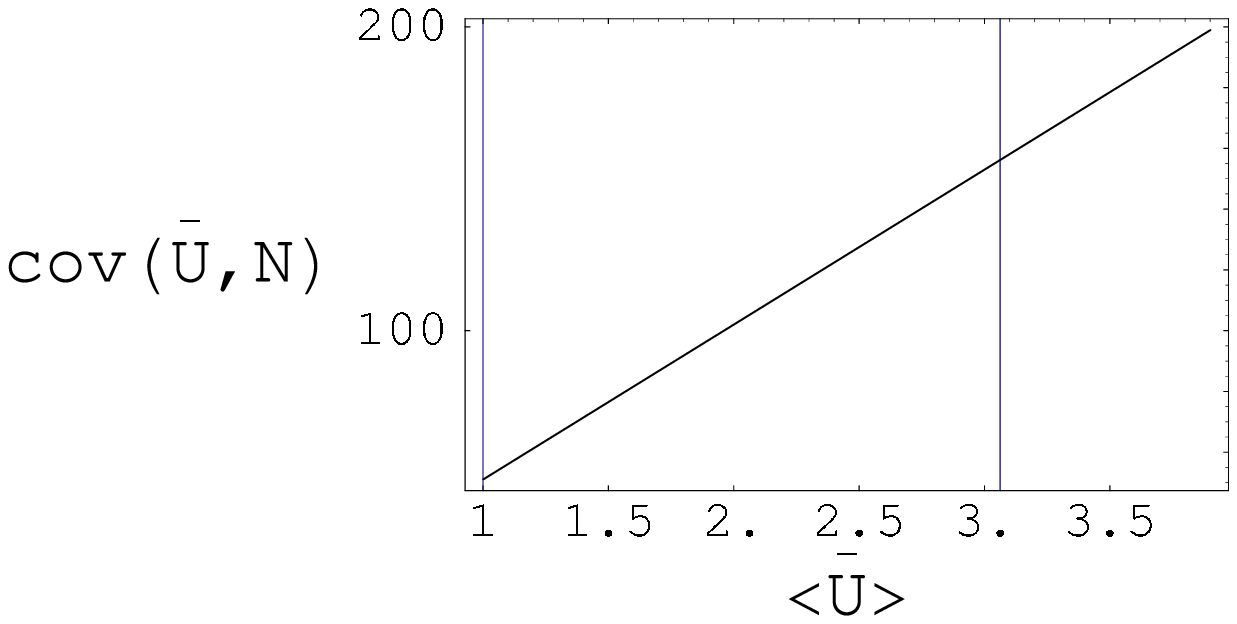}
\end{tabular}
\caption{\label{fig:fg3b} The variance of disuility {\it Var}$\,(\bar{U})$ {\it vs.} inverse temperature $\beta$ (upper left), the variance of the number of individuals {\it Var}$\,(N)$ {\it vs.} $\beta$ (upper right), {\it Var}$(\,\bar{U})$ {\it vs.} $U$ (middle left), {\it Var}$\,(N)$ {\it vs.} $\bar{U}$ (middle right), the covariance of $\bar{U}$ and $N$, {\it cov}$\,(\bar{U},N)$ {\it vs.} $\beta$ (lower left) and {\it cov}$\,(\bar{U},N)$ {\it vs.} $\bar{U}$ (lower right) for the maximum connectivity graph with 5 vertices.
The values of $\langle\bar{U}\rangle=3.062$ corresponding to $\beta=0$ are indicated by vertical lines for figures with $\bar{U}$ on the abcissae.}
\end{center}
\end{figure}

\begin{figure}[h]
\begin{center}
\begin{tabular}{cc}
\includegraphics[width=40mm,height=40mm]{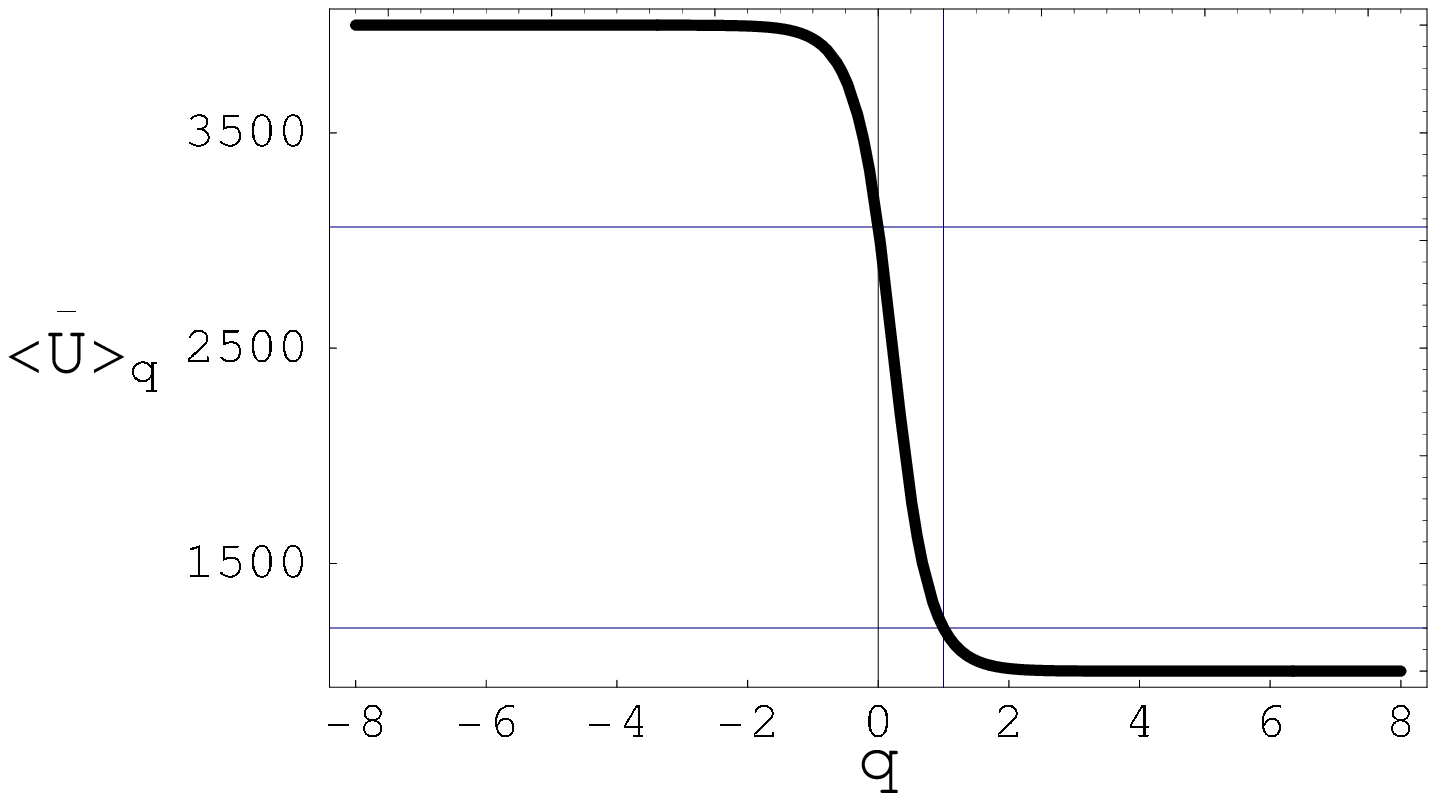} & \includegraphics[width=40mm,height=40mm]{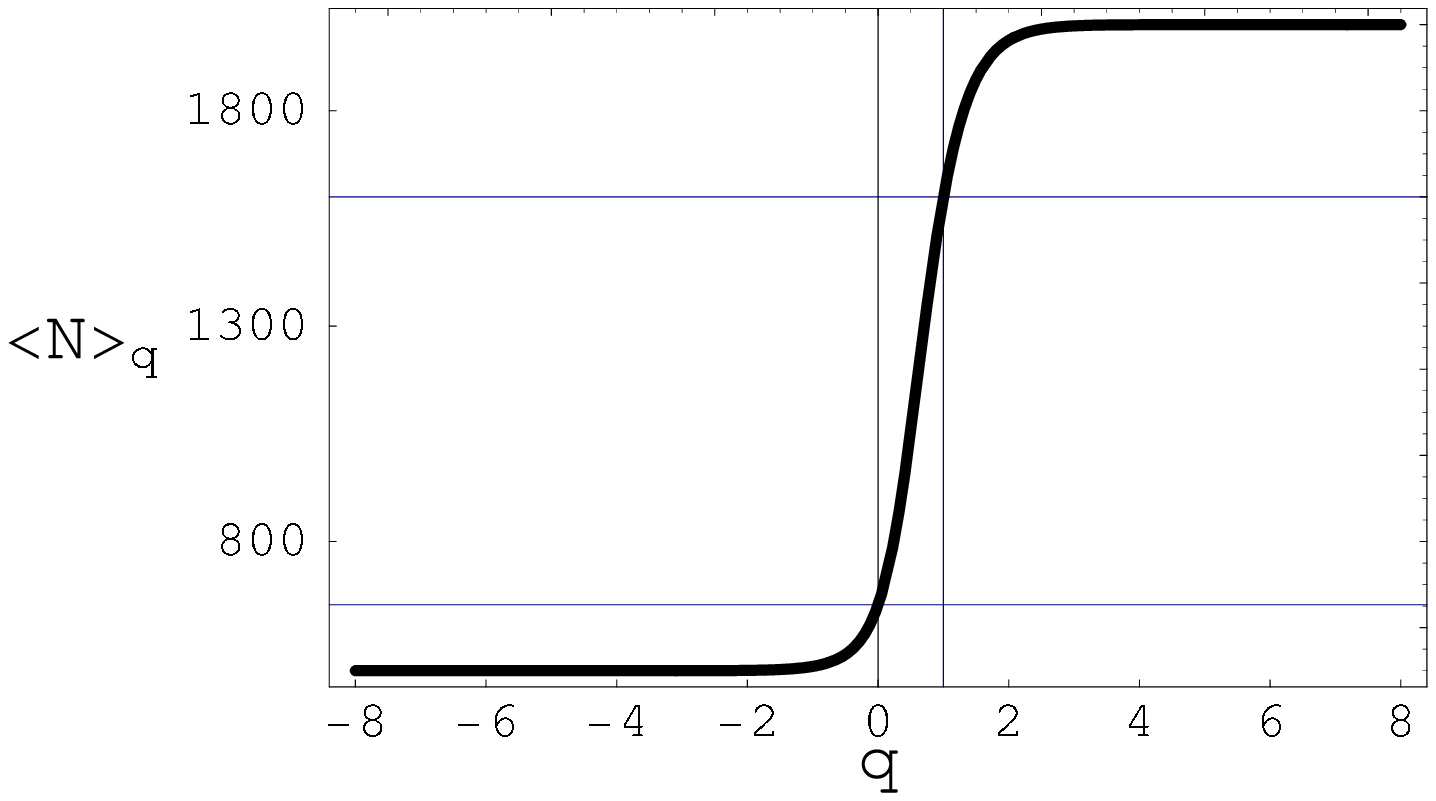} \\
\includegraphics[width=40mm,height=40mm]{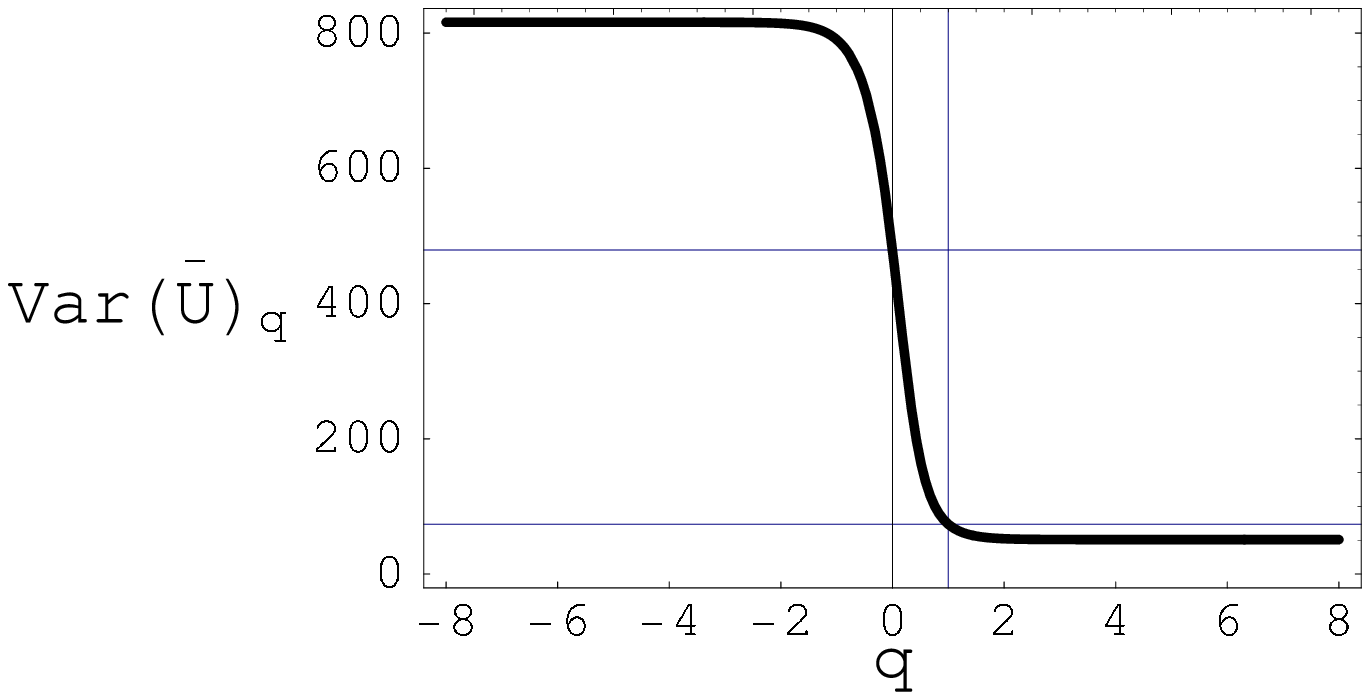} & \includegraphics[width=40mm,height=40mm]{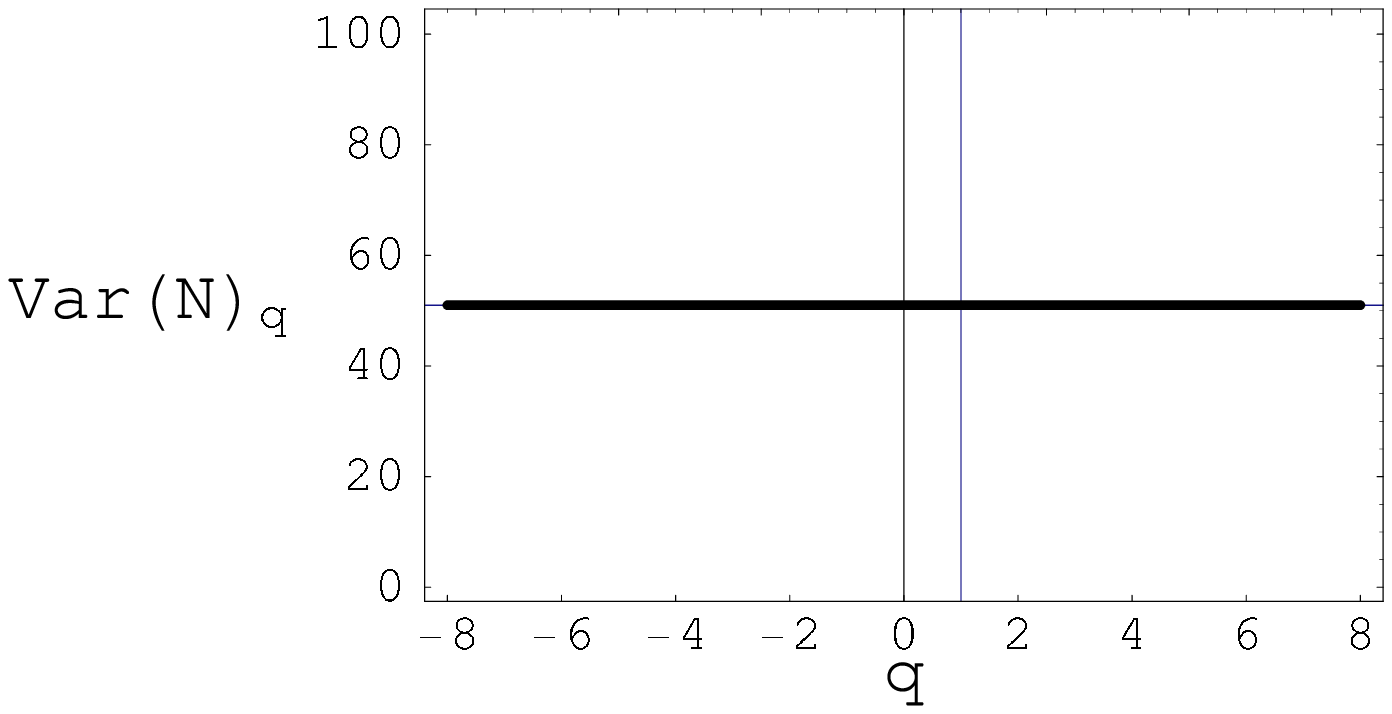} \\
\includegraphics[width=40mm,height=40mm]{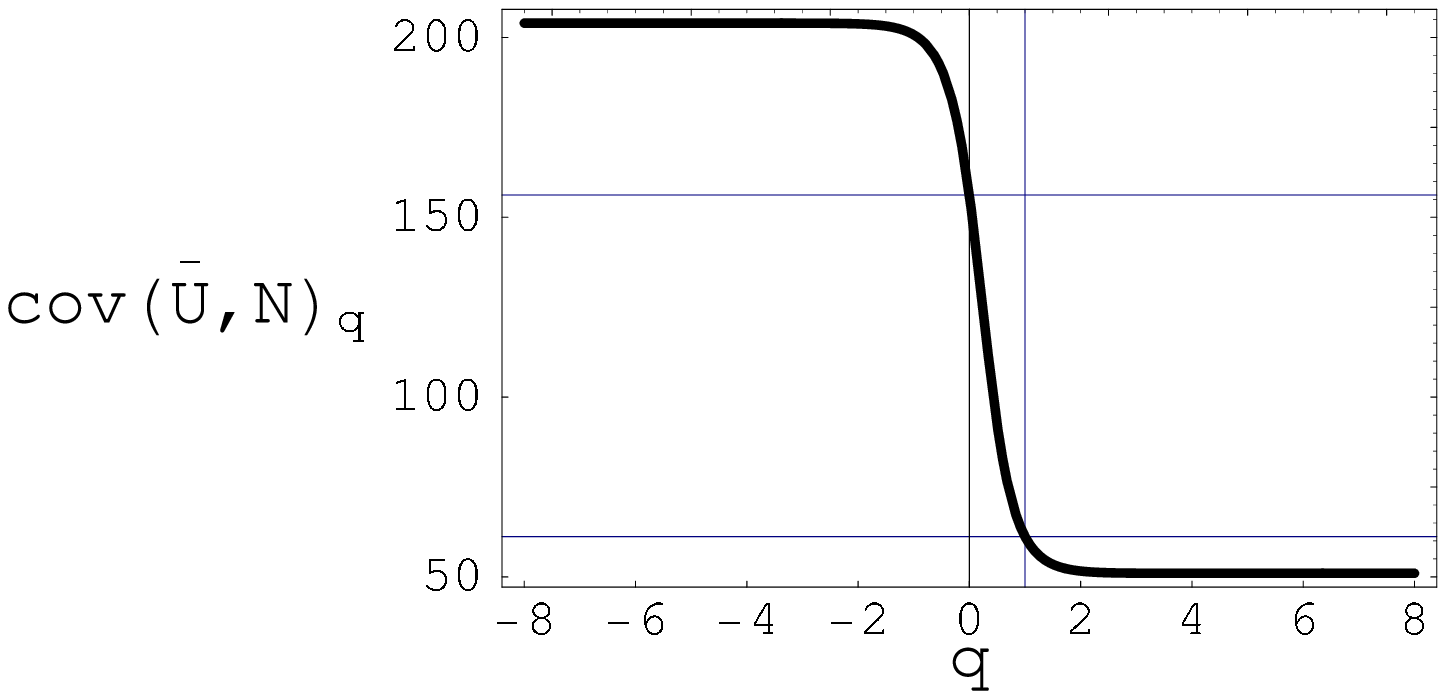} & \includegraphics[width=40mm,height=40mm]{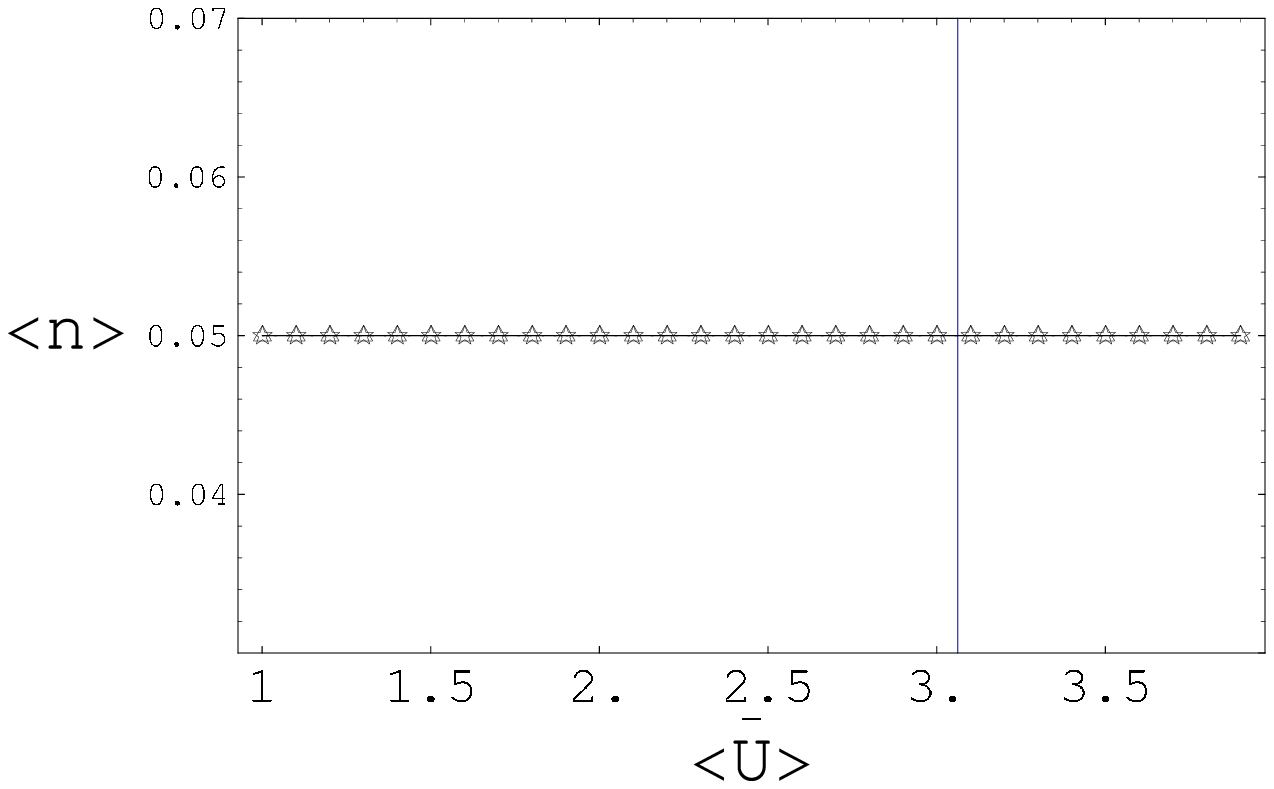}
\end{tabular}
\caption{\label{fig:fg3c} The $q$-mean disutility $\langle\bar{U}\rangle_q$ {\it vs.} $q$ (upper left) and the $q$-mean number of individuals $\langle N\rangle_q$ {\it vs.} $q$ (upper right), the $q$-variance of disutility {\it Var}$\,(\bar{U})_q$ (middle left), the $q$-variance of the number of individuals {\it Var}$\,(N)_q$ (middle right) and the $q$-covariance of $\bar{U}$ and $N$, {\it cov}$_q\,(\bar{U},N)$ (bottom left) for the maximum connectivity graph with 5 vertices. The ratio of passengers in a given type of market as a function of mean disutility is displayed in the lower right panel. The vertical line at $\langle\bar{U}\rangle=3.062$ corresponds to $\beta=0$.}
\end{center}
\end{figure}
This analogy is more than purely academic observation since the harmonic oscillator is one of the best understood non-trivial systems in physics and thus the market of Fig. \ref{fig:fg2} is an ideal testground for any concept developed within our choice model for networks.
Our definition of the system leaves some freedom for including or not the no-trip alternative to the set of routes available to the choice maker, provided one does not consider the no-trip event as a double visit in given site, which would contradict the Hamilton property of the path.
The model is perhaps more realistic excluding the no-trip choice ($n_j=0$) which means that the initial set of decision makers is restricted only to those determined to go.
Otherwise, potential passengers are indistinguishable from those who do not enter the game at all and for which the choice set ${\cal C}$ is undetermined.
Closed unit-edge loops, or {\it tadpoles}, can be easily avoided when the network topology is defined and non-random but it is known that special care in ascribing probabilities to the states is needed in case of random topologies \cite{dorogovtsev2}.

The statistical sum is trivially calculable fo both cases
\begin{eqnarray}
Z(\beta) & = & \left\{\begin{array}{ll}
               (1-e^{-\beta v})^{-1} & \;\;\;(\mbox{no-trip included}) \\
               (e^{\beta v}-1)^{-1}  & \;\;\;(\mbox{no-trip excluded})
               \end{array}\right .
\label{eq23bb}
\end{eqnarray}
The first moments of disutility in the low- and high-temperature (resp. cool and hot) limits are
\begin{eqnarray}
\langle \bar{U}\rangle & = & \frac{v}{e^{\beta v}-1} \nonumber \\
                       & \sim & \left\{\begin{array}{ll}
                         0, & \beta v\gg 1 \;\;\;(\mbox{cool}) \\
                         1/\beta, & \beta v\ll 1 \;\;\;(\mbox{hot})
                         \end{array}\right .
\label{eq23c}
\end{eqnarray}
and
\begin{eqnarray}
\langle \bar{U}\rangle & = & \frac{v}{1-e^{-\beta v}} \nonumber \\
                       & \sim & \left\{\begin{array}{ll}
			 v, & \beta v\gg 1 \;\;\;(\mbox{cool}) \\
                         1/\beta, & \beta v\ll 1 \;\;\;(\mbox{hot})
                         \end{array}\right .
\label{eq23d}
\end{eqnarray}
for {\it the no-trip included} and {\it the no-trip excluded} model, respectively.
Asymptotic behaviour is the same for both cases.

The variance of disutility, being proportional to the derivative of $\langle U\rangle$, is the same for both models:
\begin{eqnarray}
\mbox{\it Var}\,(\bar{U}) & = & \frac{v^2 e^{-\beta v}}{(1-e^{-\beta v})^2} \nonumber \\
                       & \sim & \left\{\begin{array}{ll}
                         0, & \beta v\gg 1 \;\;\;(\mbox{cool}) \\
                         1/\beta^2, & \beta v\ll 1 \;\;\;(\mbox{hot})
                         \end{array}\right .
\label{eq23e}
\end{eqnarray}

Our calculations and simulations are performed in the framework of grand canonical ensemble and we incorporate the notation
\begin{eqnarray}
X_k=e^{\beta\mu}Z_k^1(\beta),\;\;\;\;\;\;\;\;\;\;\;\;\;\;\; (k=1,\ldots,M)
\label{eq23f}
\end{eqnarray}
Then the grand partition function (\ref{eq20}) is given by
\begin{eqnarray}
\Xi(\beta,\mu)=\prod_{k=1}^M\frac{1}{1-X_k}
\label{eq23ff}
\end{eqnarray}
and formulae (\ref{eq22}) can be rewritten in more specific and simpler form:
\begin{eqnarray}
\langle\bar{U}_k\rangle & = & -\langle N_k\rangle \frac{\partial}{\partial\beta}\ln Z_k^1(\beta) \nonumber \\
                        & = & \langle N_k\rangle\langle\bar{U}_k\rangle_{\upharpoonleft}
\label{eq23g}
\end{eqnarray}
where $\langle\bar{U}_k\rangle_{\upharpoonleft}$ stands for mean disutility of the one-passenger market $k$, where
\begin{eqnarray}
\langle N_k\rangle = \frac{X_k}{1-X_k}
\label{eq23h}
\end{eqnarray}
From eqn (\ref{eq23h}) it follows that $X_k=\langle N_k\rangle/(\langle N_k\rangle+1)<1$ and the grand partition function (\ref{eq23ff}) is always non-singular.

Second moments and correlations are equal to 
\begin{eqnarray}
\mbox{{\it Var}}\,(\bar{U}) & = & \sum_{k=1}^M \mbox{{\it Var}}\,(\bar{U}_k) \nonumber \\
                            & = & \sum_{k=1}^M \big(\langle N_k\rangle\mbox{{\it Var}}\,(\bar{U}_k)_{\upharpoonleft}+\langle\bar{U}_k\rangle_{\upharpoonleft}^2\mbox{{\it Var}}\,(N_k)\big) \nonumber \\
\mbox{{\it Var}}\,(N) & = & \sum_{k=1}^M \mbox{{\it Var}}\,(N_k) \nonumber \\
                      & = & \sum_{k=1}^M \big(\langle N_k\rangle +\langle N_k\rangle^2\big) \nonumber \\
\mbox{{\it cov}}\,(\bar{U},N) & = & \sum_{k=1}^M \mbox{{\it cov}}\,(\bar{U}_k,N_k) \nonumber \\
                              & = & \sum_{k=1}^M \langle\bar{U}_k\rangle_{\upharpoonleft}\mbox{{\it Var}}\,(N_k)
\label{eq23i}
\end{eqnarray}
where $\mbox{{\it Var}}\,(\bar{U}_k)_{\upharpoonleft}$ is the variance of disutility for the one-passenger market $k$.

The canonical case is recovered for non-random $\langle N_k\rangle = N_k$ for all $k$, when $\mbox{{\it Var}}\,(N_k)=0$, as expected.

Worthwhile to note, for non-random market distility $\mbox{{\it Var}}\,(\bar{U}_k)=0$, moments of $\bar{U}$ and $N$ are the same as for the system of ideal bosons (cf. ref. \cite{majka1}).

\subsection{\label{ssec:level4}{Case study 1: the maximum connectivity network}}

\begin{figure}[h]
\begin{center}
\begin{tabular}{cc}
\includegraphics[width=40mm,height=40mm]{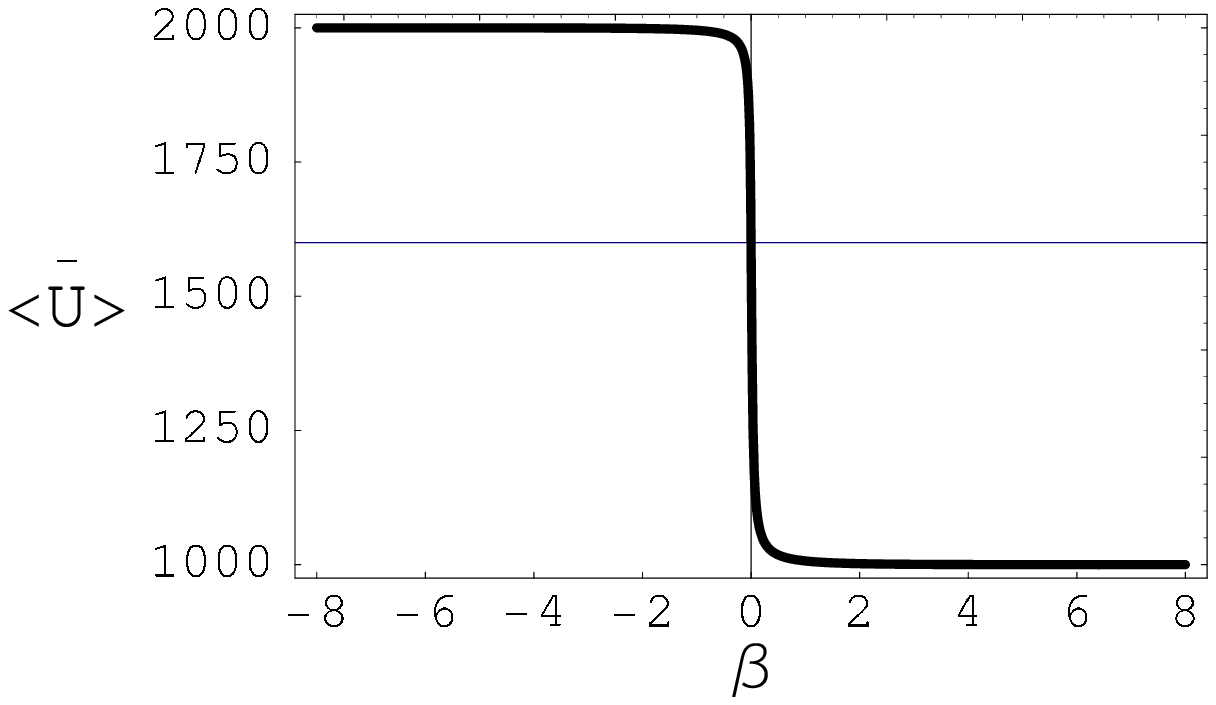} & \includegraphics[width=40mm,height=40mm]{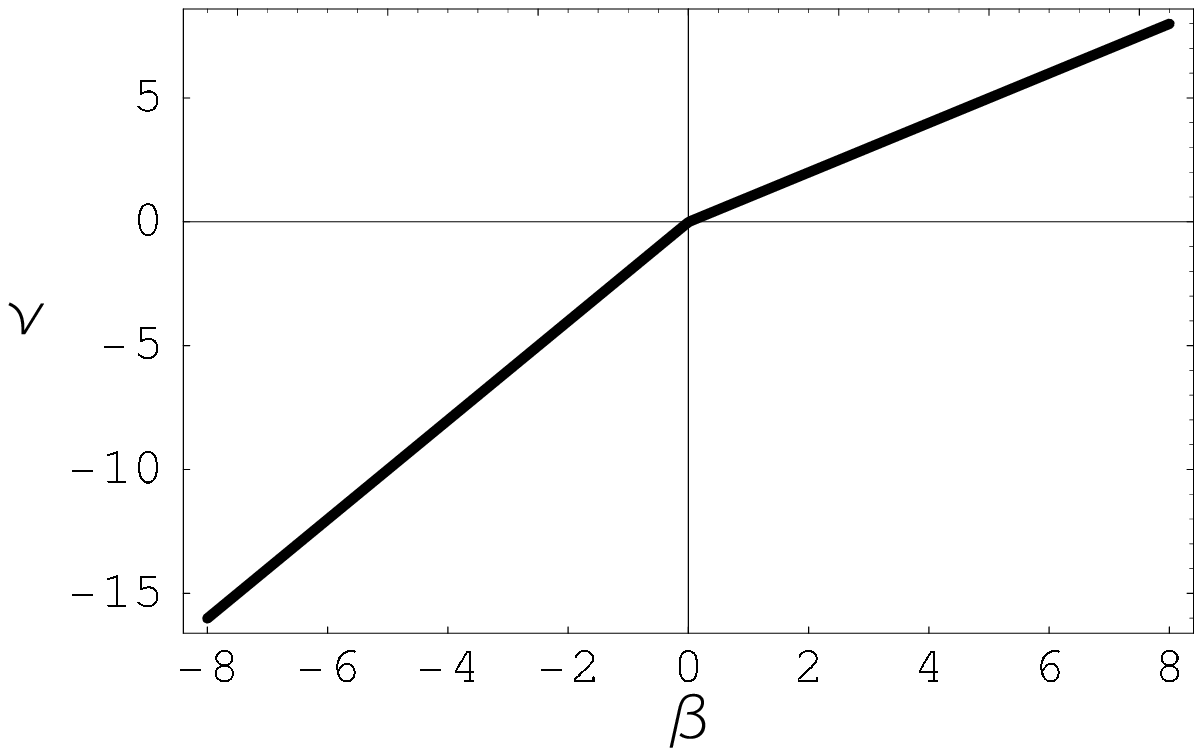} \\
\includegraphics[width=40mm,height=40mm]{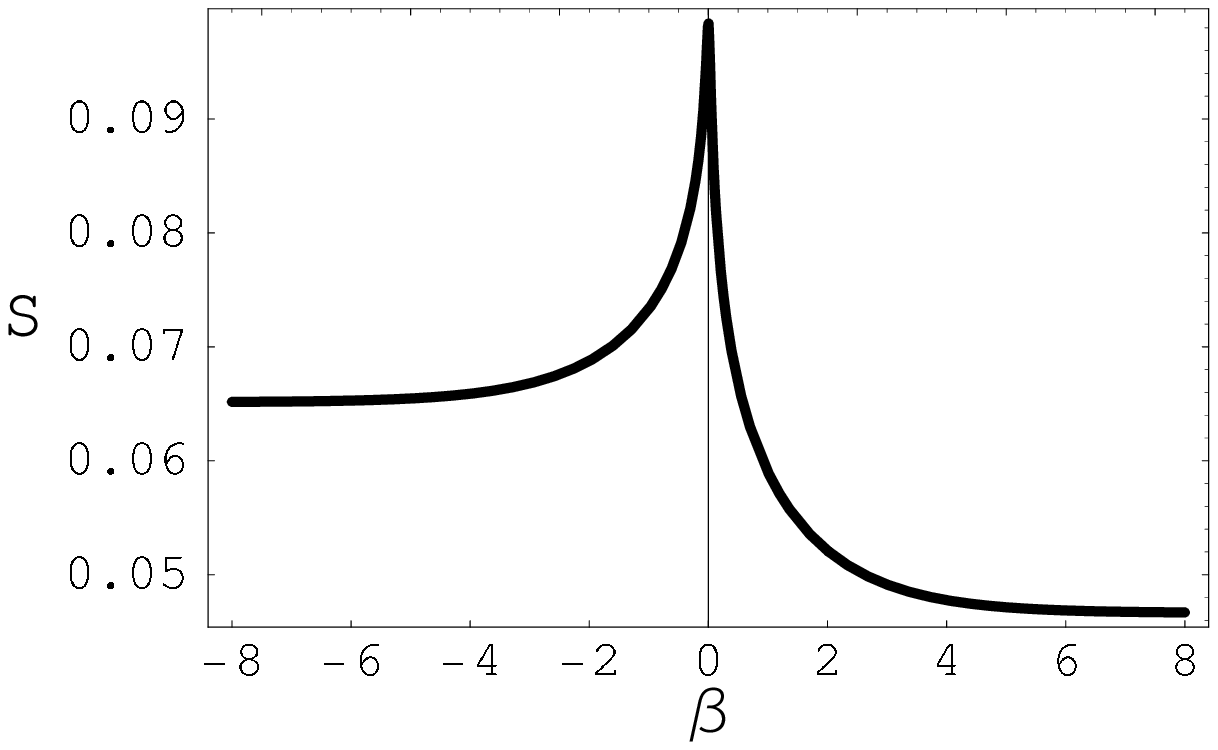} & \includegraphics[width=40mm,height=40mm]{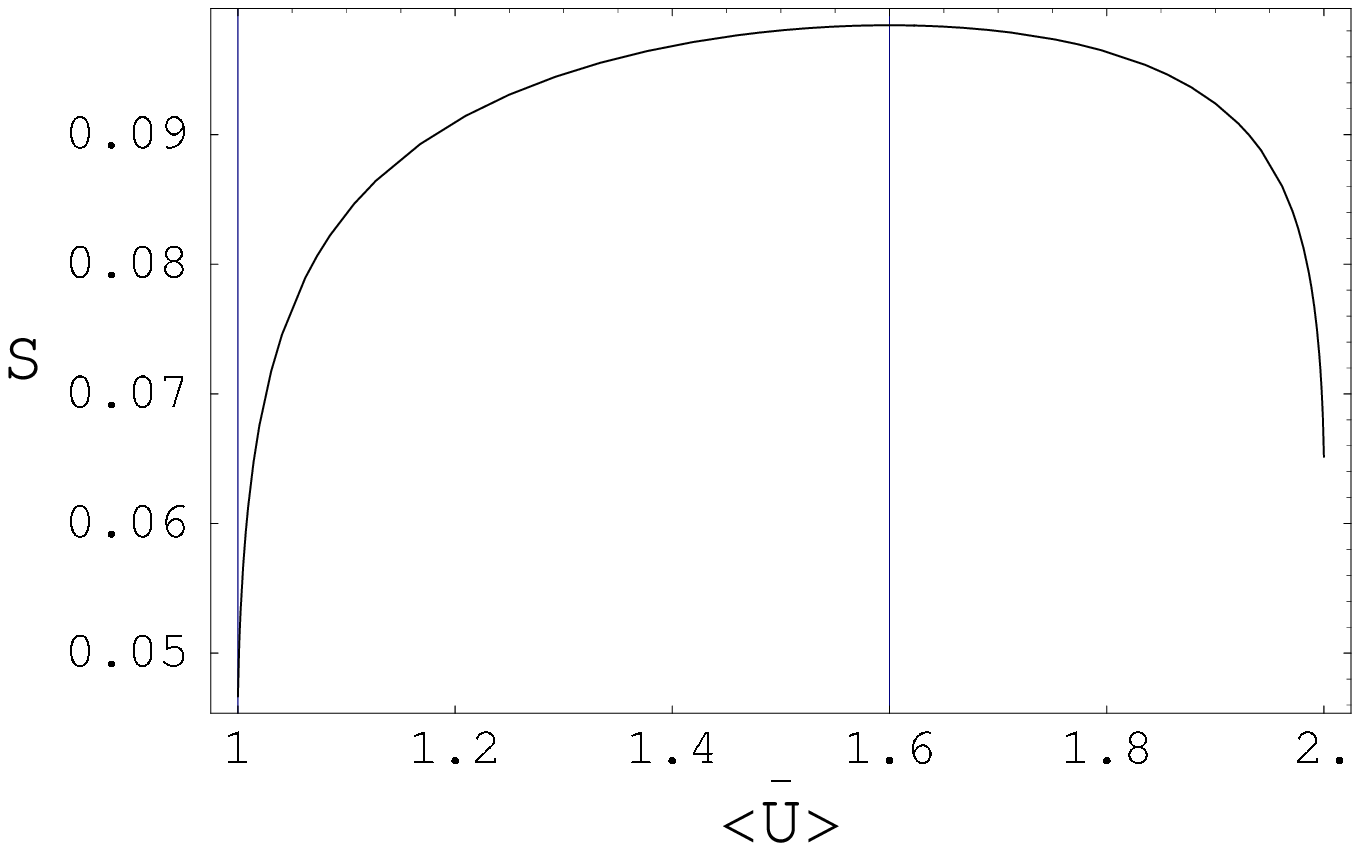}
\end{tabular}
\caption{\label{fig:fg4a} Mean disutility $\langle\bar{U}\rangle$ {\it vs.} inverse temperature $\beta$ (upper left), the $\nu=\beta\mu$ {\it vs.} $\beta$ (upper right), the entropy $S$ {\it vs.} $\beta$ (lower left) and $S$ {\it vs.} $\langle\bar{U}\rangle$ (lower right) for the hub-and-spoke graph with 1 hub and 4 spokes.
The values of $\langle\bar{U}\rangle=1.630$ corresponding to $\beta=0$ are indicated by vertical lines for figures with $\bar{U}$ on the abcissae.}
\end{center}
\end{figure}

\begin{figure}[h]
\begin{center}
\begin{tabular}{cc}
\includegraphics[width=40mm,height=40mm]{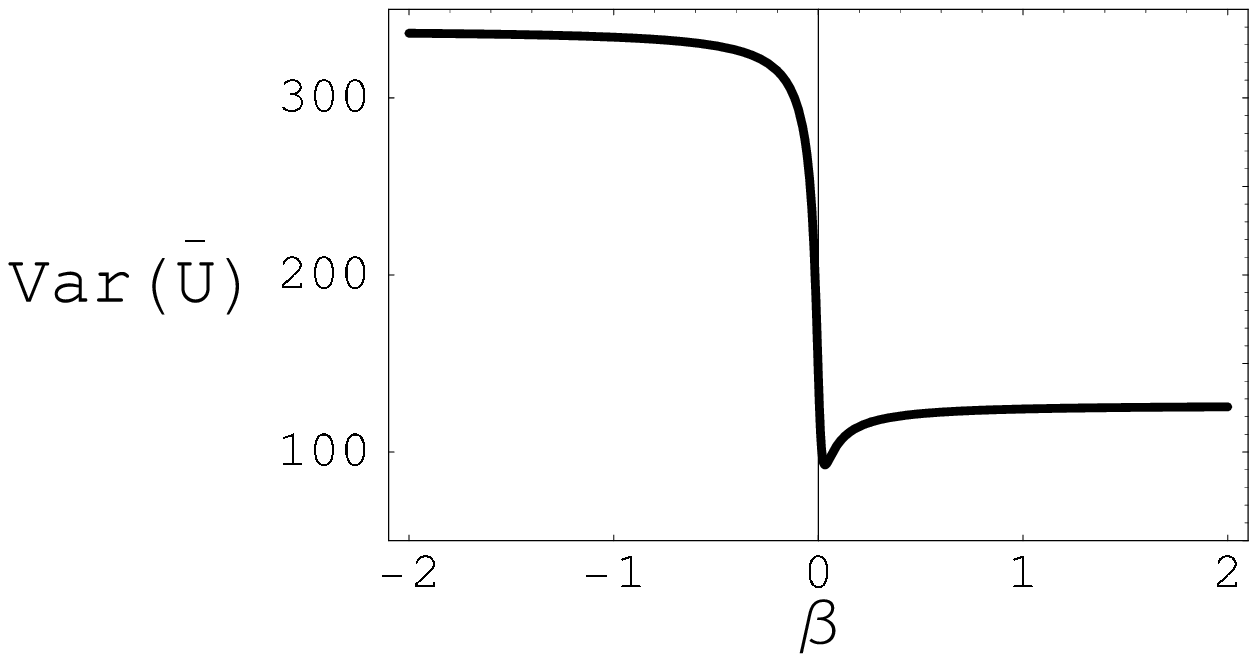} & \includegraphics[width=40mm,height=40mm]{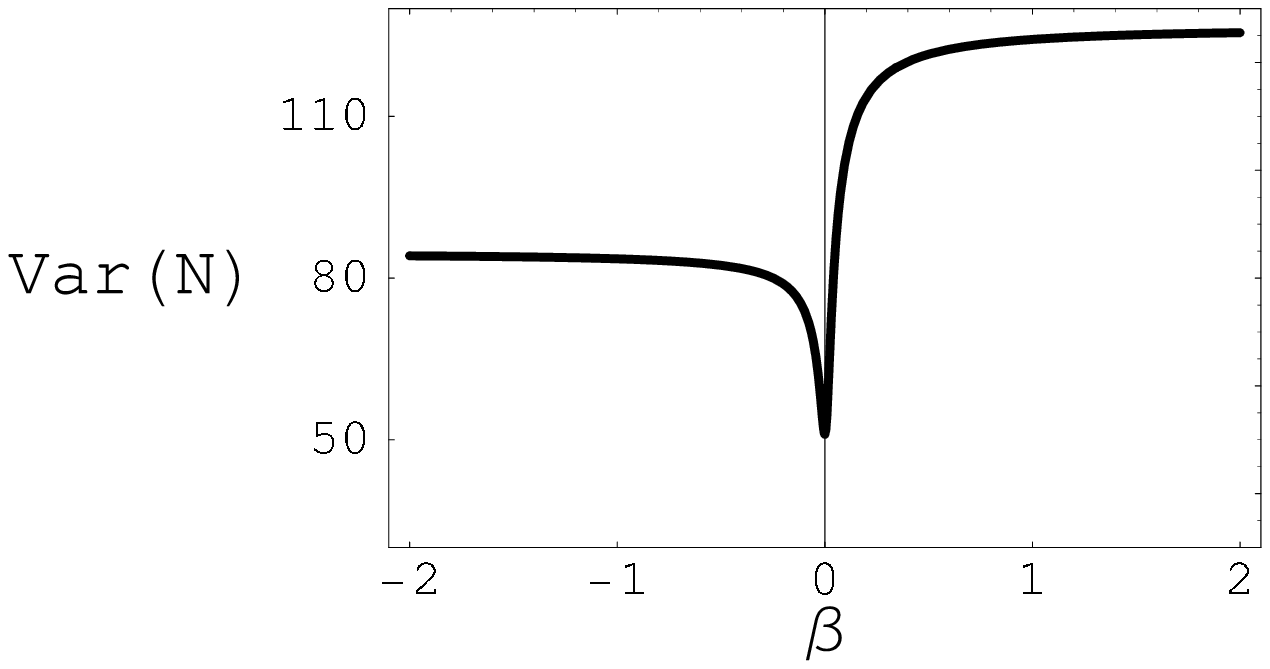} \\
\includegraphics[width=40mm,height=40mm]{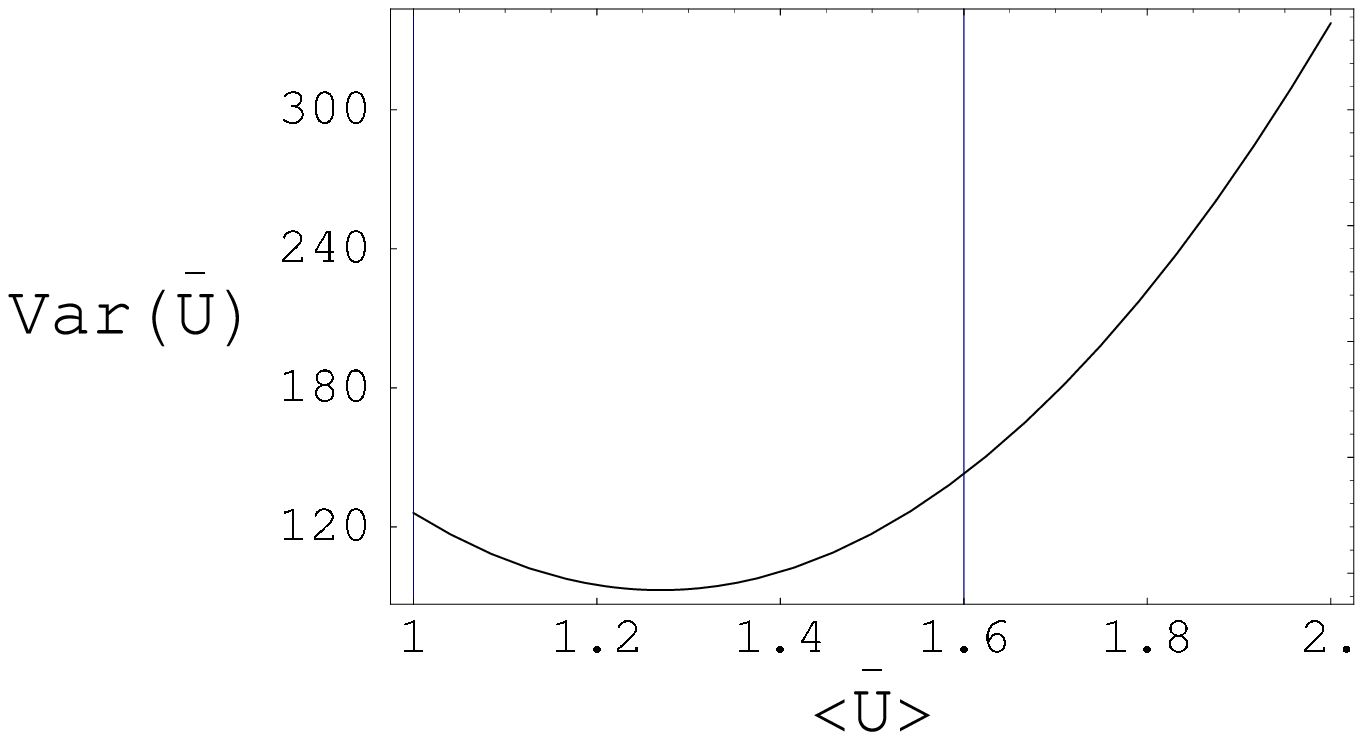} & \includegraphics[width=40mm,height=40mm]{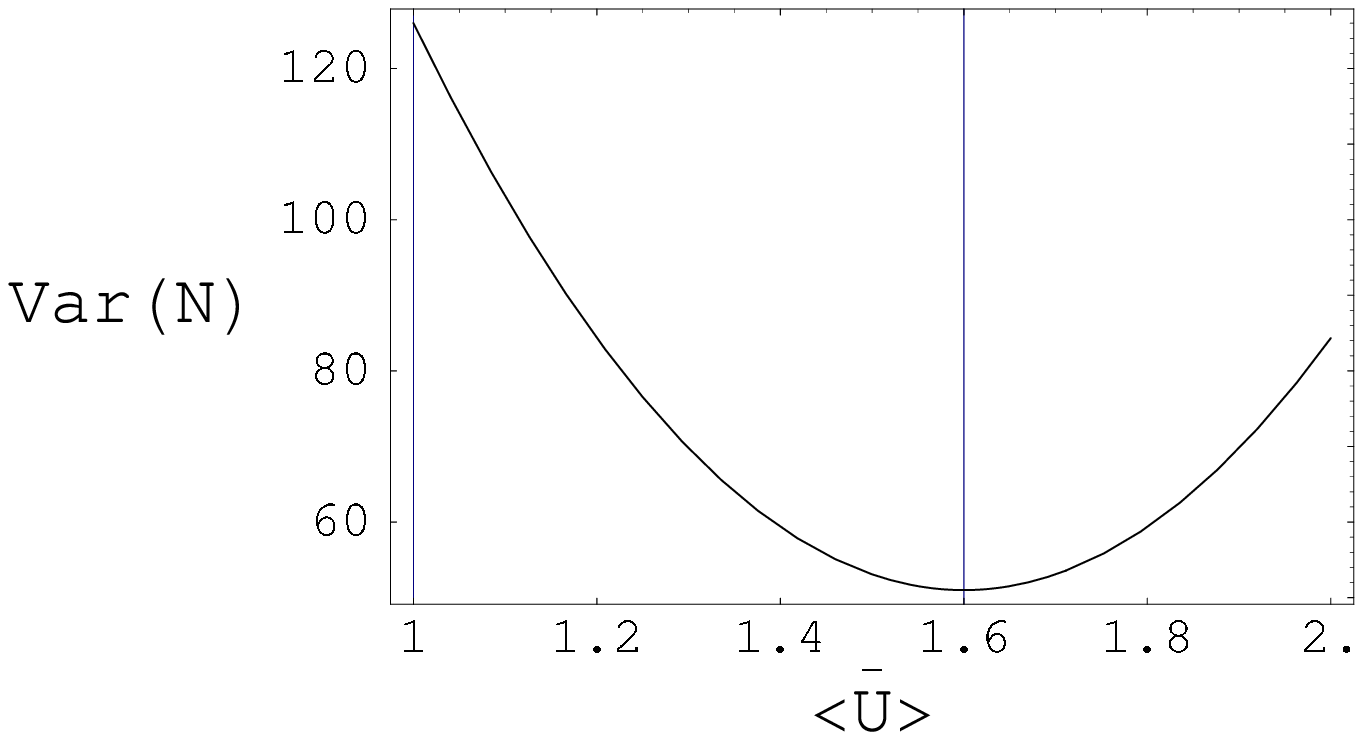} \\
\includegraphics[width=40mm,height=40mm]{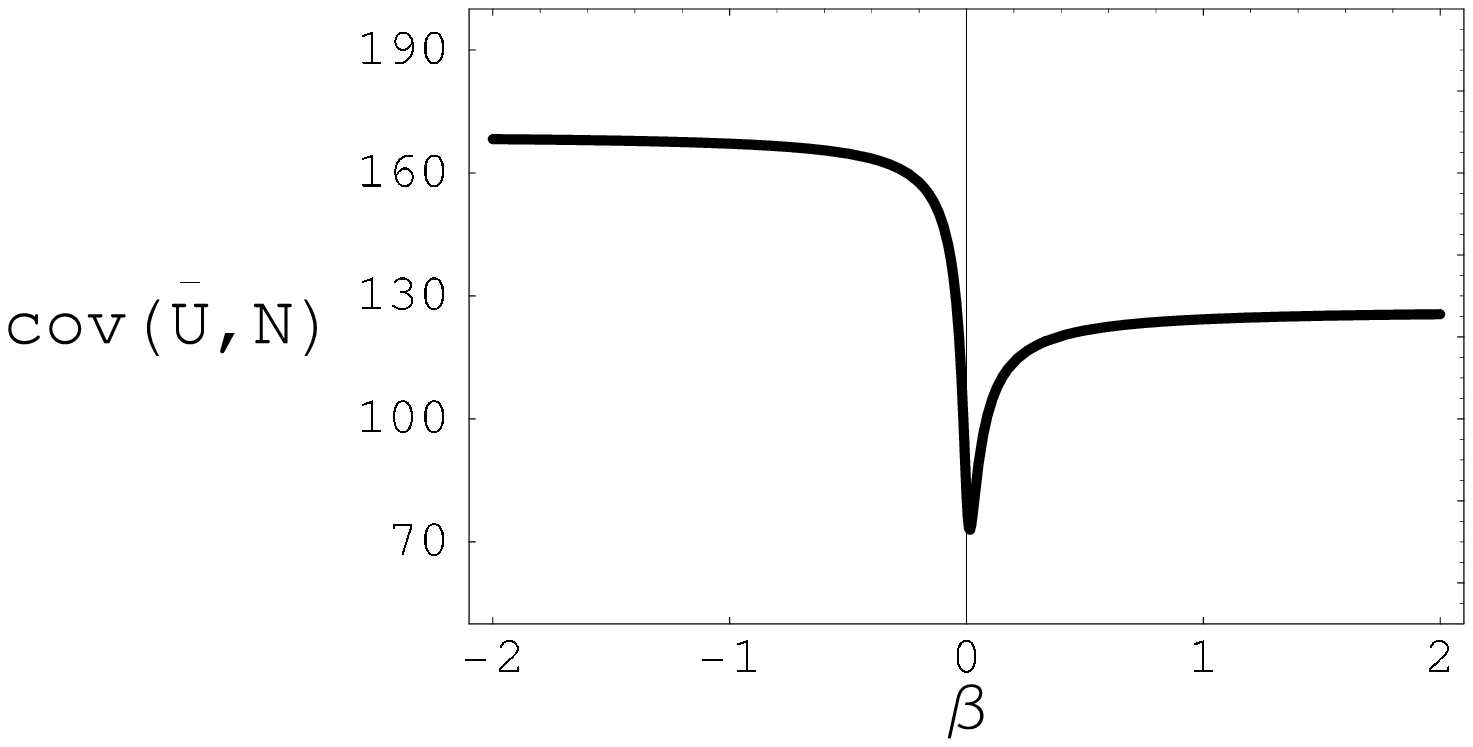} & \includegraphics[width=40mm,height=40mm]{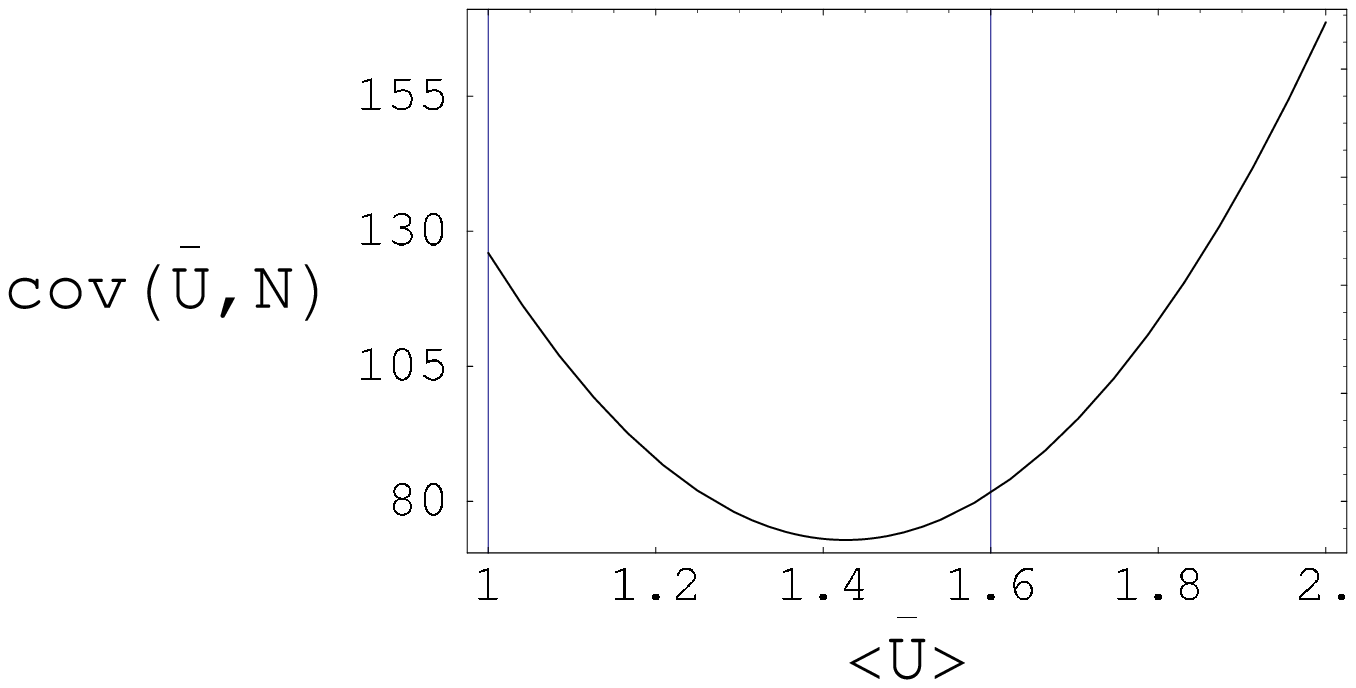}
\end{tabular}
\caption{\label{fig:fg4b} The variance of disuility {\it Var}$\,(\bar{U})$ {\it vs.} inverse temperature $\beta$ (upper left), the variance of the number of individuals {\it Var}$\,(N)$ {\it vs.} $\beta$ (upper right), {\it Var}$\,(\bar{U})$ {\it vs.} $\bar{U}$ (middle left), {\it Var}$\,(N)$ {\it vs.} $\bar{U}$ (middle right), the covariance of $\bar{U}$ and $N$ {\it cov}$\,(\bar{U},N)$ {\it vs.} $\beta$ (lower left) and {\it cov}$\,(\bar{U},N)$ {\it vs.} $U$ (lower right) for the hub-and-spoke graph with 1 hub and 4 spokes. 
The values of $\langle\bar{U}\rangle=1.630$ corresponding to $\beta=0$ are indicated by vertical lines in figures with $\langle\bar{U}\rangle$ on the abscissae.}
\end{center}
\end{figure}

\begin{figure}[h]
\begin{center}
\begin{tabular}{cc}
\includegraphics[width=40mm,height=40mm]{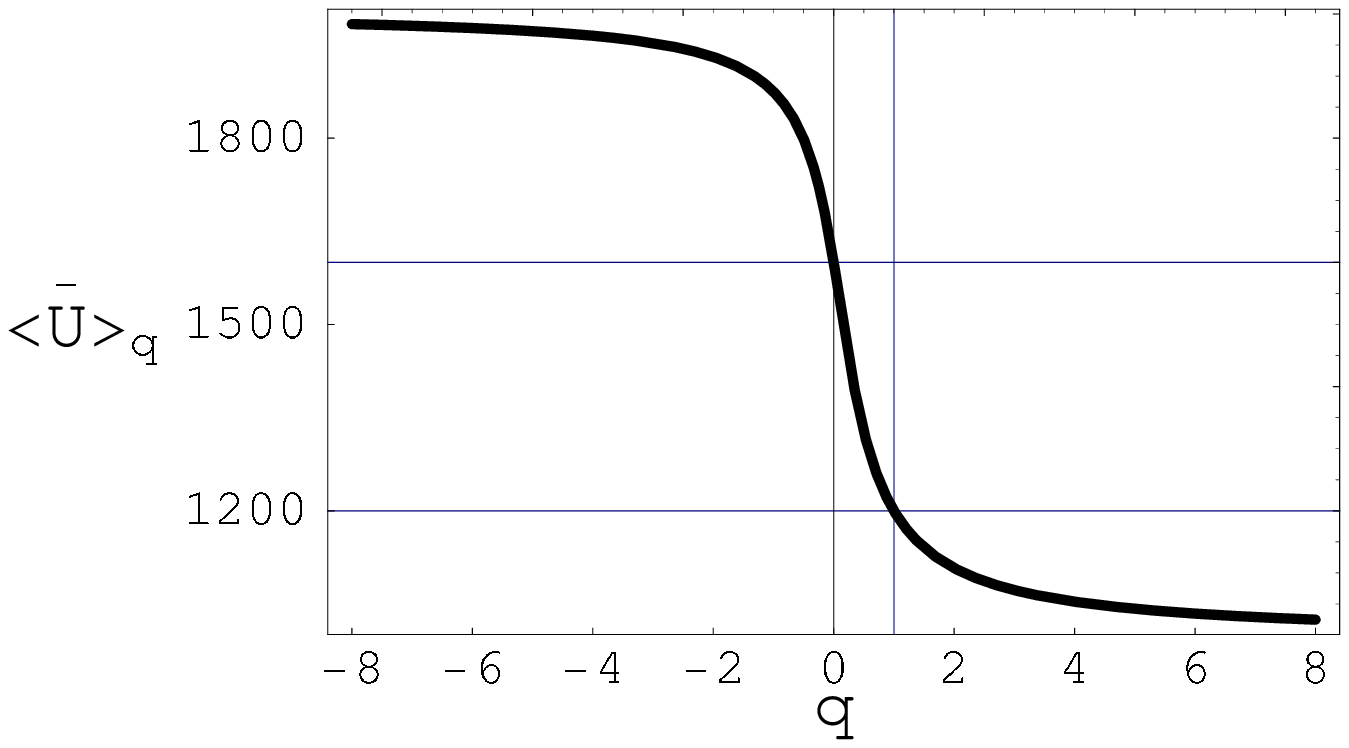} & \includegraphics[width=40mm,height=40mm]{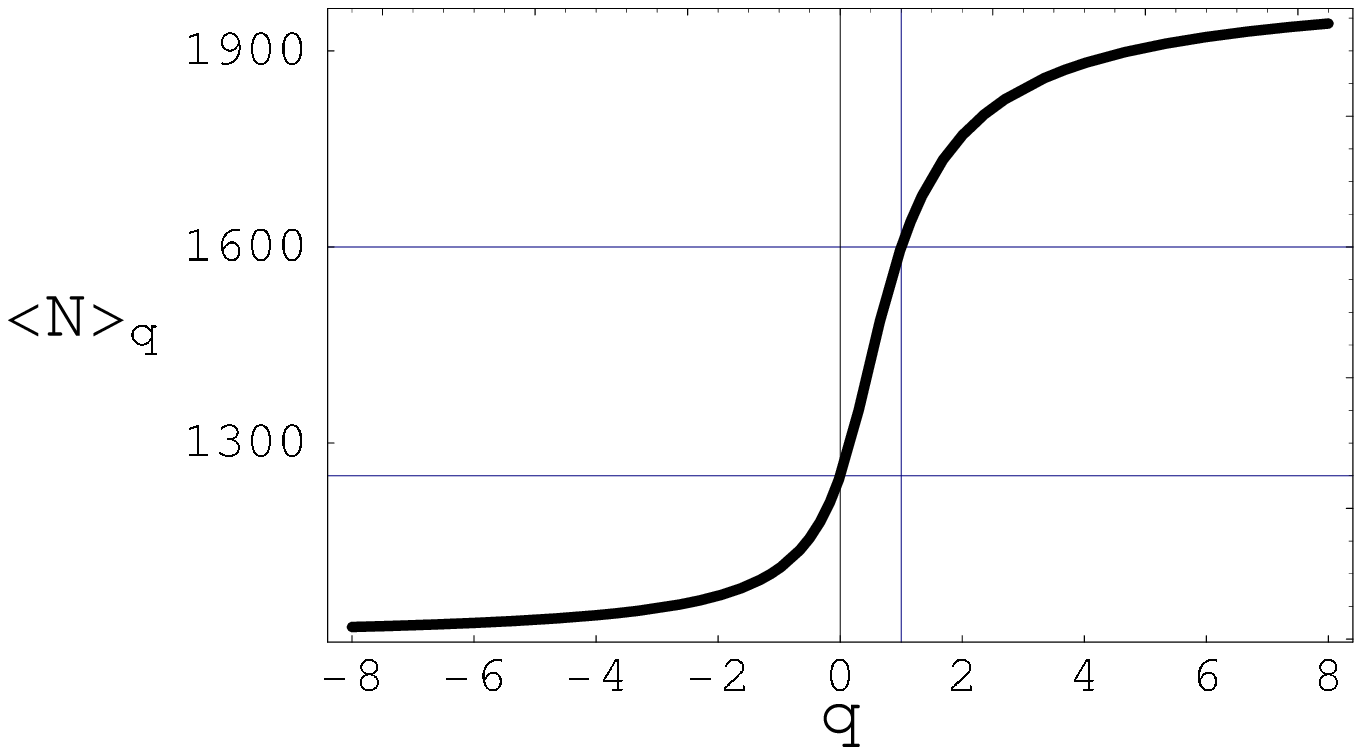} \\
\includegraphics[width=40mm,height=40mm]{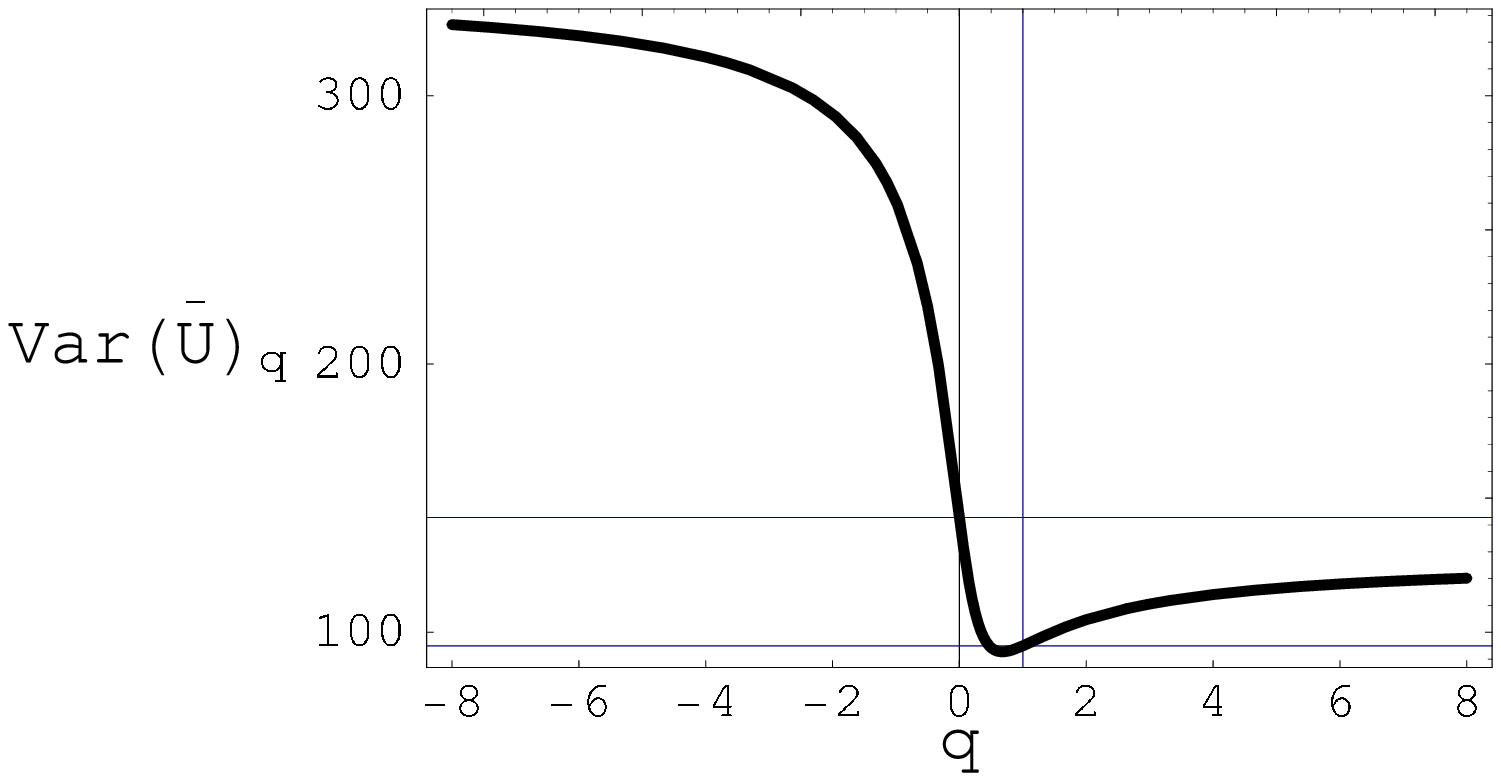} & \includegraphics[width=40mm,height=40mm]{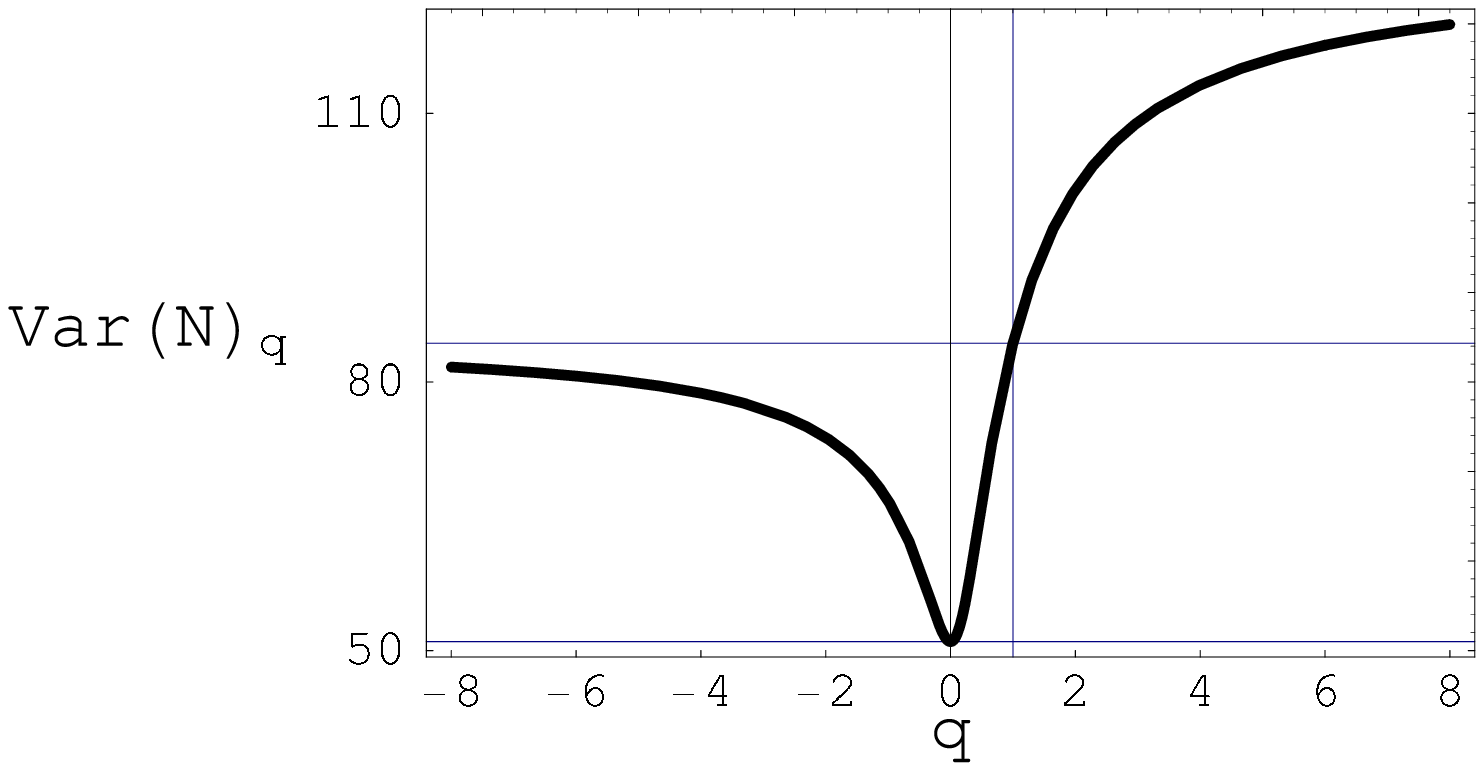} \\
\includegraphics[width=40mm,height=40mm]{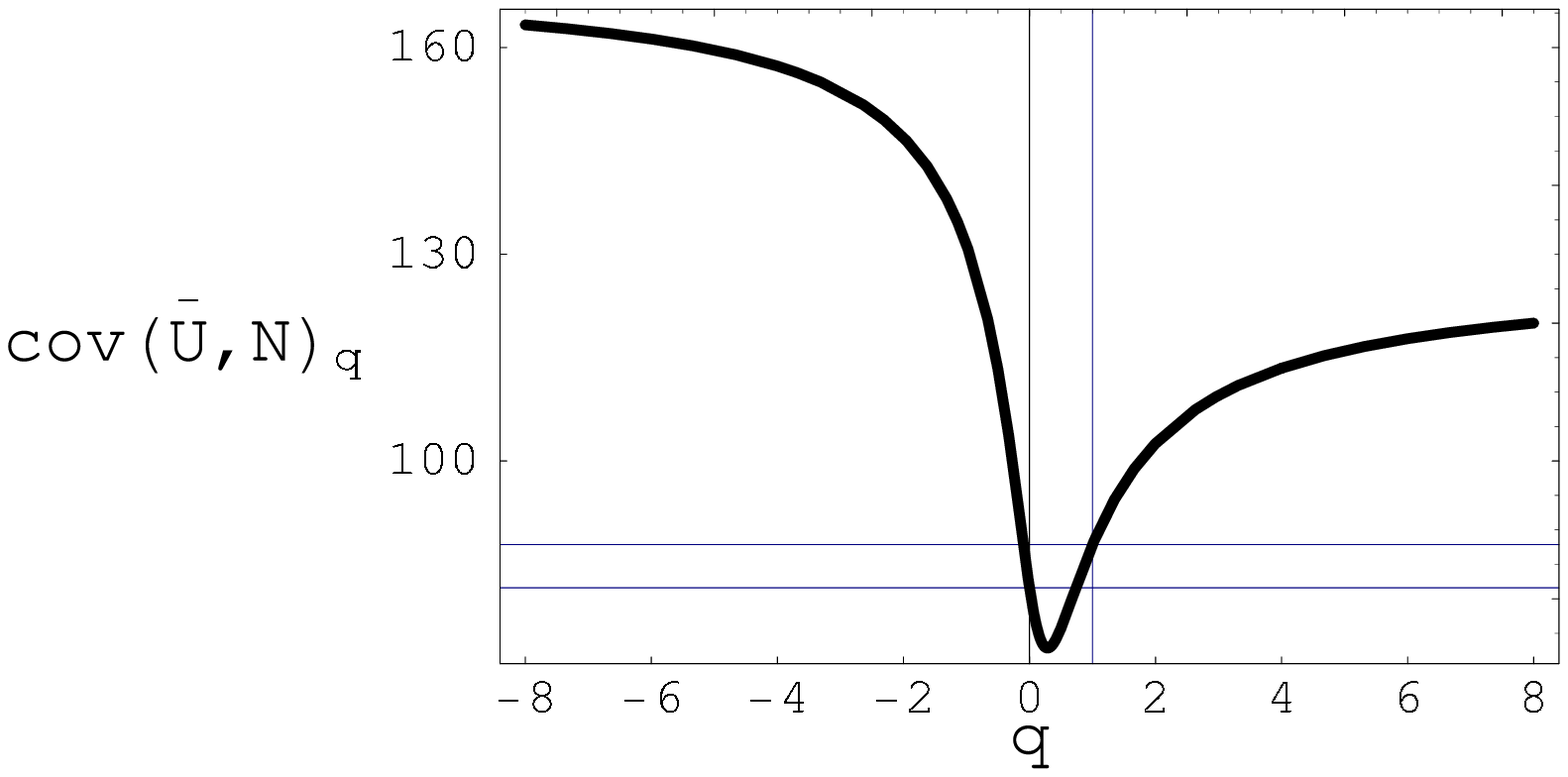} & \includegraphics[width=40mm,height=40mm]{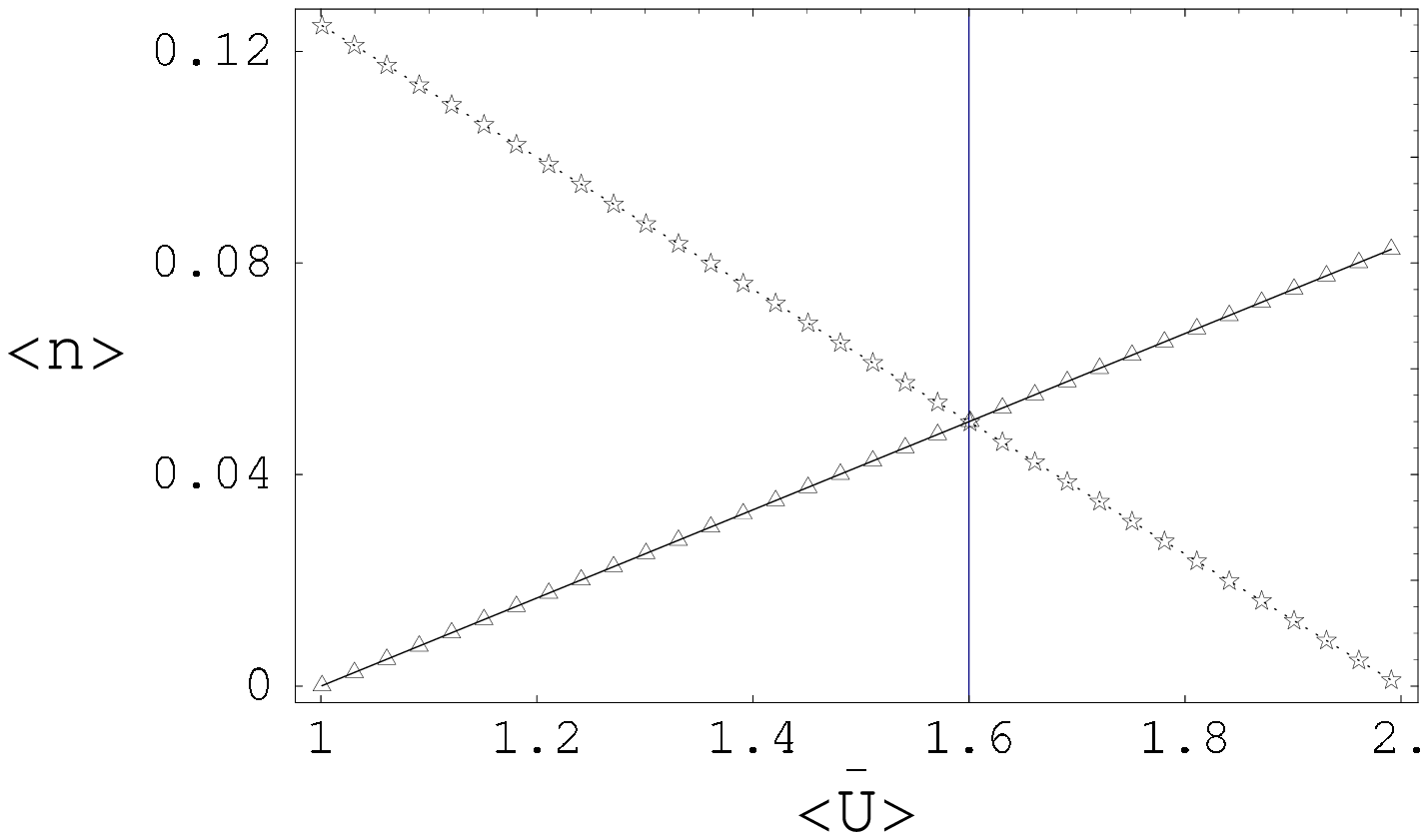}
\end{tabular}
\caption{\label{fig:fg4c} The $q$-mean disutility $\langle\bar{U}\rangle_q$ {\it vs.} $q$ (upper left) and the $q$-mean number of individuals $\langle N\rangle_q$ {\it vs.} $q$ (upper right), the $q$-variance of disutility {\it Var}$\,(\bar{U})_q$ (middle left), the $q$-variance of the number of individuals {\it Var}$\,(N)_q$ (middle right) and the $q$-covariance of $\bar{U}$ and $N$ {\it cov}$_q\,(\bar{U},N)$ (bottom left) for the hub-and-spoke graph with 1 hub and 4 spokes. The ratio of passengers in a given type of market as a function of mean disutility is displayed in the lower right panel. The vertical line at $\langle\bar{U}\rangle=1.630$ corresponds to $\beta=0$.}
\end{center}
\end{figure}

If the network graph consists of $\Gamma$ nodes, the total number of markets is $M=\Gamma^2-\Gamma$.
Using combinatorics and denoting by $v$ disutility for one segment of any route, same everywhere, one finds the partition function for the $k$-th market
\begin{eqnarray}
Z_k(\beta) & = & \Big(\sum_{l=1}^{\Gamma-1}\frac{(\Gamma-2)!}{(\Gamma-1-l)!}e^{-\beta vl}\Big)^{N_k} 
\label{eq24}
\end{eqnarray}
and $Z(\beta)$ is given by eqn. (\ref{eq11}).
The combinatorial factor accounts for the degeneration of the state of given disutility and is equal to the number of routes of the same length.
For fixed $N=\sum_{k=1}^M N_k$
\begin{eqnarray}
\langle \bar{U}\rangle & = & Nv\Big(\Gamma-1-\frac{\sum_{l=0}^{\Gamma-2}le^{\beta vl}/l!}{\sum_{l=0}^{\Gamma-2}e^{\beta vl}/l!}\Big) \nonumber \\
         & \stackrel{\Gamma\gg 1}{\longrightarrow} & Nv(\Gamma-e^{\beta v}) 
\label{eq25}
\end{eqnarray}
and 
\begin{eqnarray}
\mbox{\it Var}\,(\bar{U}) & \stackrel{\Gamma\gg 1}{\longrightarrow} & N v^2e^{\beta v} \nonumber \\
                    & = & v(Nv\Gamma-\langle \bar{U}\rangle)
\label{eq26}
\end{eqnarray}
where we used equilibrium temperature $\beta_{eq}=\frac{1}{v}\ln(\Gamma-\frac{\langle\bar{U}\rangle}{Nv})$ determined from eqn. (\ref{eq25}).

In general case we failed to find analytic formulae for moments of $N$ and correlations between $N$ and $\bar{U}$. 
The chemical potential $\mu$ can be determined, provided $\langle N\rangle$ is known.

It is instructive to discuss analytic solutions for the simplest non-trivial case of one market on the $\Gamma=3$ graph and assuming $v=1$ in order to elliminate trivial factors.
Using eqns (\ref{eq23g}) and (\ref{eq23h}), one easily finds equations for equilibrium $\beta$ and $\mu$
\begin{eqnarray}
\beta_{eq}=\ln\frac{2\langle N\rangle -\langle\bar{U}\rangle}{\langle\bar{U}\rangle-\langle N\rangle}
\label{eq27}
\end{eqnarray}
and
\begin{eqnarray}
\nu_{eq} & = & (\beta\mu)_{eq} \nonumber \\
         & = & 2\ln\frac{2\langle N\rangle -\langle\bar{U}\rangle}{1+\langle N\rangle}-\ln\frac{\langle\bar{U}\rangle-\langle N\rangle}{1+\langle N\rangle}
\label{eq28}
\end{eqnarray}
Using eqns (\ref{eq27}),(\ref{eq28}) and (\ref{eq23i}), we obtain
\begin{eqnarray} 
\mbox{{\it Var}}\,(\bar{U}) & = & -2\langle N\rangle+3\langle\bar{U}\rangle+\langle\bar{U}\rangle^2 \nonumber \\
& = & (-\langle\bar{U}\rangle^2/\langle N\rangle+3\langle\bar{U}\rangle-2\langle N\rangle) \nonumber \\
& + & (\langle\bar{U}\rangle^2/\langle N\rangle+\langle\bar{U}\rangle^2)
\label{eq29}
\end{eqnarray}
where the first term corresponds to the canonical ensemble, with fixed number of individuals $\langle N\rangle$, and the second term comes from randomization of the number of individuals in the grand canonical ensemble.
Since in our model $\langle\bar{U}\rangle$ is proportional to $\langle N\rangle$, the asymptotic behaviour of both terms can be investigated expanding $\langle\bar{U}\rangle$ around $\langle N\rangle$.
Asymptotically, we find that the first term vanishes and the second term behaves as $\langle N\rangle^2$ for $\langle N\rangle\gg 1$.

In addition, we calculate for this case
\begin{eqnarray}
\mbox{{\it Var}}\,(N)=\langle N\rangle(\langle N\rangle+1)
\label{eq30}
\end{eqnarray}
and
\begin{eqnarray}
\mbox{{\it cov}}\,(\bar{U},N)=\langle\bar{U}\rangle(\langle N\rangle+1)
\label{eq31}
\end{eqnarray}
and we see that both {\it Var}$\,(N)$ and covariance of $\langle\bar{U}\rangle$ and $\langle N\rangle$ depend quadratically on $\langle N\rangle$ in the limit of $\langle N\rangle\rightarrow\infty$.
For completness, the entropy in this case is equal to
\begin{eqnarray}
S & = & -\ln\frac{1}{1+\langle N\rangle}+(-2\langle N\rangle+\langle\bar{U}\rangle)\ln\frac{2\langle N\rangle-\langle\bar{U}\rangle}{1+\langle N\rangle} \nonumber \\
& + & (\langle N\rangle-\langle\bar{U}\rangle)\ln\frac{-\langle N\rangle+\langle\bar{U}\rangle}{1+\langle N\rangle}.
\label{eq31a}
\end{eqnarray}

We also investigated analytically higher $\Gamma$ and found qualitatively the same bahaviour as for $\Gamma=3$.

We performed numerical simulation for the maximum conectivity network with $\Gamma=5$ and $\langle N\rangle=1000$ and displayed our results in Figs \ref{fig:fg3a} and \ref{fig:fg3b}.
Both $\beta$ and $\mu$ were treated formally as Lagrange multipliers and thus we allowed for both positive and negative values of those parameters.
The behaviour of $\langle\bar{U}\rangle$, {\it Var}$\,(\bar{U})$ and $S$ with temperature is intuitively appealing for $\beta>0$.

The dependence $S(\beta)$ exhibits the same tendency for systems of bosons with inversed populations alowed, i.e. the entropy increases with temperature for both positive and nagative temperatures.
Moreover, the entropy $S$ exists in the limits of $\beta\rightarrow\pm\infty$, where $S\stackrel{\beta\rightarrow\infty}{\longrightarrow}{\it const.}$ and $S\stackrel{\beta\rightarrow -\infty}{\longrightarrow}\mbox{\it const.}\,'$.
For the canonical ensemble ${\it const.}=0$ and for random $N$, in general ${\it const.}\ne 0$.

Dependence of $\nu$ on $\beta$ exhibits steeper slope for $\beta<0$ which is related to less steep dependence of $\langle\bar{U}\rangle(\beta)$ or $\beta<0$, as compared to $\beta>0$.

It is interesting to note that {\it Var}$\,(N)$ is insensitive to neither $\beta$ nor $\langle\bar{U}\rangle$ which reflects the maximal symmetry of the maximum connectivity network.
Whatever the temperature is, relative populations for all routes are the same and the width of distribution of $N$ does not depend on temperature.

Results of simulations for $q$-moments are displayed in Figs \ref{fig:fg3c}.
One observes the same dependence of the $q$-moments on $q$ and explicit $\beta$-dependence of the ordinary moments in Figs \ref{fig:fg3b}. 
It clearly confirms interpretation of $q$ as a temperature scaling parameter, as mentioned in chapt. III.E.
Consistently, the values of ordinary moments for $\beta=0$ are the same as $q$-moments for $q=0$.

Fig. \ref{fig:fg3b} (bottom right) presents the ratio of passengers in a given market as a function of utility, where each point correspons to one overall mean disutility, or temperature. 
Clearly, there is only one type of markets because of the maximal symmetry of this network.
Vertical line at $\langle\bar{U}\rangle=3.062$ divides the domain of $\langle\bar{U}\rangle$ into subregions corresponding to $\beta>0$ (to the left) and $\beta<0$ (to the right).

\subsection{\label{ssec:level4}{Case study 2: the hub-and-spoke network}}

\begin{figure}[h]
\begin{center}
\begin{tabular}{cc}
\includegraphics[width=40mm,height=40mm]{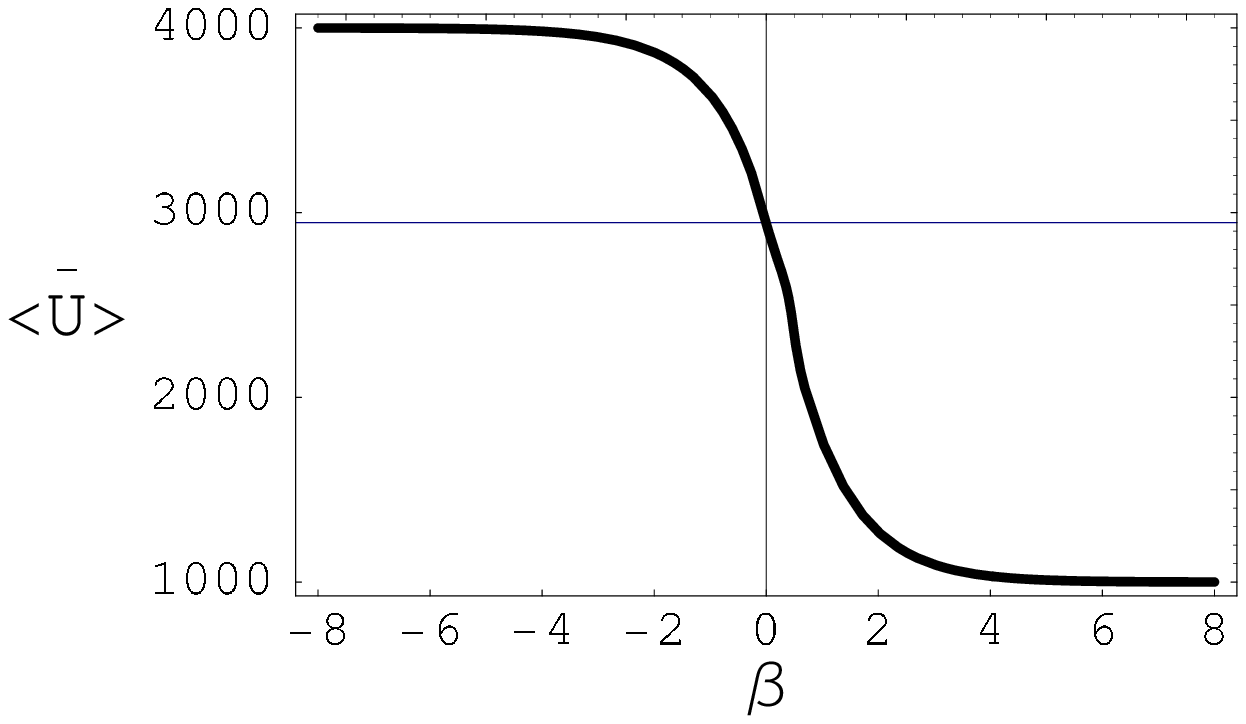} & \includegraphics[width=40mm,height=40mm]{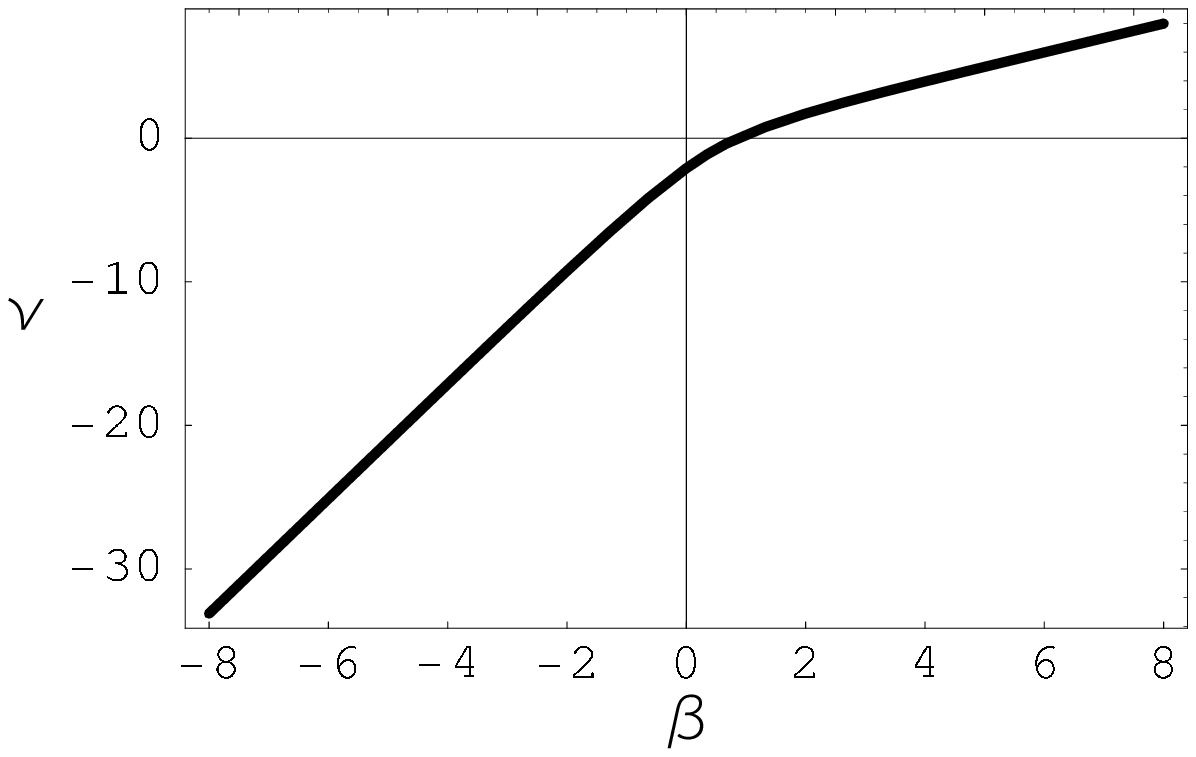} \\
\includegraphics[width=40mm,height=40mm]{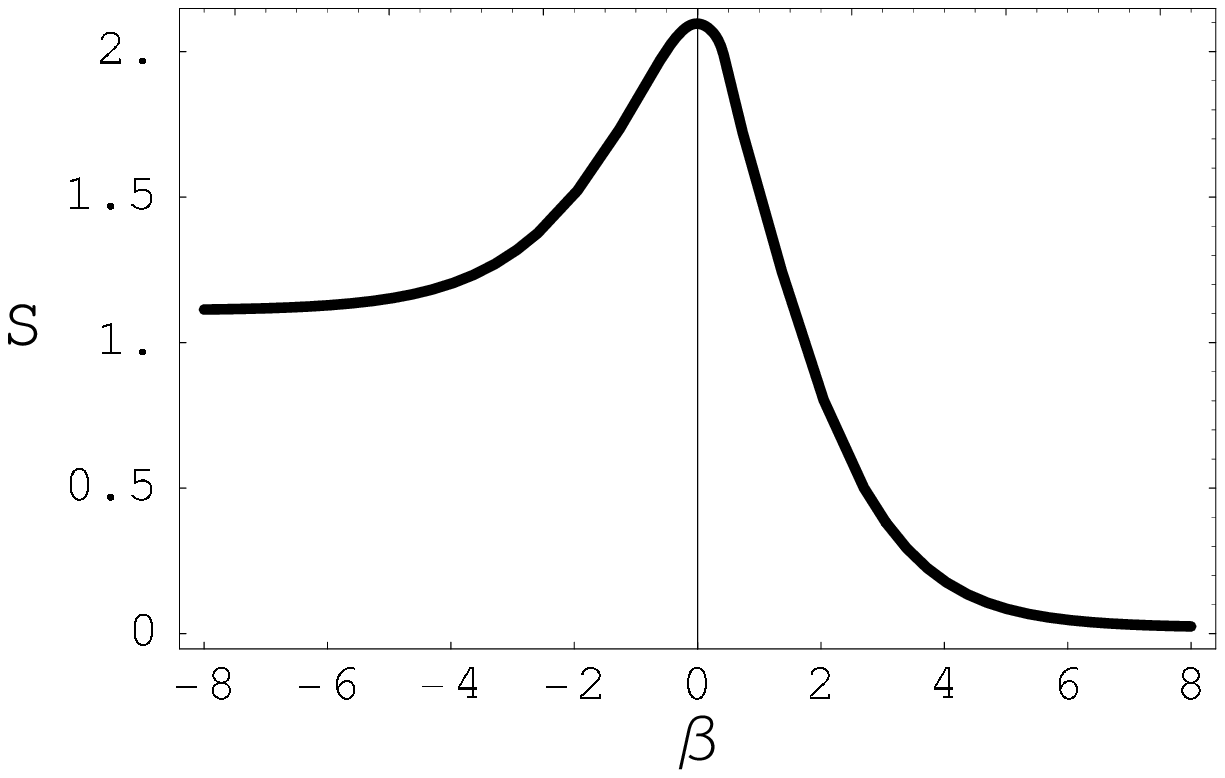} & \includegraphics[width=40mm,height=40mm]{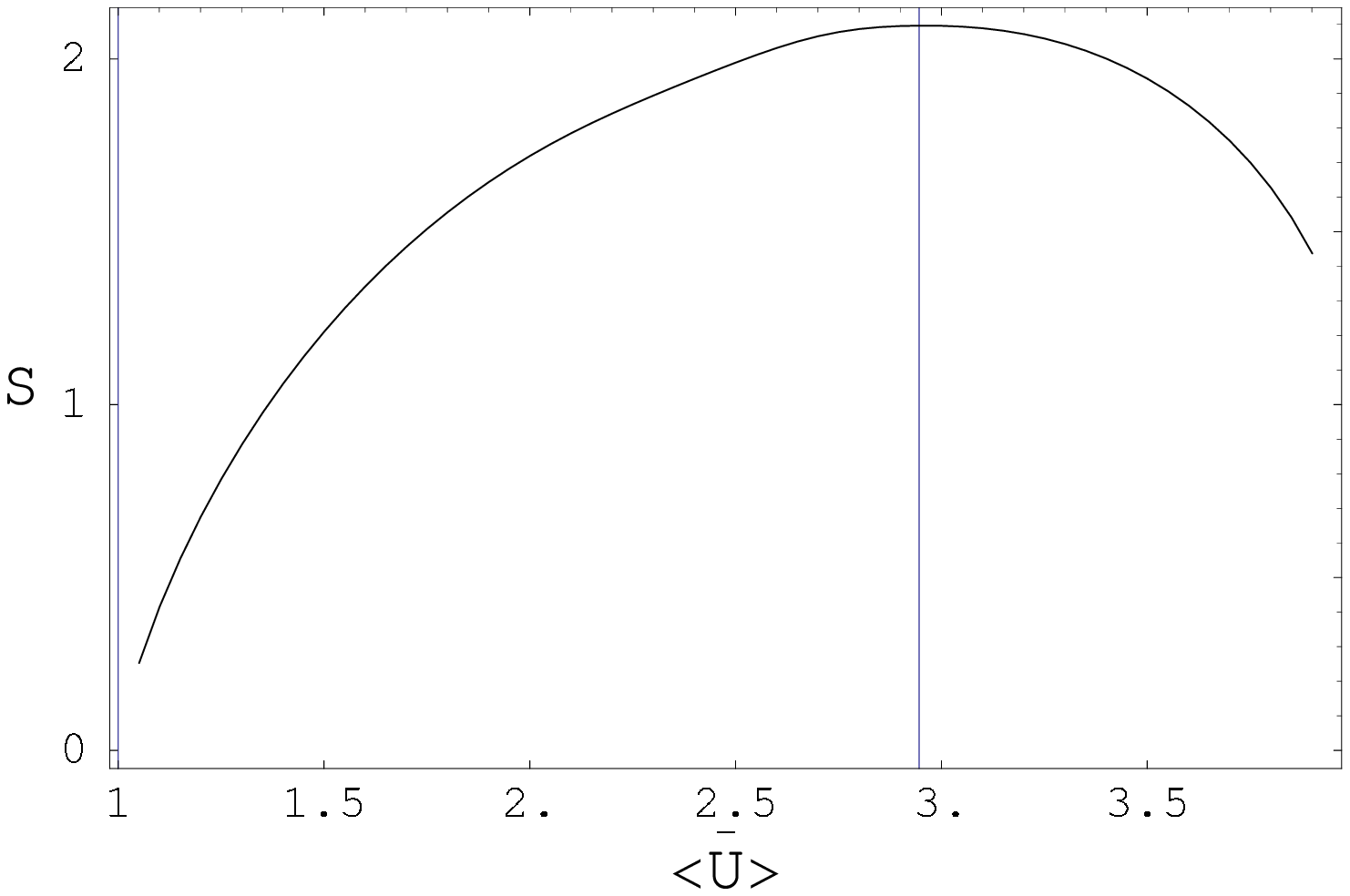}
\end{tabular}
\caption{\label{fig:fg5a} Mean disutility $\langle\bar{U}\rangle$ {\it vs.} inverse temperature $\beta$ (upper left), the $\nu=\beta\mu$ {\it vs.} $\beta$ (upper right), the entropy $S$ {\it vs.} $\beta$ (lower left) and $S$ {\it vs.} $\langle\bar{U}\rangle$ (lower right) for the spider-web graph with 1 hub and 4 spokes.
The values of $\langle\bar{U}\rangle=2.954$ corresponding to $\beta=0$ are indicated by vertical lines for figures with $\bar{U}$ on the abcissae.}
\end{center}
\end{figure}

\begin{figure}[h]
\begin{center}
\begin{tabular}{cc}
\includegraphics[width=40mm,height=40mm]{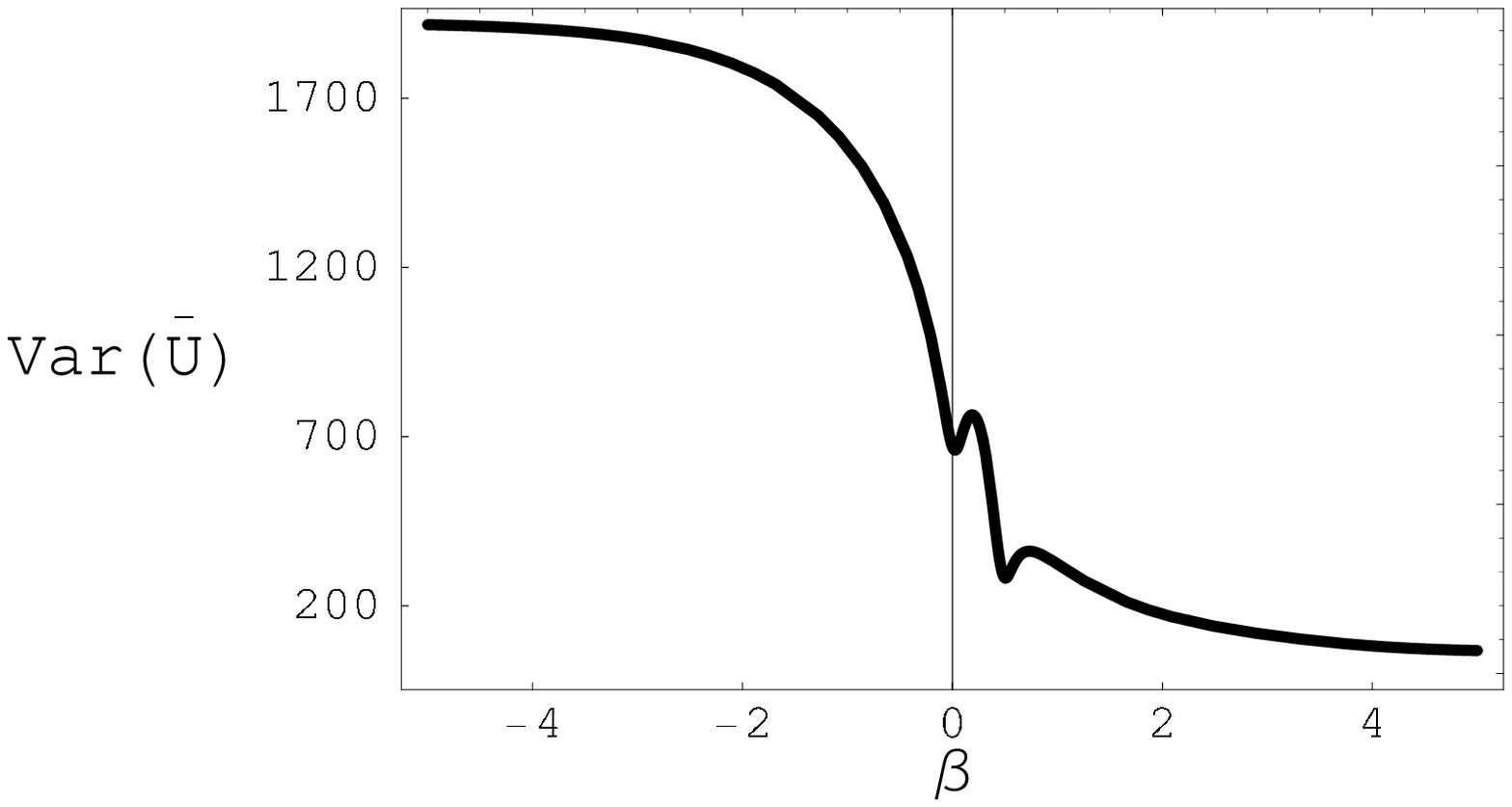} & \includegraphics[width=40mm,height=40mm]{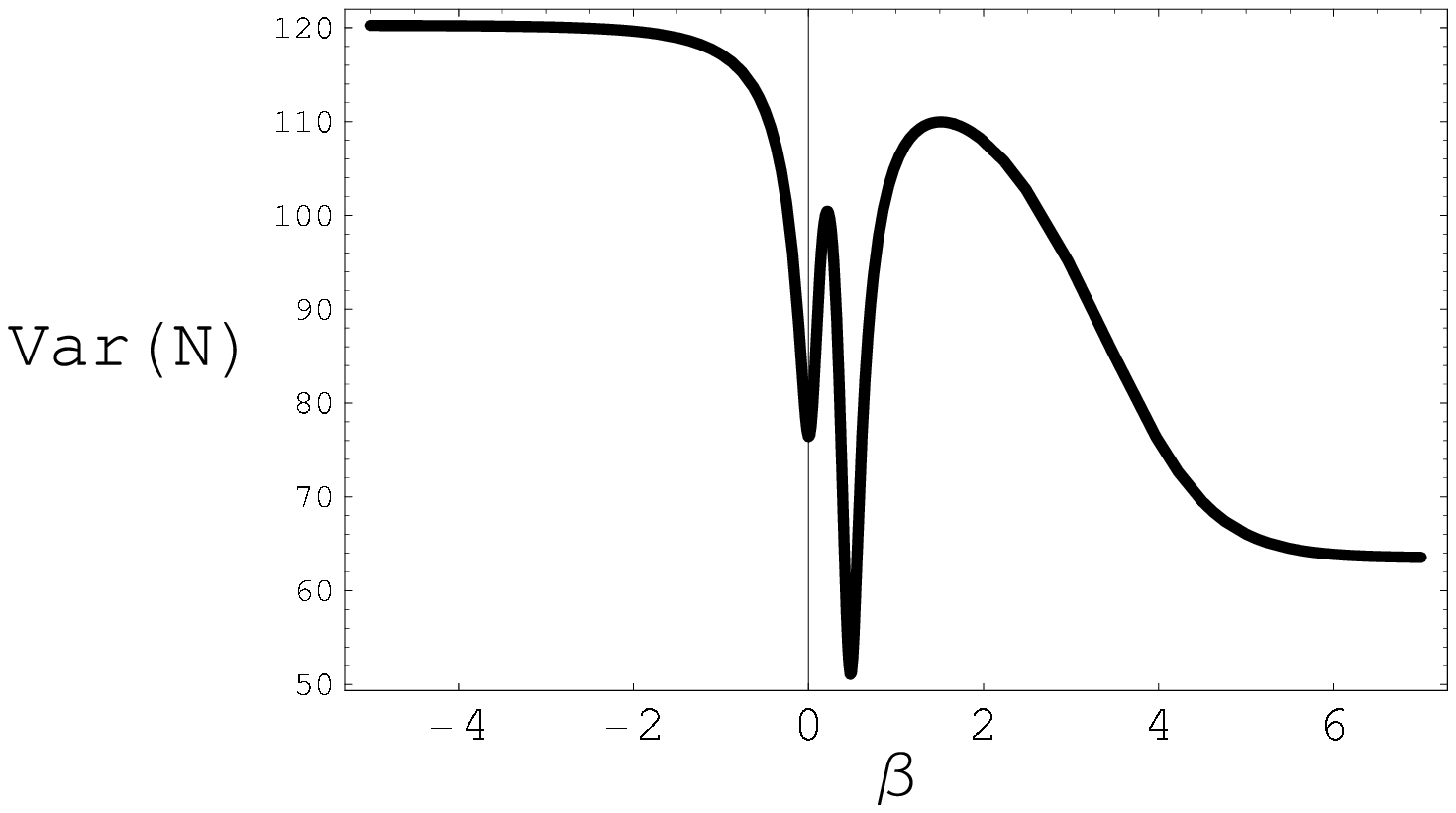} \\
\includegraphics[width=40mm,height=40mm]{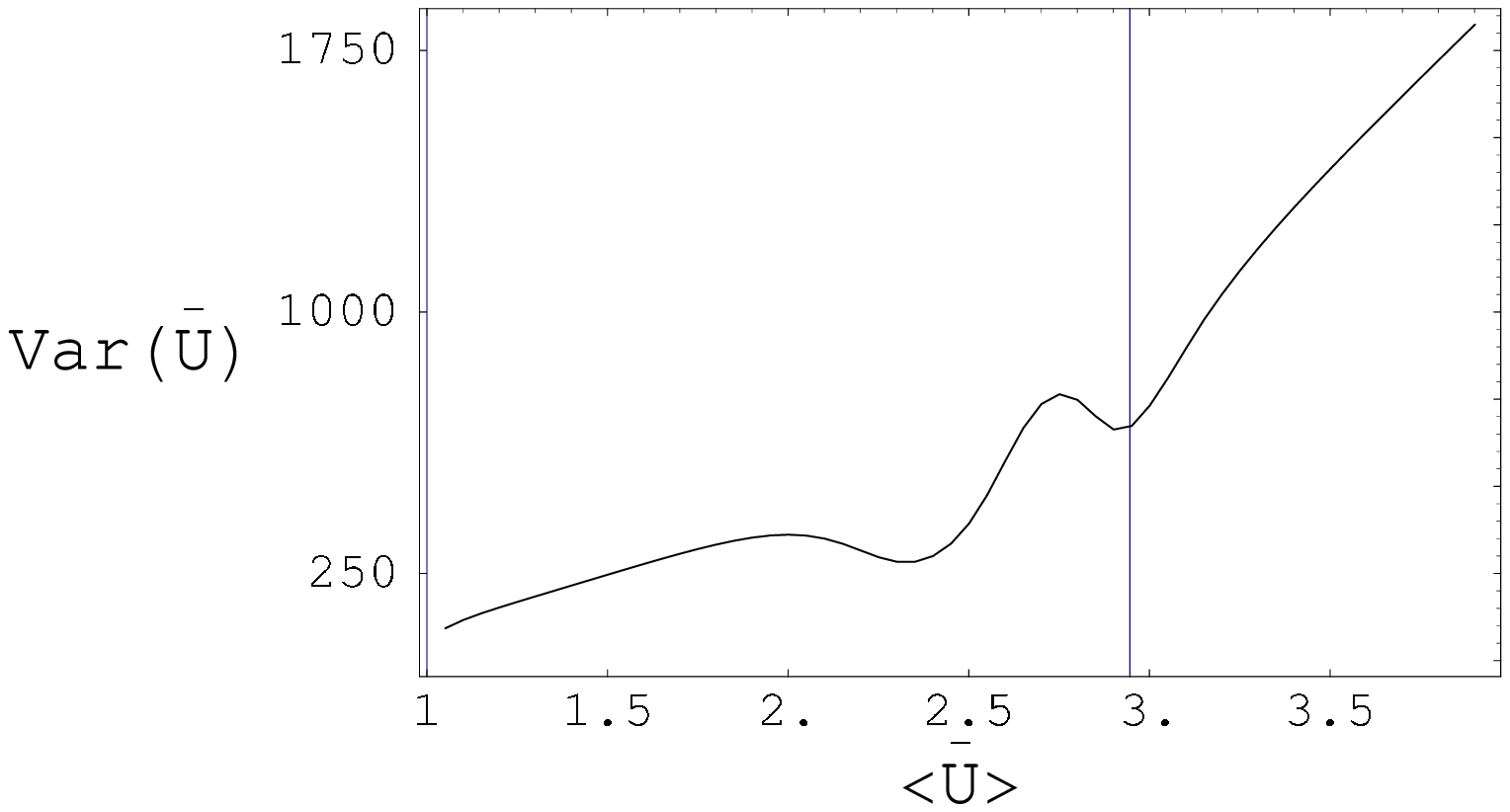} & \includegraphics[width=40mm,height=40mm]{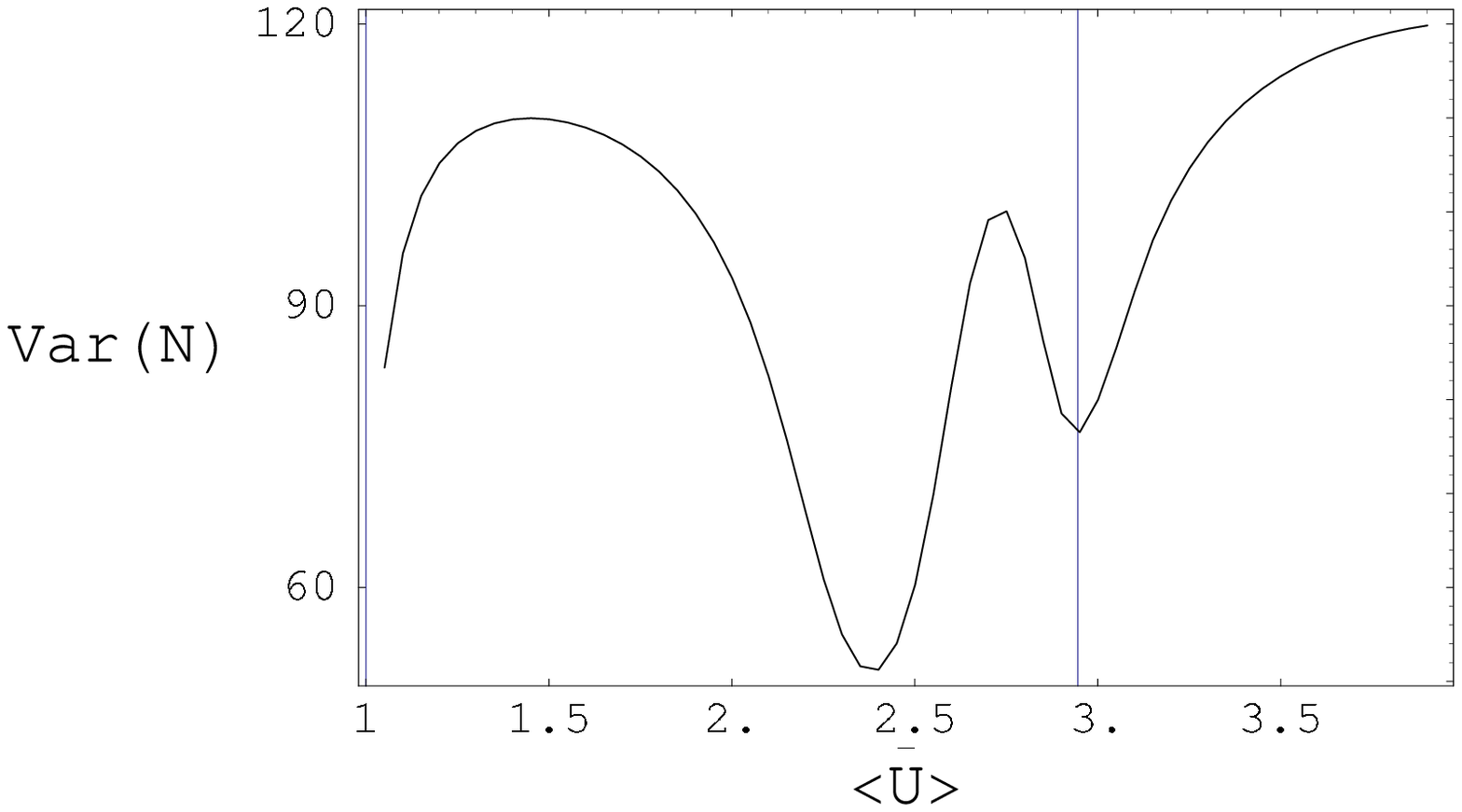} \\
\includegraphics[width=40mm,height=40mm]{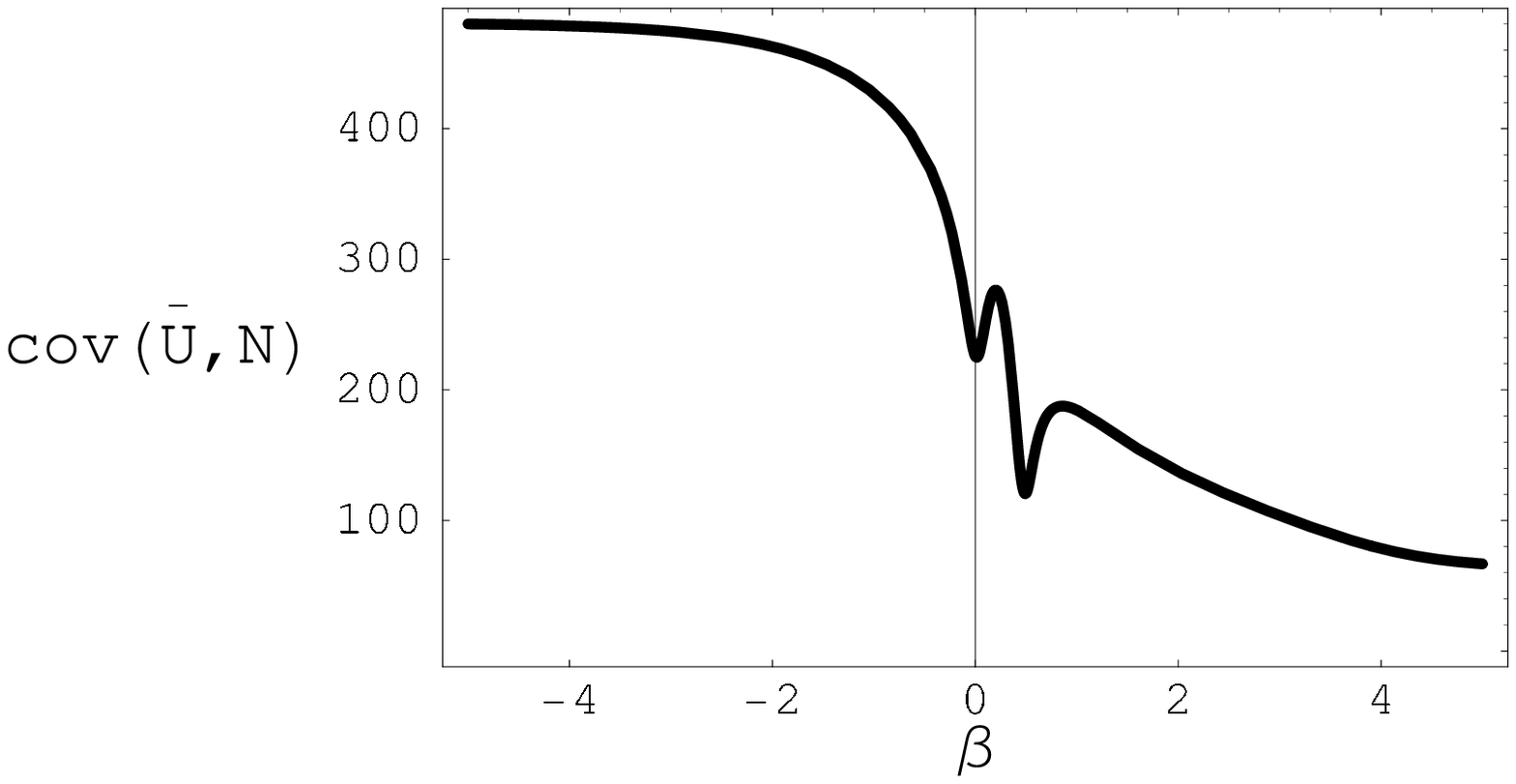} & \includegraphics[width=40mm,height=40mm]{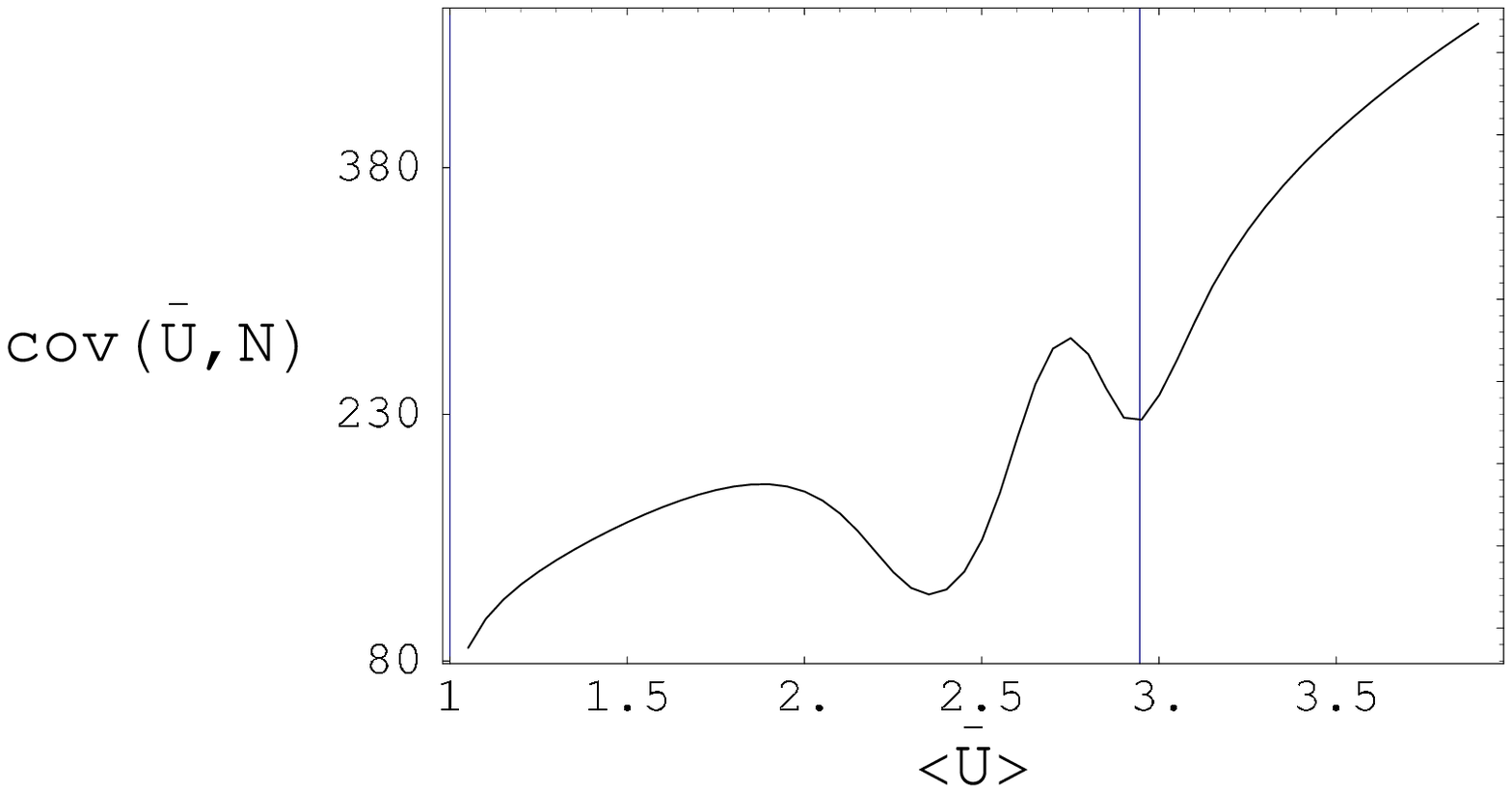}
\end{tabular}
\caption{\label{fig:fg5b} The variance of disuility {\it Var}$\,(\bar{U})$ {\it vs.} inverse temperature $\beta$ (upper left), the variance of the number of individuals {\it Var}$\,(N)$ {\it vs.} $\beta$ (upper right), {\it Var}$\,(\bar{U})$ {\it vs.} $\bar{U}$ (middle left), {\it Var}$\,(N)$ {\it vs.} $\bar{U}$ (middle right), the covariance of $\bar{U}$ and $N$, {\it cov}$\,(\bar{U},N)$ {\it vs.} $\beta$ (lower left) and {\it cov}$\,(\bar{U},N)$ {\it vs.} $\bar{U}$ (lower right) for the spider-web graph with 1 hub and 4 spokes.
The values of $\langle\bar{U}\rangle=2.954$ corresponding to $\beta=0$ are indicated by vertical lines for figures with $\bar{U}$ on the abcissae.}
\end{center}
\end{figure}

\begin{figure}[h]
\begin{center}
\begin{tabular}{cc}
\includegraphics[width=40mm,height=40mm]{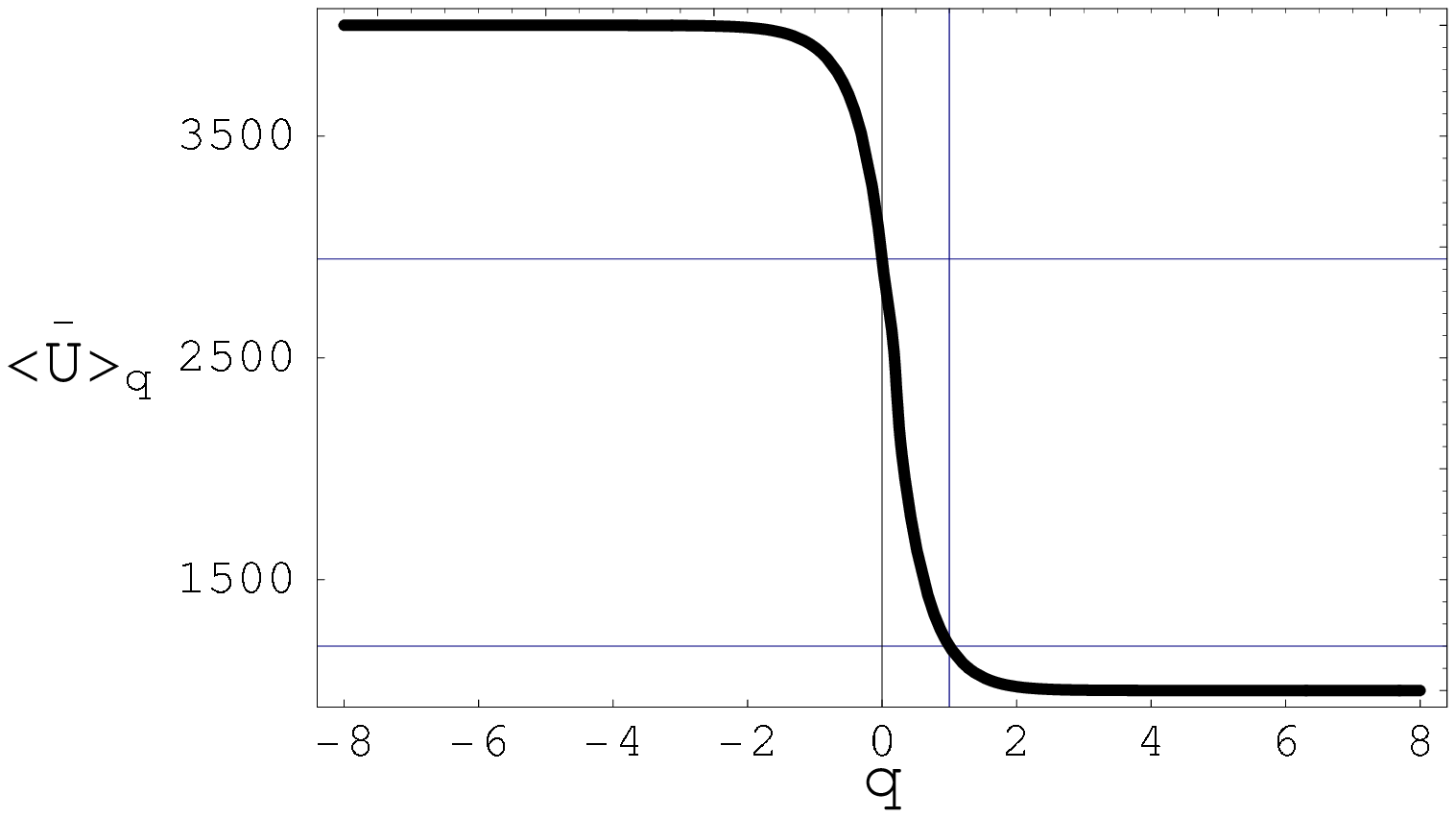} & \includegraphics[width=40mm,height=40mm]{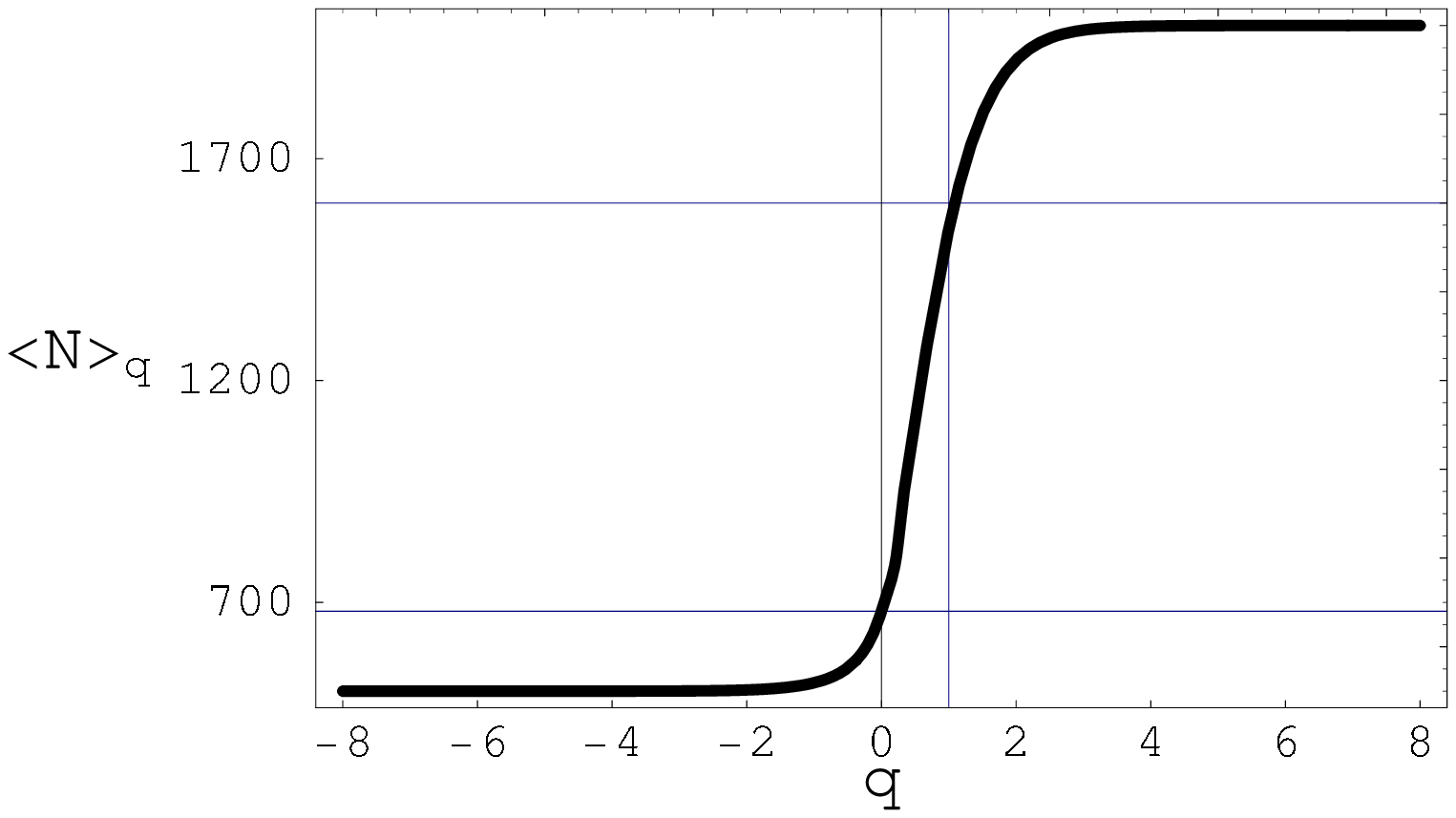} \\
\includegraphics[width=40mm,height=40mm]{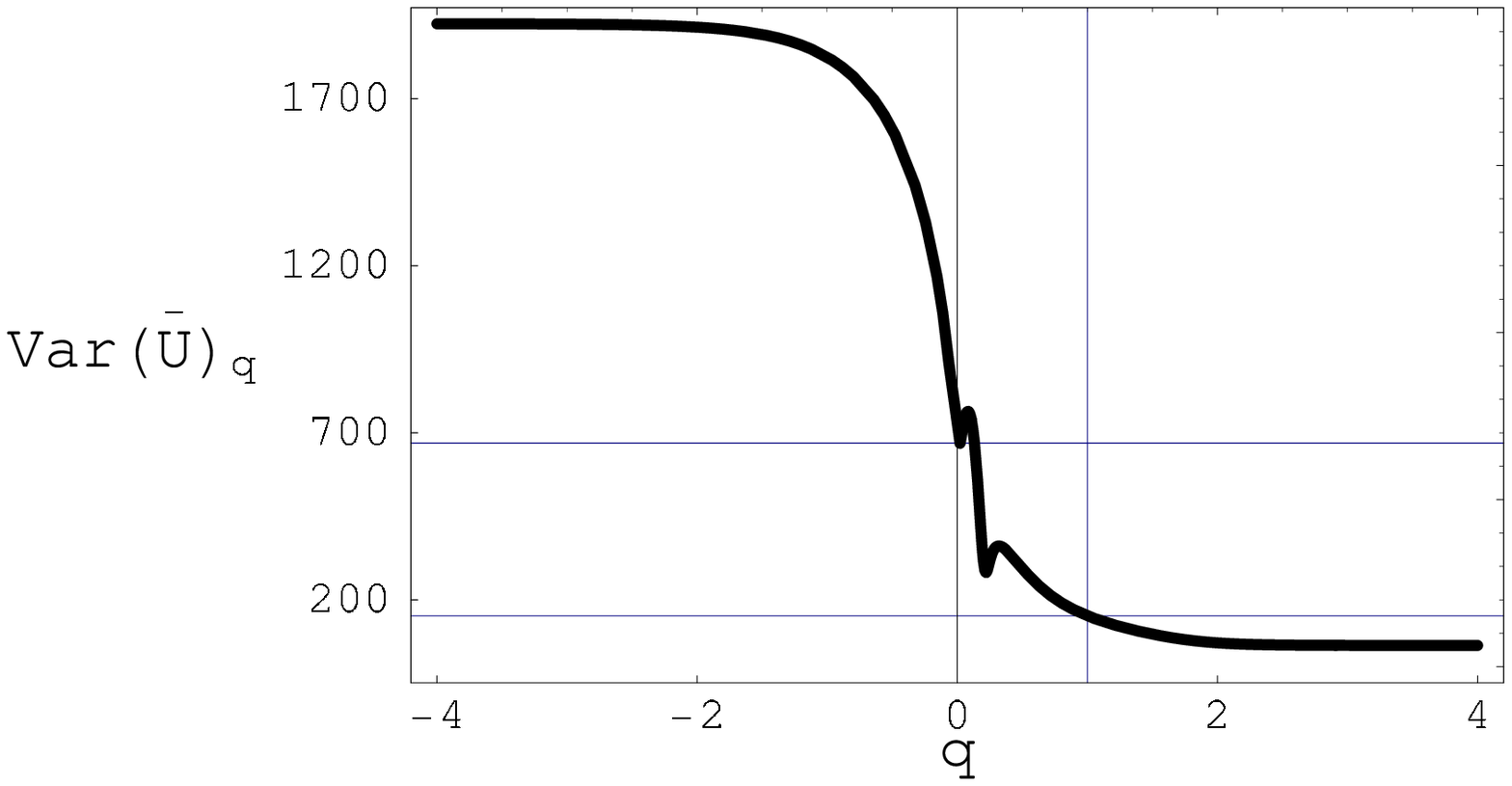} & \includegraphics[width=40mm,height=40mm]{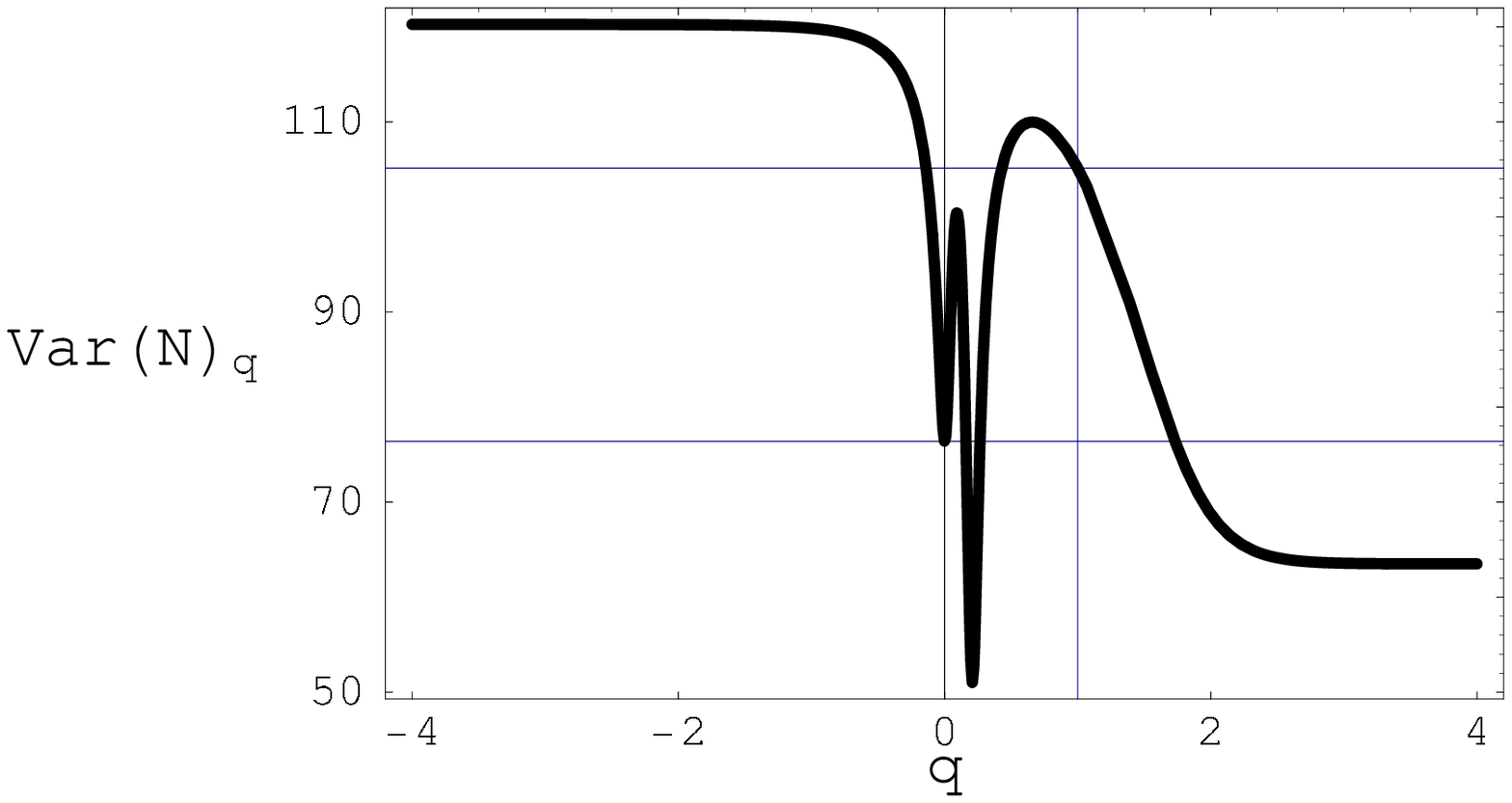} \\
\includegraphics[width=40mm,height=40mm]{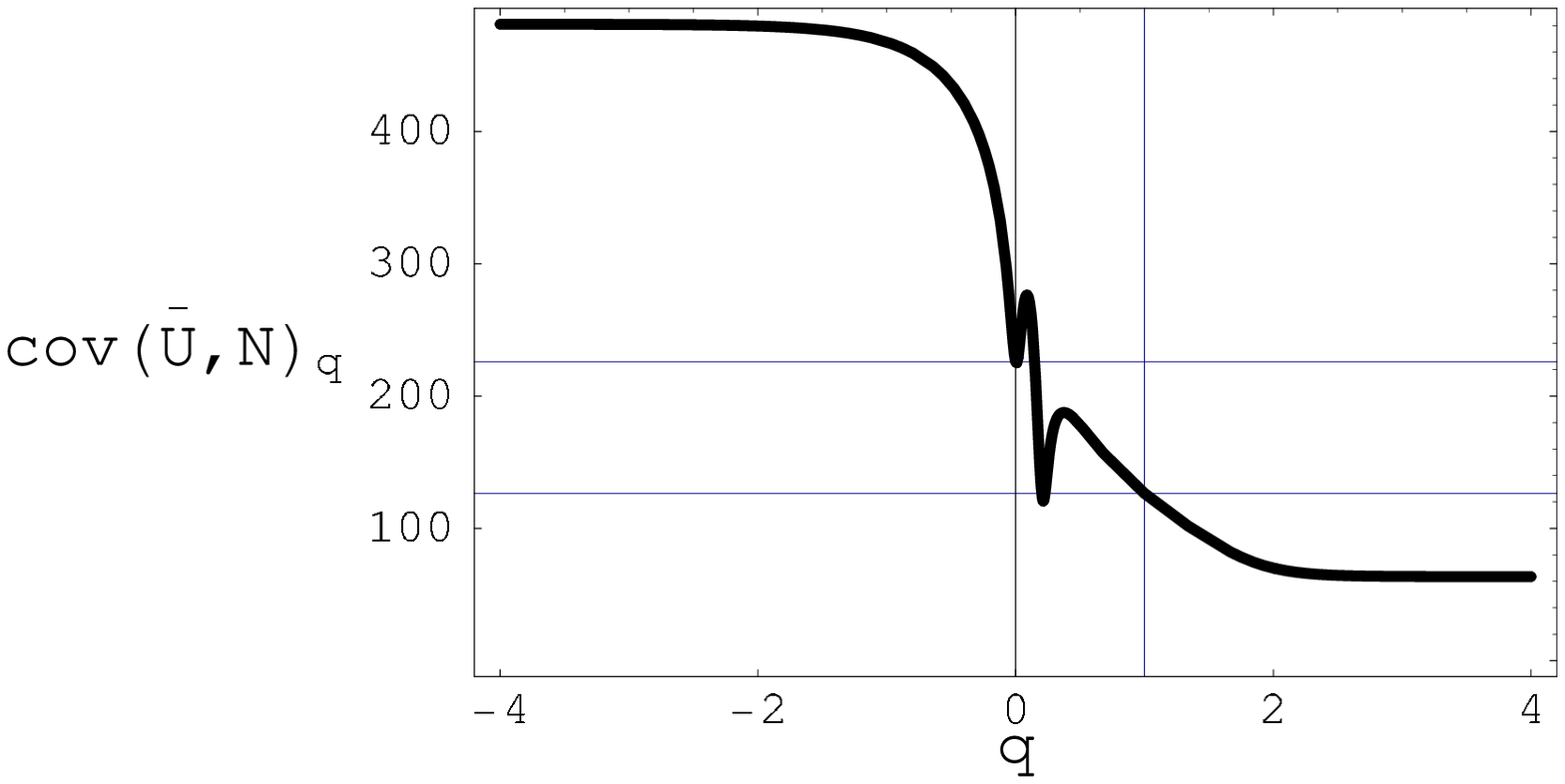} & \includegraphics[width=40mm,height=40mm]{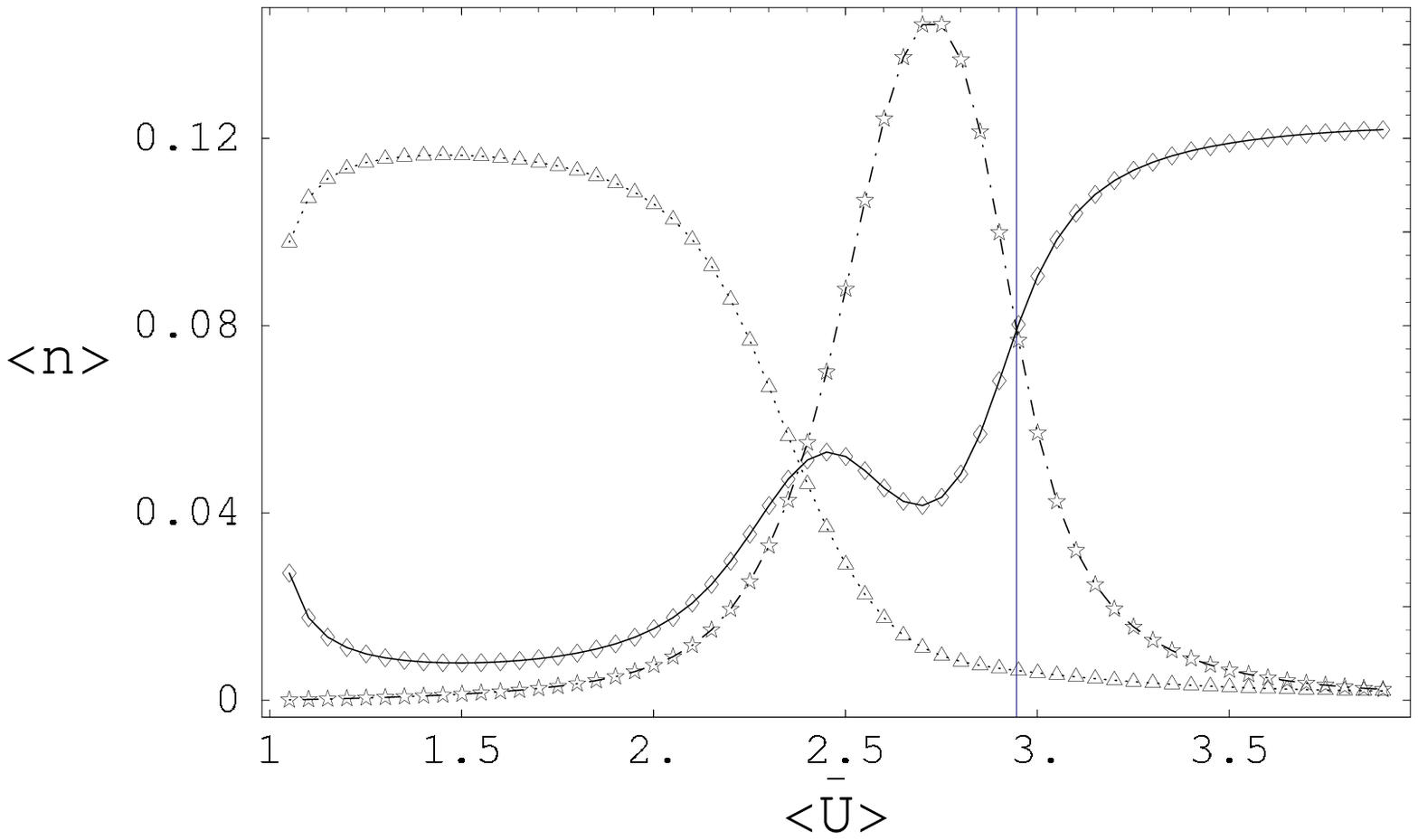}
\end{tabular}
\caption{\label{fig:fg5c} The $q$-mean disutility $\langle\bar{U}\rangle_q$ {\it vs.} $q$ (upper left) and the $q$-mean number of individuals $\langle N\rangle_q$ {\it vs.} $q$ (upper right), the $q$-variance of disutility {\it Var}$\,(\bar{U})_q$ (middle left), the $q$-variance of the number of individuals {\it Var}$\,(N)_q$ (middle right) and the $q$-covariance of $\bar{U}$ and $N$ {\it cov}$_q\,(\bar{U},N)$ (bottom left) for the spider-web graph with 1 hub and 4 spokes. The ratio of passengers in a given type of market as a function of mean disutility is displayed in the lower right panel. The vertical line at $\langle\bar{U}\rangle=2.954$ corresponds to $\beta=0$.}
\end{center}
\end{figure}

For this topology, the one-passenger disutility in the $k$-th market is equal to
\begin{eqnarray}
\bar{U}_{k_i} & = & \left\{\begin{array}{lc}
    2v, & \mbox{$O$ and $D$ spokes $\;\;\;\;$ (case (i))} \\
    v,  & \mbox{otherwise $\;\;\;\;\;\;\;\;\;\;\;\;\;\;\;$ (case (ii))}
             \end{array}\right .
\label{eq32}
\end{eqnarray}
Denoting by $M_{1(2)}$ the number of markets for the case (i) and (ii), which for the total numbers of spokes $\Gamma-1$ are equal to $\Gamma^2-3\Gamma+2$ and $\Gamma(\Gamma-1)$, respectively, and using $(\ref{eq13})$, one gets for the partition function
\begin{eqnarray}
Z(\beta)=\prod_{k=1}^{M_1}e^{-2\beta vN_k}\prod_{k=M_1}^{M_1+M_2}e^{-\beta vN_k}
\label{eq33}
\end{eqnarray}
and, for the canonical enesemble, the mean disutility is equal to
\begin{eqnarray}
\langle\bar{U}\rangle=(2N_O+N_H)v
\label{eq34}
\end{eqnarray}
where $N_O=\sum_{k=1}^{M_1}N_k$ and $N_H=\sum_{k=M_1+1}^{M_1+M_2}N_k$ are total numbers of individuals for cases (i) and (ii), respectively.

For large $N_k$s, the grand partition function is equal to
\begin{eqnarray}
\Xi(\beta,\mu)=\frac{1}{(1-e^{\beta(\mu-2v)})^{M_1}}\cdot\frac{1}{(1-e^{\beta(\mu-v)})^{M_2}}
\label{eq35}
\end{eqnarray}
and the first moments of $\bar{U}$ and $N$ are equal to
\begin{eqnarray}
\langle\bar{U}\rangle & = & \frac{2vM_1}{e^{-\beta(\mu-2v)}-1}+\frac{vM_2}{e^{-\beta(\mu-v)}-1} \nonumber \\
\langle N\rangle & = & \frac{M_1}{e^{-\beta(\mu-2v)}-1}+\frac{M_2}{e^{-\beta(\mu-v)}-1}
\label{eq36}
\end{eqnarray}
Following the same procedure as for the maximum connectivity network and assuming $v=1$, one finds the variances:
\begin{eqnarray}
{\it Var}\,(\bar{U}) & = & \langle N\rangle\big(-2+\frac{5}{6}\langle N\rangle\big) \nonumber \\
                     & + & \langle\bar{U}\rangle\big(3-\frac{7}{6}\langle N\rangle+\frac{11}{24}{\langle\bar{U}\rangle}\big) \nonumber \\
{\it Var}\,(N) & = & \langle N\rangle\big(1-\frac{7}{12}\langle N\rangle\big) \nonumber \\
               & - & \langle\bar{U}\rangle\big(\frac{2}{3}\langle N\rangle-\frac{5}{24}\langle\bar{U}\rangle\big)
\label{eq37}
\end{eqnarray}
the covariance
\begin{eqnarray}
{\it cov}\,(\bar{U},N)=\frac{2}{3}\langle N\rangle^2+\langle\bar{U}\rangle\big(1-\frac{5}{6}\langle N\rangle+\frac{7}{24}\langle\bar{U}\rangle\big)
\label{eq38}
\end{eqnarray}
and the entropy
\begin{eqnarray}
S & = & \langle\bar{U}\rangle\ln\frac{(-12+\langle N\rangle-\langle\bar{U}\rangle)(2\langle N\rangle-\langle\bar{U}\rangle)}{(\langle N\rangle-\langle\bar{U}\rangle)(8+2\langle N\rangle-\langle\bar{U}\rangle)} \nonumber \\
  & - & \langle N\rangle\ln\frac{(-12+\langle N\rangle-\langle\bar{U}\rangle)(2\langle N\rangle-\langle\bar{U}\rangle)^2}{(\langle N\rangle-\langle\bar{U}\rangle)(8+2\langle N\rangle-\langle\bar{U}\rangle)^2} \nonumber \\
  & - &8\ln\frac{8}{8+2\langle N\rangle-\langle\bar{U}\rangle} \nonumber \\
  & - &12\ln\frac{12}{12-\langle N\rangle+\langle\bar{U}\rangle}.
\label{eq39}
\end{eqnarray}

As for the previous case, we performed numerical simulation for the hub-and-spoke network with $\Gamma=5$ (one hub and four spokes) and $\langle N\rangle=1000$ and show the results, together with analytic curves, in Figs \ref{fig:fg4a} and \ref{fig:fg4b}.
Quantitatively, the $\langle\bar{U}\rangle$, $\nu$ and $S$ with $\beta$ exhibit the same type of behaviour, including asymptotics, as observed for the maximal graph.
Dependence of $\langle\bar{U}\rangle$ on $\beta$ for the hub-and-spoke for small $\beta$ is steeper than for the maximum connectivity graph.
The maximum of entropy at $\beta=0$ is sharper for the hub-and-spoke than for the maximum connectivity graph.

Contrary to the maximum connectivity graph, the variance of $N$ does depend on temperature (Fig. \ref{fig:fg4b} upper and middle right) and decreases with temperature for $\beta>0$.
The same tendency is observed for variance of $\bar{U}$ (Fig. \ref{fig:fg4b} upper and middle left).
This behaviour reflects lower symmetry of the hub-and-spoke network, as compared to the maximum connectivity, which means that not all markets are identical.
There are two types of markets, spoke-spoke and spoke-hub (cases (i) and (ii) in eqn (\ref{eq32})) and relative occupancy of the second one increases with temperature, making the market probability distribution narrower, eventually reaching its minimum for $\beta\rightarrow 0$ ($T\rightarrow\infty$).
The mirror-like behaviour is observed for negative $\beta$ but the plateau values for {\it Var}$\,(\bar{U})$, {\it Var}$\,(N)$ and {\it cov}$\,(\bar{U},N)$ are different than for positive $\beta$.

Simulations of $q$-moments are displayed in Figs \ref{fig:fg4c}.
Again, simulating the $q$-dependence one reproduce all features of explicit $\beta$-dependences of moments and correlations observed in previous figures.

Fig. \ref{fig:fg4c} (bottom right) presents the ratio of passengers in a given market as a function of utility, where each point corresponds to one mean overall disutility.
Asterisks correspond to the market with disutility $v$, i.e. the spoke-hub market, and triangles to the spoke-spoke market with disutility $2v$.
The region of $\beta>0$ extends to the left of the vertical line at $\langle\bar{U}\rangle=1.630$ and $\beta<0$ - to the right.

\subsection{\label{ssec:level4}{Case study 3: the spider-web network}}

\begin{figure}[h]
\begin{center}
\includegraphics[width=80mm,height=80mm]{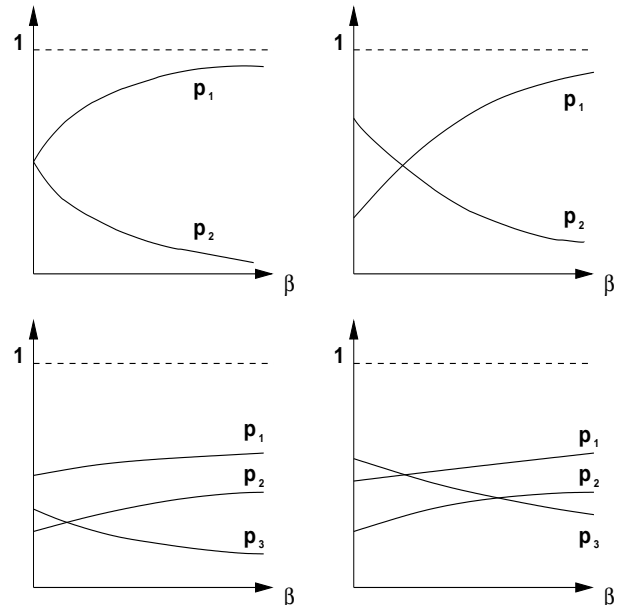}
\caption{\label{fig:fg6} Various $\beta$-dependent probability distributions leading to single or multiple extrema for second moments of disutilities and numbers of individuals, located at different values of $\beta$. Upper panels present two types of probability distributions for binary random variable: the maximum variance at $\beta=0$ (upper left) and the maximum variance at $\beta>0$ (upper right). Lower panels show two types of ternary random variables: one with maxima at $\beta=0$ and at $\beta>0$ (lower left) and with three local maxima, two of them for $\beta>0$ (lower right).}
\end{center}
\end{figure}

The spider-web network represents an intermediate case between the maximum connectivity and the hub-and-spoke toplogies, discussed above. 
But concerning symmetry, the spider topology is less symmetric than previous two, in a sense that there are three different types of markets in this network. 
We found analytic formulae also for this case but they are very lenghty and we do not print them here.
All the results shown in Figs \ref{fig:fg5a}, \ref{fig:fg5b} and \ref{fig:fg5c} come from analytic formulae and from direct simulations with the same conditions as for previous cases.
Many qualitative characteristics of the expected values and $q$-expected values of $\bar{U}$ and $N$, and also of the entropy $S$, are the same as for more symmetric topologies, but inspection of Figs \ref{fig:fg5b} and \ref{fig:fg5c} reveals also some interesting structures in $\beta$-dependence of the ($q$-)variances and ($q$-)covariances, deserving more attention.
Contrary to more symmetric topologies, one observes two minima: one at $\beta=0$ or $q=0$, and another one for $\beta>0$ or $q>0$, and two maxima at non-zero values.  
Qualitatively, such behaviour can be understood by following temperature dependence of probabilities, as illustrated in Fig. \ref{fig:fg6}, where probability distributions for binary and ternary random variables are shown as functions of $\beta$.
For the binary case (cf. Figs \ref{fig:fg6} upper), the widest distribution corresponds to equal probabilities $p_1=p_2$ which can happen either for $\beta=0$ or $\beta>0$. 
More complex structure of maxima for variance can easily arise for multinomial probability density function where probablilities intersect pairwise in more than one regions, or points of intersections clusterize.
An example of ternary random variable is presented in lower Figs \ref{fig:fg6}, where distributions for multiple maxima are illustratively drawn.

In order to find equilibrium values $\beta_{eq}$ and $\nu_{eq}$, one has to solve equations for $\langle\bar{U}\rangle$ and $\langle N\rangle$. 
We observe that degree of those equations depends on the symmetry of the network and is equal to the number of topologically different markets for given networks: for the maximum connectivity all markets are the same, for hub-and-spoke there are two types of markets and for the spider-web there are three types of markets.
We further observe that the number of local minima is less by one than the degree of those equations.
We strongly suspect that this is true for higher numbers of different markets.
At the moment we do not have any proof for the general case and leave this statement as a hypothesis.

Fig. \ref{fig:fg5c} (bottom right) presents the ratio of passengers in a given market as a function of utility, where each point corresponds to one mean overall disutility, or one temperature.
Triangles represent market where $O$ is located on the rim and $D$ in the hub, such that the set of possible routes for this market can be represented by the set of disutilities $\{v,2v,2v,3v,3v,4v,4v\}$.
Asterisks correspond to the market with $O$ and $D$ on the rim and non-adjacent, such that the set of routes is $\{2v,2v,2v,3v,3v,3v,4v,4v\}$.
Finally, diamonds are for the market of adjacent non-hub endpoints, where the routes are $\{v,2v,3v,3v,3v,4v,4v,4v\}$.
The region of $\beta>0$ extends to the left of the vertical line at $\langle\bar{U}\rangle=2.954$ and $\beta<0$ - to the right.
It is interesting to observe that for certain values of $\langle\bar{U}\rangle$, where the hub-rim market occupancy increases with temperature, two other sorts of markets tend to decrease, reaching local or global maxima.
Such non-trivial behaviour demonstrates potential power of the model developed here, for estimation of market occupancies.

\section{\label{sec:level5} Conclusions and final remarks}

Summarizing, we proposed a formalism, based on the equilibrium statistical thermodynamics, describing complex communication systems, consisting of sets of communicating nodes and individuals able to make decisions on how they travel.
In classical approach to the problem of choice making, based on game theory, the key concept is the utility function which quantifies choice probabilities.
Utility function was defined axiomatically, according to neoclassical economic theory.
We defined our system as a sum of markets, each market represented by a directed graph with two communicating endpoints and set of Hamilton paths connecting them.
Our networks were not embeded in metric spaces and thus the model deals with topologies and not with geometry.
The state of the system is defined by ascribing each individual to the market and specifying its choice of the comunication route.  
Disutility of the whole system, ascribed to each state of it, is defined as a sum of individual disutilities, integrated over all markets.
We have shown that such disutility is an additive and extensive stochastic function and therefore the probability measure on the state space is of Boltzmann type, thus reproducing the multinomial logit choice model, known from decision theory, using quite general and fundamental reasoning.
Further, we considered statistical ensembles: the microcanonical, canonical, grand canonical and super canonical.
The first three are commonly known in statistical mechanics and are rather straightforward to construct and use, whereas the super canonical ensemble incorporates random network topology and needs much further and more detailed specifications.
We did not randomize intensive variables and kept them as nonrandom variables or control parameters.
Generalizations in this direction and using random intensive fields of temperature, pressure etc. seem quite natural and are certainly ahead of us in further developments. 
Incorporation of geometry, i.e. using metric characteristics of connections like distances, shapes etc., are quite straightforward but not equally relevant in all applications.
In particular, for airline networks, topology plays principal role, as long as intercontinental connections are not concerned.

Following usual procedures, known from classical equilibrium therodynamics, we find the entropy and first moments of extensive random variables of the model and their correlations.
We also propose using escort distributions for monitoring of temperature evolution of moments of stochastic variables.

We applied our model to networks with specified topologies and we obtained analytical results for two topologies: the maximum connectivity and the hub-and-spoke, using rather simplistic disutility function, proportional to the number of route segments.
For this disutility we found physical analogy, as the hierarchy of routes of lengths increasing in steps of one unit can be directly mapped onto the one-dimensional quantum harmonic oscillator and exhibits close analogy to systems of bosons.
For the spider-web topology the formulae can be also found but are very complex.

Any network can be in principle simulated numerically.
We performed such simulations, in the framework of the grand canonical ensemble for all topologies, in order to crosscheck our analytical findings.
Our simulations, however, were done so far for rather small networks consisting of five nodes and average number of thousand decision makers. 
Evaluation of statistical sums requires summing over all possible distributions of decision makers over all Hamiltonian paths on the graph and this is in general very demanding computationally.
Since the world air traffic integrated over one day incorporates millions of passengers, its realistic simulations, even with tight constraints not accounted so far, will require more efficient software tools. 

We see considerable economic potential of our thermodynamic approach to the process of decision making in communicating population.
After complementing the model with constraints making it more suitable for real life applications in air transportation, as bunching of passengers in finite-size planes, realistic flight schedules, availability of air space and time slots, finite node transmittance, accounting for delays and other perturbations, and finally, using realistic disutility function, and many other factors, it can be used for monitoring and forecasting the state of the network.
Predictions for market occupancies and fluctuations of numbers of individuals depending on passenger's disutility and all types of correlations between disutilities and occupancies are useful for management of resources. 

We see three principal directions of development of our model:
\begin{itemize}
\item Network description in the framework of super-canonical ensemble, i.e. using random topology, and incorporation of geometric characteristics of the network by embeding the graph in metric space,
\item Monitoring of collective phenomena in the communication network using entropies, response functions and an approach based on the phase space uniformity, developed in ref. \cite{wislicki1},
\item Transforming our choice model into the full game, with airline operators and carriers being also active players in such game. 
This needs finding a coupling mechanism between utility functions of all parties, being e.g. typical predator-prey coupling of the Lotka-Volterra type, as a good starting point for further developments.
\end{itemize}

\begin{acknowledgments}
We acknowledge support granted to this work by Boeing Company in the framework of joint Boeing--ICM UW {\it Capstone} project.
\end{acknowledgments}

\appendix


\begin{thebibliography}{99}
\bibitem{neumann1} J. von Neumann and O. Morgenstern, \textit{Theory of Games and Economic Behaviour}, (Wiley, 1967, orig. ed. 1944)
\bibitem{hennings1} K. Hennings and W.J. Samuels, \textit{Neoclassical Economic Theory}, (Kluwer Academic Publishers, Boston, 1990)
\bibitem{bierlaire1} M. Bierlaire, http://roso.epfl.ch/mbi/ papers/ discretechoice/ paper.html, 1997
\bibitem{benakiva1} M.E. Ben-Akiva and S.R. Lerman, \textit{Discrete Choice Analysis: Theory and Applications to Travel Demand}, (MIT Press, Cambridge, Massachussetts, 1985) \\
                    S.P. Anderson, A. de Palma and J.-F. Thisse, \textit{Discrete Choice Theory of Product Differentiation}, (MIT Press, Cambridge, Massachussetts, 1992)
\bibitem{louviere1} J.J. Louviere, D.A. Hensher and J.D. Swait, \textit{Stated Choice Methods: Analysis and Applications}, (Cambridge University Press, Cambridge, New York, Melbourne, Madrid, 2000)
\bibitem{dobson1} G. Dobson and F.J. Lederer, Transp. Science {\bf 27}, 281 (1993)
\bibitem{straffin1} P.D. Straffin, \textit{Game Theory and Strategy}, (The Mathematical Association of America, Washington D.C., 1993)
\bibitem{touchette1} H. Touchette, Physica {\bf A305}, 84 (2002)
\bibitem{tsallis1} C. Tsallis, Braz. J. Phys. {\bf 29}, 1 (1999)
\bibitem{beck1} C. Beck and C. Cohen, Physica {\bf A322}, 267 (2003)
\bibitem{train1} K. Train, \textit{Discrete choice methods with simulations}, (Cambridge University Press, Cambridge, New York, Melbourne, Madrid, 2002), http://elsa.berkeley.edu/books/choice2.html
\bibitem{erdos1} P. Erd\"os and A. R\'enyi, Publ. Math. Debrecen {\bf 6}, 290 (1959), Publ Math. Inst. Hung. Acad. Sci. {\bf 5}, 17 (1960)
\bibitem{dorogovtsev1} S.N. Dorogovtsev, J.F.F. Mendes and A.N. Samukhin, http://arXiv.org, cond-mat/0204111
\bibitem{burda1} Z. Burda, J.D. Correia and A. Krzywicki, Phys Rev. {\bf E64}, 046118 (2001)
\bibitem{dorogovtsev2} S.N. Dorogovtsev and J.F.F. Mendes, \textit{Evolution of Networks: From Biological Nets to the Internet and WWW}, (Oxford University Press Inc., New York, 2003)
\bibitem{barabasi1} L. Albert and A.-L. Barab\'asi, Rev. Mod. Phys. {\bf 74}, 47 (2002)
\bibitem{beck2} C. Beck and A. Schl\"ogl, \textit{Thermodynamics of chaotic systems}, (Cambridge University Press, Cambridge, 1993)
\bibitem{wislicki1} W. Wi\'slicki, J. Phys. {\bf A34}, 4663 (2001)
\bibitem{majka1} A. Majka and W. Wi\'slicki, Physica {\bf A322}, 313 (2003)
\bibitem{renyi2} A. R\'enyi, \textit{Proceedings of the Fourth Berkeley Symposium on Mathematical Statistics and Probability, Vol. 1}, p. 547 (University of California Press, Berkeley, Los Angeles, 1961) \\
                 A. R\'enyi, \textit{Probability Theory}, (Akademiai Kiado, Budapest, 1970)
\bibitem{jackson} F. Jackson, Mess. Math. {\bf 38}, 57 (1909)
\end{thebibliography}
\end{document}